\documentclass[aps,preprint,nofootinbib,hyperref,include graphics]{revtex4}%
\usepackage{amsfonts}
\usepackage{amsmath}
\usepackage{amssymb}
\usepackage{hyperref}
\usepackage{graphicx}%
\providecommand{\U}[1]{\protect\rule{.1in}{.1in}}
\providecommand{\U}[1]{\protect\rule{.1in}{.1in}}
\providecommand{\U}[1]{\protect\rule{.1in}{.1in}}
\providecommand{\U}[1]{\protect\rule{.1in}{.1in}}
\providecommand{\U}[1]{\protect\rule{.1in}{.1in}}
\providecommand{\U}[1]{\protect\rule{.1in}{.1in}}
\providecommand{\U}[1]{\protect\rule{.1in}{.1in}}

\begin{document}
\preprint{ }
\title{Extended Rindler Spacetime and a New Multiverse Structure}
\author{Ignacio J. Araya}
\affiliation{Departamento de Ciencias F\'{\i}sicas, Universidad Andr\'{e}s Bello, Sazi\'{e}
2212, Santiago, Chile}
\author{Itzhak Bars}
\affiliation{Department of Physics and Astronomy, University of Southern California, Los
Angeles, CA, 90089-0484, USA,}
\affiliation{Perimeter Institute for Theoretical Physics, Waterloo, ON, N2L 2Y5, Canada}

\begin{abstract}
This is the first of a series of papers in which we use analyticity properties
of quantum fields propagating on a spacetime to uncover a new multiverse
geometry when the classical geometry has horizons and/or singularities. The
nature and origin of the \textquotedblleft multiverse\textquotedblright\ idea
presented in this paper, that follow from the standard model coupled to
gravity, are different from other notions of a multiverse. Via analyticity we
are able to establish definite relations among the universes. In this paper we
illustrate these properties for the extended Rindler space, while black hole
spacetime and the cosmological geometry of mini-superspace (see Appendix B)
will appear in later papers. In classical general relativity, extended Rindler
space is equivalent to flat Minkowski space; it consists of the union of the
four wedges in $(u,v)$ light-cone coordinates as in Fig.(1). In quantum
mechanics, the wavefunction is an analytic function of $(u,v)$ that is
sensitive to branch points at the horizons $u=0$ or $v=0,$ with branch cuts
attached to them. The wavefunction is uniquely defined by analyticity on an
infinite number of sheets in the cut analytic $(u,v)$ spacetime. This
structure is naturally interpreted as an infinite stack of identical Minkowski
geometries, or \textquotedblleft universes\textquotedblright, connected to
each other by analyticity across branch cuts, such that each sheet represents
a different Minkowski universe when $(u,v)$ are analytically continued to the
real axis on any sheet. We show in this paper that, in the absence of
interactions, information doesn't flow from one Rindler sheet to another. By
contrast, for an eternal black hole spacetime, which may be viewed as a
modification of Rindler that includes gravitational interactions, analyticity
shows how information is \textquotedblleft lost\textquotedblright\ due to a
flow to other universes, enabled by an additional branch point and cut due to
the black hole singularity.

\end{abstract}

\maketitle
\tableofcontents

\newpage

\section{Extended Rindler spacetime \label{Rindler}}

A massive particle moving in a background spacetime with metric $g_{\mu\nu
}\left(  x\right)  $ is described by a worldline action
\begin{equation}
S=\int d\tau\left[  \frac{1}{2e\left(  \tau\right)  }g_{\mu\nu}\left(
x\left(  \tau\right)  \right)  \partial_{\tau}x^{\mu}\left(  \tau\right)
\partial_{\tau}x^{\nu}\left(  \tau\right)  -\frac{e\left(  \tau\right)  }%
{2}\mu^{2}\right]  . \label{Sg}%
\end{equation}
The einbein $e\left(  \tau\right)  $ is the gauge field for $\tau
$-reparametrization symmetry. Its equation of motion is a constraint that may
be written in terms of the canonical conjugate momentum $p_{\mu}\left(
\tau\right)  $ as, $g^{\mu\nu}\left(  x\right)  p_{\mu}p_{\nu}+{\mu}^{2}=0.$
When the system is quantized, the wavefunction in position space
$\varphi\left(  x^{\mu}\right)  $ must satisfy the quantum-ordered constraint
that takes the form of the Klein-Gordon equation in a curved background%
\begin{equation}
\left(  -\nabla^{2}+\mu^{2}\right)  \varphi\left(  x\right)  =0,\text{ with
}\nabla^{2}\varphi\equiv\frac{1}{\sqrt{-g}}\partial_{\mu}\left(  \sqrt
{-g}g^{\mu\nu}\left(  x\right)  \partial_{\nu}\varphi\left(  x\right)
\right)  . \label{laplace}%
\end{equation}

The case of $g_{\mu\nu}\left(  x\right)  $ for Rindler spacetime commonly
refers to the coordinate frame of an observer undergoing constant proper
acceleration in an otherwise flat spacetime \cite{rindler}. Using lightcone
coordinates $\left(  u,v\right)  $ in flat spacetime, Rindler spacetime
corresponds to just region-I in Fig.(1), namely $u>0,v<0$, bounded by horizons
at $u=0$ or $v=0.$ This wedge of flat spacetime can be re-parametrized in
terms of Rindler coordinates, $y>0,~-\infty<t<\infty,$ as in
Eq.(\ref{uv-tyRegions}).%
\begin{center}
\includegraphics[
height=2.0116in,
width=1.9969in
]%
{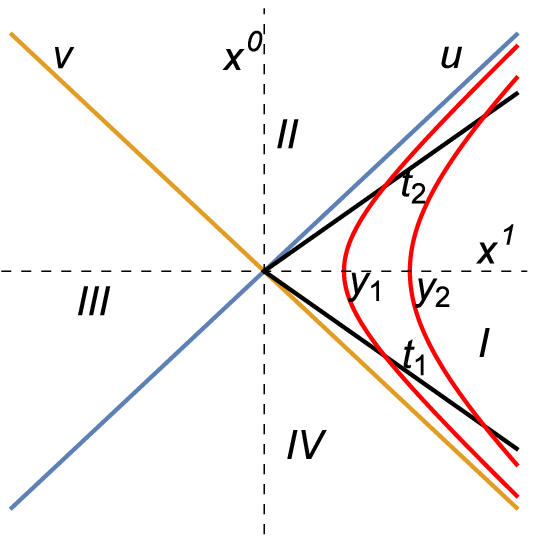}%
\\
{\protect\scriptsize Fig.(1)- Four regions of the map (}$u,v$%
{\protect\scriptsize ) to (}$t,y${\protect\scriptsize ).}%
\label{Fig-4regions}%
\end{center}

By extended Rindler spacetime we mean the union of the four regions I-IV shown
in Fig.1, which seems to be equivalent to the full Minkowski space. We will
motivate the study of the union of the four regions and will find new features
beyond just Minkowski space that are not apparent at the classical level (such
as geodesics). The new aspects emerge only at the quantum level as properties
of the first quantized wavefunction $\varphi$, or equivalently a property of
fields $\varphi$ that satisfy the Klein-Gordon equation $\left(  -\nabla
^{2}+\mu^{2}\right)  \varphi=0$ in extended Rindler spacetime.

We were motivated to study extended Rindler spacetime because we found that
the wavefunctions for the cases of cosmology as well as black hole physics
have the same features. These applications are consequences of the standard
model (SM) coupled to general relativity (GR) that includes a modest
modification that lifts the conventional theory to a locally scale invariant
(Weyl symmetric) version of GR+SM \cite{BST}. The conventional GR+SM at low
energies is recovered by fixing a Weyl gauge that introduces the dimensionful
parameters, the Newton constant $G_{N}$, dark energy $\Lambda$ and electroweak
scale $v_{EW}$, all coming from the same source \cite{BST}. This explains that
all dimensionful constants are the same everywhere in the observed spacetime
because they come from the same field that, when frozen to a constant by a
gauge fixing, fills the entire universe of the conventional GR+SM. The Weyl
symmetry geodesically completes the universe of the conventional theory at
high energies, beyond cosmological or black hole singularities, by including
previously missing patches of spacetime in a way analogous to enlarging the
Rindler patch in Fig.(1) to the extended Rindler spacetime. In cosmological
studies, using the Friedman equation at the classical level or the Wheeler
deWitt equation at the quantum level, it is found that the effective geometry
of mini-superspace - as a \textit{geometry in field space} that includes the
scale factor, curvature, anisotropy, and matter in the form of radiation and
the Higgs field - is closely related to the geometry of the extended Rindler
spacetime discussed in this paper, including some interactions that are not
part of the discussion here. In certain limits of the interactions the
mini-superspace geometry reduces mathematically exactly to the extended
Rindler space. Then a wedge in mini-superspace (region II) is related to the
expanding spacetime after the big bang, while the other regions I,III, and IV
play a role in determining a geodesically complete history of the universe.
These comments are amplified in Appendix-\ref{mini} to which the interested
reader may turn anytime without having to read the rest of the paper. Full
details will appear in separate papers \cite{barsMiniSuper}\cite{barsArayaBH}.
Until then, we will discuss the mathematical properties of the familiar
Rindler space and its extensions without any reference to mini-superspace,
cosmology or black holes. The applications outlined in this paragraph are
motivating factors, otherwise we emphasize that, this paper stands on its own
to discuss mainly the new quantum aspects of extended Rindler spacetime.

As seen by a traditional Rindler observer in region I, during the entire time
span of the Rindler universe, $-\infty<t<\infty$, geodesics of moving
particles remain only within the Rindler wedge (see section (\ref{geodesics}%
)). However, region-I is a geodesically incomplete spacetime from the
perspective of other observers, such as a Minkowski observer that uses $x^{0}
$ rather than $t$ as "time", or more generally a proper observer that uses
proper time $\tau$. So even though physical particles may escape/enter through
the horizons, and physical phenomena may exist in all the four regions in
Fig.(1), a Rindler observer is incapable of detecting such phenomena from
his/her own perspective. Explorers that wish to understand the deeper nature
of space-time beyond their own limited observational capabilities must
therefore consider all possible observers, not only those observers limited by
information available in some chosen coordinate system. Examples of observers
with limited capabilities of observation due to geodesically incomplete
coordinate systems include an observer outside of a black hole that is similar
to a Rindler observer. With this thought in mind, in this paper we are
interested in the "extended Rindler space" that consists of the geodesically
complete union of the four regions in Fig.(1). This means that, in the absence
of interactions, extended Rindler space is essentially flat Minkowski space.
Indeed this is true in classical physics. However, in quantum physics, we will
show that the wavefunctions of particles are sensitive to aspects of extended
Rindler space that classical physics cannot capture even with geodesically
complete spacetime. Wavefunctions for particles in first quantization amount
to fields. Therefore, as a first exercise, we study here scalar fields in the
background of extended Rindler space.

Rindler geometry has a long history of applications including the Unruh effect
\cite{unruh}-\cite{higuchi}, therefore, it is inescapable that some of our
discussion below overlaps old analyses. But for completeness, as well as for
establishing notation and conceptual background, we include in this paper some
familiar material along with our newer ideas to help the reader follow our
views on the multiverse aspects of extended Rindler spacetime that becomes
apparent only at the quantum level. The same approach will be used in future
papers to make similar cases for black holes and cosmology for which the
discussion and results in this paper are a prelude toward the more complicated
multiverse nature of geodesically complete cosmological spacetimes
\cite{barsMiniSuper} and eternal black hole spacetimes \cite{barsArayaBH}.
Therefore, in the present paper we wish to provide sufficient details to build
up the ideas through the simpler case of the extended Rindler spacetime
without interactions.

Minkowski spacetime in 1+1 dimensions\footnote{We focus on 1+1 dimensions for
simplicity; this is easily generalized to any number of dimensions.}, $\left(
x^{0},x^{1}\right)  ,$ may be re-written in terms of lightcone coordinates
$\left(  u,v\right)  ,$
\[
u\equiv x^{0}+x^{1},\;v\equiv x^{0}-x^{1},\text{ or }x^{0}=\frac{u+v}%
{2},\;x^{1}=\frac{u-v}{2}.
\]
Rindler coordinates $\left(  t,y\right)  ,$ that are convenient to describe
each region separately, are given by a coordinate transformation
\begin{equation}
2y=-uv\;\;\;\text{and\ \ \ }e^{2t}\text{sign}\left(  y\right)  =-\frac{u}{v},
\label{tyRindleruv}%
\end{equation}
with $-\infty<t<\infty$ and $-\infty<y<\infty$. In the $\left(  t,y\right)  $
coordinates, the flat Minkowski metric takes the appearance of a curved
metric, $ds^{2}=g_{\mu\nu}dx^{\mu}dx^{\nu},$ with its corresponding Laplacian
as in Eq.(\ref{laplace}),
\begin{equation}%
\begin{array}
[c]{l}%
ds^{2}=-dudv=-\left(  2y\right)  dt^{2}+\left(  2y\right)  ^{-1}dy^{2}=\pm
e^{2\xi}\left(  -dt^{2}+d\xi^{2}\right)  ,\\
\nabla^{2}\varphi=-4\partial_{u}\partial_{v}\varphi=-\frac{1}{2y}\partial
_{t}^{2}\varphi+\partial_{y}\left(  2y\partial_{y}\varphi\right)  =\pm
e^{-2\xi}\left(  -\partial_{t}^{2}\varphi+\partial_{\xi}^{2}\varphi\right)  ,
\end{array}
\label{Laplace}%
\end{equation}
where $e^{2\xi}\equiv\left\vert 2y\right\vert ,$ and the $\left(  \pm\right)
=$sign$\left(  y\right)  $ refer to regions I\&III versus II\&IV. For the
transformation of Eq.(\ref{tyRindleruv}) it is useful to distinguish four
regions, I,II,III,IV, as indicated in Fig.(1). In various regions $\left(
t,y\right)  $ is related to $\left(  u,v\right)  $ as follows
\begin{equation}%
\begin{array}
[c]{lllll}%
I_{\left(  u>0,v<0,y>0\right)  }: & u=+\sqrt{2y}e^{t}=e^{t+\xi}, &
v=-\sqrt{2y}e^{-t}=-e^{-t+\xi}, & 2y=-uv, & e^{2t}=-\frac{u}{v},\\
II_{\left(  u>0,v>0,y<0\right)  }: & u=+\sqrt{-2y}e^{t}=e^{t+\xi}, &
v=+\sqrt{-2y}e^{-t}=+e^{-t+\xi}, & 2y=-uv, & e^{2t}=+\frac{u}{v},\\
III_{\left(  u<0,v>0,y>0\right)  }: & u=-\sqrt{2y}e^{t}=-e^{t+\xi}, &
v=+\sqrt{2y}e^{-t}=+e^{-t+\xi}, & 2y=-uv, & e^{2t}=-\frac{u}{v},\\
IV_{\left(  u<0,v<0,y<0\right)  }: & u=-\sqrt{-2y}e^{t}=-e^{t+\xi}, &
v=-\sqrt{-2y}e^{-t}=-e^{-t+\xi}, & 2y=-uv, & e^{2t}=+\frac{u}{v}.
\end{array}
\label{uv-tyRegions}%
\end{equation}
The sign of the square root, $\pm^{\prime}\sqrt{\left\vert 2y\right\vert }$
(which agree with the signs of $u$ and $v$), distinguishes region I from III
and II from IV. The square roots $\pm^{\prime}\sqrt{\left\vert 2y\right\vert
}$ appear in both the classical and quantum solutions of the extended Rindler
system. In particular, continuity of the solutions in the $\left(  t,y\right)
$ coordinates across the horizons in Fig.(1), require the inclusion of all
four Rindler regions.

An intuitive description of the extended Rindler geometry in classical physics
is partially conveyed by the following comments. The horizons, that form the
boundaries of the four regions, occur at either $u=0$ or $v=0.$ The $u=0$
horizons are indicated as the orange line in Fig.(1), where $-\infty<v<\infty$
and $t=-\infty,$ $y=0$; the $v=0$ horizons are indicated as the blue line in
Fig.(1), where $-\infty<u<\infty$ and $t=\infty,$ $y=0$. A foliation of the
$\left(  u,v\right)  $ plane is provided by either fixed values of $y$ or
fixed values of $t$ within each Rindler region separately. The case of
$y=-\frac{1}{2}uv=$fixed corresponds to hyperbolas in each $\left(
u,v\right)  $ region; red curves labeled by $y_{1,2}$ in Fig.(1), with
$0<y_{1}<y_{2}<\infty,$ are examples shown only in region I. The case of
$t=\frac{1}{2}\ln\left\vert u/v\right\vert $=fixed correspond to straight rays
that extend from the origin to infinity within each $\left(  u,v\right)  $
region; black rays in Fig.(1), labeled by $-\infty<t_{1}<t_{2}<+\infty,$ are
examples shown only in region I. In all regions $\left\vert y\right\vert $
increases uniformly from the center or horizons ($\left\vert y\right\vert =0$)
to the outer boundaries of the region at infinity ($\left\vert y\right\vert
=\infty$). On the other hand, going around in the counterclockwise direction
in Fig.(1), the Rindler $t$ that labels the rays increases from $-\infty$ to
$+\infty$ in region I, followed by a decrease from $+\infty$ to $-\infty$ in
region II, followed by an increase from $-\infty$ to $+\infty$ in region III,
and followed by a decrease from $+\infty$ to $-\infty$ in region IV.

It is important to emphasize that in region $I,$ the Minkowski time,
$x^{0}=\left(  u+v\right)  /2,$ \textit{increases}, while the Rindler time
(the $t$ that labels the rays) also increases counterclockwise from $-\infty$
to $+\infty$; however in region III the Minkowski time $x^{0}$
\textit{decreases} while the Rindler time $t$ increases counterclockwise from
$-\infty$ to $+\infty.$ This difference between regions I and III is important
in the interpretation of particle versus antiparticle quantum waves, and it
leads to an interchange of creation/annihilation symbols, $a\leftrightarrow
b^{\dagger},$ in the construction of the field in region-I versus region-III,
as exhibited later in Eq.(\ref{f100}) versus Eq.(\ref{f300}).

The rest of this paper is organized as follows. In section (\ref{geodesics})
we discuss the geodesics in the classical extended Rindler space. In section
(\ref{fieldMR}) we discuss the complete and orthonormal set of modes of the
Klein-Gordon equation in the extended Rindler background, construct the
general first quantized wavepackets and the second quantized quantum field,
insuring that these are continuous across horizons of the four Rindler
quadrants in Fig.(1). In section (\ref{analy}) we determine the analyticity
properties of the first quantized wavepackets and quantum field and show how,
by analytic continuation, these naturally take values in an infinite stack of
Minkowski sheets labeled by two integers, $\left(  n,m\right)  ,$ that
constitute the multiverse. In section (\ref{bc}) we impose boundary conditions
at the horizons of the four quadrants in the $\left(  0,0\right)  $ universe
to require that on this sheet the extended Rindler space is equivalent to
Minkowski space. By analyticity, this determines the boundary conditions on
all $\left(  n,m\right)  $ sheets of the multiverse, and we show that the
quantum oscillators at various sheets are related to each other by a specific
canonical transformation determined by analyticity. In section
(\ref{canonMink}) we display the multiverse directly in the Minkowski basis
and derive a very non-trivial canonical transformation that relates the
general level $\left(  n,m\right)  $ Minkowski field to the level-$\left(
0,0\right)  $ Minkowski field. This canonical transformation represents in the
Minkowski basis the analytic continuation of the field in the Rindler basis,
and it could not be obtained without going through the Rindler basis$.$ In
section (\ref{prob}) we study charge (or information) conservation and
unitarity and show that, even though there is a flux of information (or
charge) across the horizons of neighboring quadrants, charge is conserved
within each Rindler quadrant separately at each $\left(  n,m\right)  $
universe. From this we conclude that there is no leakage of information among
levels of the Rindler multiverse. In section (\ref{discuss}) we summarize the
essential message of this paper and then suggest that the multiverse structure
discussed here in the simple context of extended Rindler space is more general
and also emerges in any spacetime that has horizons, such as black holes,
including the Schwarzchild black hole and others. Furthermore, we argue that
in the presence of interactions, such as gravitational interactions
represented by a black hole, big bang and others, the levels of the multiverse
are no longer isolated from each other, and charge/information/probability do
leak from one level of the modified multiverse to any other level, as
discussed in other papers including the case of the Rindler-like geometry of
mini-superspace with interactions \cite{barsMiniSuper} and the case of an
eternal black hole \cite{barsArayaBH}. Appendix-\ref{appendix} gives details
of computations of information conservation and information fluxes across the
horizons and at asymptotic regions in each Rindler wedge. Appendix-\ref{mini}
is included to clarify and amplify the physically motivating factors outlined
at the beginning of this section, in particular in the case of cosmology where
the new multiverse idea should be relevant to the cyclic universe scenario.

\section{Geodesics \label{geodesics}}

Before discussing the first quantized wavefunction or equivalently the field,
in this section we study the geodesics in extended Rindler space. The purpose
is to first understand the motion of particles in the classical geometry. This
will provide a background to better understand the flux of charge or
information from the perspective of wavepackets. We will see in section
(\ref{analy}) that the wavefunction reveals a far richer geometry involving an
infinite stack of $\left(  u,v\right)  $ sheets with each sheet related to the
classical geometry.

The geodesics in a curved spacetime with metric $g_{\mu\nu}\left(  x\right)  $
can be computed by solving the equations of motion of a massive or massless
particle on a worldline $x^{\mu}\left(  \tau\right)  $ moving in the curved
background. The action on the worldline in the first order formalism is given
by, $S\left(  x\right)  =\int d\tau\{\dot{x}^{\mu}\left(  \tau\right)  p_{\mu
}\left(  \tau\right)  -\frac{e\left(  \tau\right)  }{2}[g^{\mu\nu}x\left(
\tau\right)  p_{\mu}\left(  \tau\right)  p_{\nu}\left(  \tau\right)  +\mu
^{2}]\}.$ The equation of motion for varying the einbein $\delta e\left(
\tau\right)  $ gives the on-shell constraint, and the equation of motion for
varying $\delta p_{\mu}$ gives the relation between the velocity and momentum.
After the variations, choosing the gauge $e\left(  \tau\right)  =1$ (due to
$\tau$-reparametrization), these equations take the form
\begin{equation}
g^{\mu\nu}\left(  x\right)  p_{\mu}p_{\nu}+\mu^{2}=0,\;\dot{x}^{\mu}=g^{\mu
\nu}\left(  x\right)  p_{\nu}. \label{eoms}%
\end{equation}
The equation of motion for varying $\delta x^{\mu}$ gives an expression for
$\dot{p}_{\mu}$ which amounts to a second order differential equation for
$x^{\mu}\left(  \tau\right)  .$ This is the geodesic equation. A first
integral of the geodesic equation is already contained in the constraint
equation, therefore it can be ignored and concentrate on solving just the
equations above in order to find the geodesic solution for $x^{\mu}\left(
\tau\right)  $ as a function of $\tau.$ Note that $\tau$ is invariant under
target spacetime reparametrizations so, unlike observer-dependent choices of
\textquotedblleft time\textquotedblright\ in target space-time, $\tau$ is an
unambiguous choice of \textquotedblleft time\textquotedblright\ as the
evolution parameter for the motion of the particle from the perspective of a
proper observer in the frame of the particle itself.

In the case of the flat 2D Minkowski metric, $ds^{2}=-dudv$, or $g_{\mu\nu
}=\eta_{\mu\nu}=\left(
\genfrac{}{}{0pt}{}{0}{-1/2}%
\genfrac{}{}{0pt}{}{-1/2}{0}%
\right)  ,$ the \textit{inverse} metric is $g^{\mu\nu}=\eta^{\mu\nu}=\left(
\genfrac{}{}{0pt}{}{0}{-2}%
\genfrac{}{}{0pt}{}{-2}{0}%
\right)  ,$ and the equations to be solved (\ref{eoms}) are, $\dot{u}%
=-2p_{v}=p^{u},$ $\dot{v}=-2p_{u}=p^{v},$ where $p^{u}\left(  \tau\right)
=k^{+}$ and $p^{v}\left(  \tau\right)  =k^{-}$ are constants of motion (due to
translation invariance of the action, or the $\dot{p}$ equations of motion),
while the constraint is, $-4p_{u}p_{v}+\mu^{2}=0=-k^{+}k^{-}+\mu^{2}$. So, the
geodesic solution is%
\begin{equation}
u\left(  \tau\right)  =k^{+}\tau+u_{0},\;v\left(  \tau\right)  =\frac{\mu^{2}%
}{k^{+}}\tau+v_{0},\;\text{and }-\infty<k^{+}<\infty, \label{u+v+tau}%
\end{equation}
where $\left(  u_{0},v_{0}\right)  $ is the initial position in the $\left(
u,v\right)  $ plane. By eliminating $\tau$ between the first two equations
this solution is rewritten as a straight line in the flat $\left(  u,v\right)
$ spacetime, $u\left(  \tau\right)  =\left(  k^{+}/\mu\right)  ^{2}v\left(
\tau\right)  +$constant$.$ Equivalently, the solution is plotted as a
parametric plot that amounts to a timelike straight line whose direction in
$\left(  u,v\right)  $ space is set by the timelike on-shell momentum,
$k^{\mu}=\left(  k^{+},k^{-}\right)  $, as shown in the upper left corners of
Figs.(2,3). Note that $k^{+}>0$ corresponds to a particle (both $k^{+}$ and
$k^{-}$ positive, so upward arrow) while $k^{+}<0$ corresponds to an
antiparticle (both $k^{+}$ and $k^{-}$ negative, so downward arrow).

In the case of the Extended Rindler metric, $ds^{2}=-\left(  2y\right)
dt^{2}+\left(  2y\right)  ^{-1}dy^{2}$, the \textit{inverse} metric is
$g^{\mu\nu}=\left(
\genfrac{}{}{0pt}{}{-1/2y}{0}%
\genfrac{}{}{0pt}{}{0}{2y}%
\right)  ,$ and the equations to be solved (\ref{eoms}) are, $\dot{t}\left(
\tau\right)  =-\frac{p_{t}\left(  \tau\right)  }{2y\left(  \tau\right)  },$
$\dot{y}\left(  \tau\right)  =2y\left(  \tau\right)  p_{y}\left(  \tau\right)
,$ where $p_{t}\left(  \tau\right)  =\omega$ is a constant of motion (due to
translation invariance, $t\left(  \tau\right)  \rightarrow t\left(
\tau\right)  +c,$ of the action, or the $\dot{p}_{t}$ equations of motion),
while the constraint is, $-\frac{p_{t}^{2}}{2y}+2yp_{y}^{2}+\mu^{2}=0.$ We
rewrite this constraint by substituting the expressions for the momenta in
terms of velocities,
\begin{equation}
-\frac{\omega^{2}}{2y\left(  \tau\right)  }+\frac{\dot{y}^{2}\left(
\tau\right)  }{2y\left(  \tau\right)  }+\mu^{2}=0. \label{constraintR}%
\end{equation}
So, $y\left(  \tau\right)  $ is given by the solution of a simple first order
differential equation while $t\left(  \tau\right)  $ is an integral over
$\frac{-\omega}{2y\left(  \tau\right)  },$
\begin{equation}
\dot{y}\left(  \tau\right)  =\pm\sqrt{\omega^{2}-2\mu^{2}y\left(  \tau\right)
},\;t\left(  \tau\right)  =t_{\ast}+\int_{\tau_{\ast}}^{\tau}d\tau^{\prime
}\frac{-\omega}{2y\left(  \tau^{\prime}\right)  }, \label{ttau}%
\end{equation}
where the sign change $\pm$ for the velocity $\dot{y}$ occurs at a specific
time, $\tau=\tau_{\ast},$ when $\dot{y}\left(  \tau\right)  $ vanishes, namely
at $y_{\ast}=y\left(  \tau_{\ast}\right)  =\frac{\omega^{2}}{2\mu^{2}}.$ The
solution is,
\begin{equation}
y\left(  \tau\right)  =-\frac{\mu^{2}}{2}\left(  \tau-\tau_{\ast}\right)
^{2}+\frac{\omega^{2}}{2\mu^{2}},\;\;t\left(  \tau\right)  =\frac{1}{2}%
\ln\left\vert \frac{\tau-\tau_{\ast}-\omega/\mu^{2}}{\tau-\tau_{\ast}%
+\omega/\mu^{2}}\right\vert +t_{\ast}. \label{geodRind}%
\end{equation}
where $\left(  \omega,\tau_{\ast},t_{\ast}\right)  $ are integration constants
determined by initial conditions.

Of course, the geodesics written in terms of $\left(  t\left(  \tau\right)
,y\left(  \tau\right)  \right)  $ in the extended Rindler space must be the
same as those written in terms of Minkowski space $\left(  u\left(
\tau\right)  ,v\left(  \tau\right)  \right)  $ given in Eq.(\ref{u+v+tau}).
Therefore, a more elegant solution is to compute $\left(  t\left(
\tau\right)  ,y\left(  \tau\right)  \right)  $ by using the map between the
Rindler and Minkowski coordinates in Eqs.(\ref{tyRindleruv},\ref{uv-tyRegions}%
) and inserting the geodesics in Eq.(\ref{u+v+tau}), as follows%
\begin{equation}%
\begin{array}
[c]{l}%
y\left(  \tau\right)  =-\frac{u\left(  \tau\right)  v\left(  \tau\right)  }%
{2}=-\frac{1}{2}\left(  \mu^{2}\tau^{2}+\left(  k^{+}+\mu^{2}/k^{-}\right)
\tau+u_{0}v_{0}\right)  ,\\
t\left(  \tau\right)  =\frac{1}{2}\ln\left\vert \frac{u\left(  \tau\right)
}{v\left(  \tau\right)  }\right\vert =\frac{1}{2}\ln\left\vert \frac{k^{+}%
\tau+u_{0}}{\frac{\mu^{2}}{k^{+}}\tau+v_{0}}\right\vert .
\end{array}
\label{geodRindler}%
\end{equation}
By comparing Eqs.(\ref{geodRind},\ref{geodRindler}) one can establish the
relation between the integration parameters in the two versions $\left(
\omega,\tau_{\ast},t_{\ast}\right)  $ versus $\left(  k,u_{0},v_{0}\right)  $
that provide different physical insights.

The parametric plots of the explicit Minkowski and Rindler solutions are given
in Figs.(2,3). The Minkowski plots appear in the upper left corner of these
figures while the Rindler plots appear in the main body of these figures.
These are each other's images according to the maps in Eq.(\ref{uv-tyRegions}%
). In both the Minkowski and Rindler plots the Rindler regions (I-IV)
traversed by the geodesics are also shown. The bending point $\left(  y_{\ast
},t_{\ast}\right)  $ where the Rindler plot turns around in region-I (or III)
occurs at $\tau=\tau_{\ast}.$

In these figures the arrows show the direction of motion as the proper time
$\tau$ increases uniformly from $\tau=-\infty$ to $\tau=+\infty$. Proper time
is the time used by an observer that travels in the frame of the particle.
Minkowski observers use $x^{0}\left(  \tau\right)  $ as \textquotedblleft
time\textquotedblright\ as measured by clocks in a static laboratory, while
Rindler observers use $t\left(  \tau\right)  $ for that purpose noting that
this is the clock that ticks in the frame of a laboratory experiencing
constant proper acceleration \cite{rindler}. A series of events that occur
sequentially according to proper time $\tau$, may have different
interpretations when they are rearranged according to one choice of
\textquotedblleft time\textquotedblright\ versus another. To see this in the
present case, first focus on the Minkowski plots in the upper left-hand corner
of each figure (2,3), where the upward (downward) trajectory indicates that
the Minkowski time $x^{0}\left(  \tau\right)  $ increases (decreases) as
proper time $\tau$ increases; hence the upward (downward) trajectory is for a
Minkowski particle (antiparticle) that has positive (negative) energy
$E=\left(  k^{+}+k^{-}\right)  /2$, since both $k^{\pm}$ are positive
(negative). In the Rindler images of these same geodesics, note that in
Fig.(2), $t\left(  \tau\right)  $ \textit{increases }during passage of a
Minkowski particle (antiparticle) through region I (III), so these are
interpreted as Rindler particles by Rindler observers in regions I and III. By
contrast, in Fig.(3), $t\left(  \tau\right)  $ \textit{decreases} during
passage of a Minkowski particle (antiparticle) through region-III (I), so
these are interpreted as antiparticles by Rindler observers in regions I or
III. So a Rindler observer's particle is a mixture of Minkowski particles and
antiparticles, and vice-versa. As is well known, this is expressed as a
Bogoliubov transformation for the corresponding particle creation/annihilation
operators, as re-derived below in Eqs.(\ref{bogol1},\ref{bogol2}).%
\[%
\begin{array}
[c]{cc}%
{\parbox[b]{2.5815in}{\begin{center}
\includegraphics[
height=2.3021in,
width=2.5815in
]%
{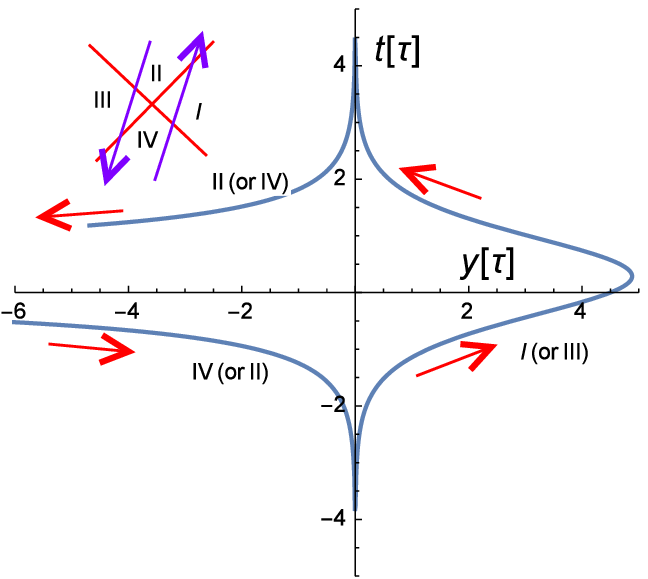}%
\\
{\protect\scriptsize Fig.(2)- Increasing }$t\left(  \tau\right)
${\protect\scriptsize \ in I or III.}%
\end{center}}}
&
{\parbox[b]{2.8487in}{\begin{center}
\includegraphics[
height=2.3021in,
width=2.8487in
]%
{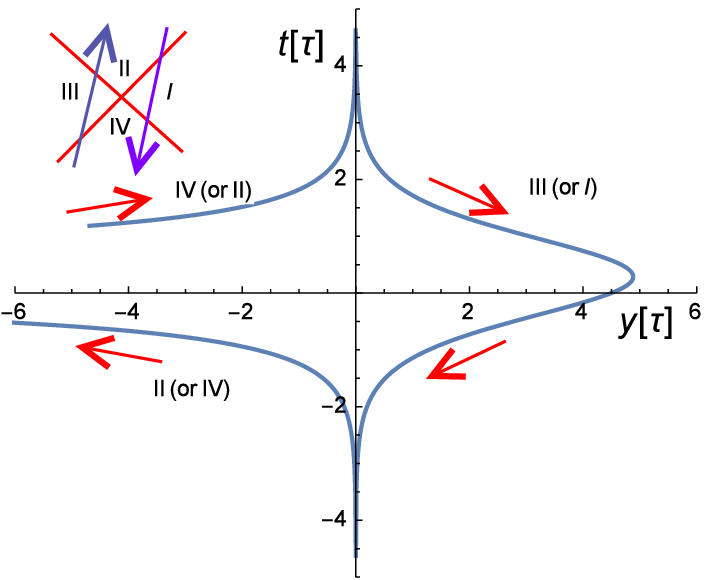}%
\\
{\protect\scriptsize Fig.(3)- Decreasing }$t\left(  \tau\right)
${\protect\scriptsize \ in I or III.}%
\end{center}}}
\end{array}
\]

The plots clearly show that the geodesics passing through regions I or III
take an infinite amount of Rindler time $t\left(  \tau\right)  .$ Hence, only
this portion of the complete geodesic is measurable by the Rindler observer in
region-I (or III). Meanwhile, the complete geodesic, that includes regions
beyond I (or III) is measurable by the Minkowski observer as shown in the
upper left corner of each figure. Of course, the proper time $\tau$ captures
the full geodesics in all curved spacetimes, so $\tau$ will be our preferred
choice of evolution parameter to discuss complete geodesics when analyzing the
geometry of more complicated cases, such as black holes.

Using $\tau$ as the evolution parameter, there is another way to intuitively
determine the complete trajectory for $y\left(  \tau\right)  $ without solving
it explicitly. The method is exhibited here because this approach can be
applied generally to any spacetime with a timelike Killing vector when an
explicit solution is not available. Consider the constraint in
Eq.(\ref{constraintR}) that the trajectory $y\left(  \tau\right)  $ must
satisfy and write it in the form of a vanishing non-relativistic Hamiltonian
\begin{equation}
\frac{\dot{y}^{2}}{2\mu^{2}}+V\left(  y\right)  =0,\;\text{with }V\left(
y\right)  =\left(  y-\frac{\omega^{2}}{2\mu^{2}}\right)  . \label{yconstraint}%
\end{equation}
In this form the constraint describes the dynamics of a non-relativistic
particle in 1-dimension with some potential energy $V\left(  y\right)  $, such
that its total energy (kinetic + potential) is constrained to be zero. This is
shown in Fig.(4) where the potential energy $V\left(  y\right)  $ in the
current case is plotted as a straight blue line.
\begin{center}
\includegraphics[
trim=0.025919in -0.020731in -0.025920in 0.020731in,
height=2.2468in,
width=2.2468in
]%
{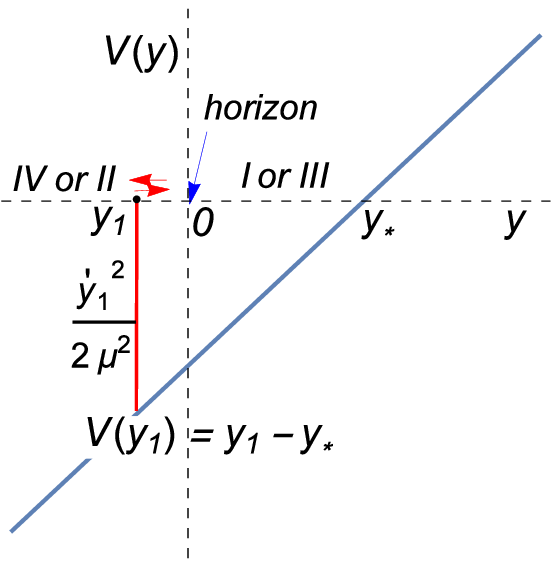}%
\\
{\protect\scriptsize Fig,(4)-}$V\left(  y\right)  ${\protect\scriptsize \ and
kinetic energy. The particle cannot move to the region }$y>y_{\ast}.$%
\label{figv(y)}%
\end{center}
For a more general case, such as a black hole, the potential $V\left(
y\right)  $ is a more general curve. The $0$ total energy level, which is
conserved throughout the motion for all $\tau$ (because of the timelike
Killing vector) is represented by the horizontal axis, and the evolving
kinetic energy at any point $y_{1}$ is constrained to be, $\dot{y}_{1}%
^{2}/2\mu^{2}=-V\left(  y_{1}\right)  ,$ corresponds to the length of the
vertical red segment that connects the 0 energy level and the value of the
potential at $y_{1}$. Without solving any equations, from this figure we see
intuitively that, as $y_{1}\left(  \tau\right)  $ evolves dynamically, a
classical particle/antiparticle is confined to the region, $y\left(
\tau\right)  \leq y_{\ast}=\omega^{2}/2\mu^{2}=-\frac{u_{\ast}v_{\ast}}{2},$
because its kinetic energy is positive, $\frac{\dot{y}^{2}}{2\mu^{2}%
}=-V\left(  y\right)  >0$. Its total-energy-conserving motion proceeds in the
direction of the velocity (sign$\left(  \dot{y}\left(  \tau\right)  \right)
$) indicated by the red arrows on the real axis$.$ The particle (antiparticle)
approaches from region $y\left(  \tau\right)  <0$ which is region-IV
(region-II); reaches the horizon at $y=0$ and proceeds to $y\left(
\tau\right)  >0$ which is region-I (region-III); it reaches a maximum at
$y_{\ast}$ within region-I (region-III); bounces back at $y_{\ast}$ at time
$\tau=\tau_{\ast}$, and then moves toward the horizon, to proceed to $y\left(
\tau\right)  <0$ which is region-II (region-IV), in the future of region-I (region-III).

This physical description of the trajectory, obtained only from the physical
interpretation of Fig.(4), clearly matches the behavior of the Rindler plots
in Figs.(2,3). This intuitive guide for the complete geodesics can of course
be complemented by analytic methods or approximations if necessary. This
approach has been applied to the case of the Schwarzchild black hole in
\cite{barsArayaJames} and will feature also in our upcoming work on black
holes \cite{barsArayaBH}.

Finally we comment on the geodesics of a massless particle. The zero mass
limit of the constraint in Minkowski basis, $k^{+}k^{-}=\mu^{2}\rightarrow0,$
has solutions $k^{\mu}=\left(  k^{+},\left(  k^{-}=0\right)  \right)  $ or
$\left(  \left(  k^{+}=0\right)  ,k^{-}\right)  .$ Therefore the massless
Minkowski geodesics are, $u\left(  \tau\right)  =k^{+}\tau+u_{0}$ and
$v\left(  \tau\right)  =v_{0},$ or $u\left(  \tau\right)  =u_{0}$ and
$v\left(  \tau\right)  =k^{-}\tau+v_{0}.$ These correspond to lines parallel
to either the $u$ or the $v$ axis that replace the slanted lines in the upper
left hand corners of Figs.(2,3). As for the massive case, for $k^{\pm}%
\gtrless0$ these are particle/antiparticle trajectories. Their images in
Rindler coordinates are,
\begin{equation}%
\begin{array}
[c]{l}%
y\left(  \tau\right)  =-\frac{u\left(  \tau\right)  v\left(  \tau\right)  }%
{2}=\left[  -\frac{1}{2}v_{0}\left(  k^{+}\tau+u_{0}\right)  \text{ ~or
}-\frac{1}{2}u_{0}\left(  k^{-}\tau+v_{0}\right)  \right]  ,\\
t\left(  \tau\right)  =\frac{1}{2}\ln\left\vert \frac{u\left(  \tau\right)
}{v\left(  \tau\right)  }\right\vert =\left[  \frac{1}{2}\ln\left\vert
\frac{k^{+}\tau+u_{0}}{v_{0}}\right\vert \text{ }\;\text{or\ \ }\frac{1}{2}%
\ln\left\vert \frac{u_{0}}{k^{-}\tau+v_{0}}\right\vert \right]  .
\end{array}
\end{equation}
By eliminating $\tau$ between these two equations one finds the geodesic
relation, $t\left(  \tau\right)  =\pm\frac{1}{2}\ln\left\vert y\left(
\tau\right)  \right\vert +c_{\pm}$, where the constants $c_{\pm}$ are fixed
with some initial conditions. The new parametric plots of $\left(  y\left(
\tau\right)  ,t\left(  \tau\right)  \right)  $ produce a deformation of the
curves in Figs.(2,3) to two possible branches such that the remaining branch
contains either the $t\rightarrow+\infty$ or the $t\rightarrow-\infty$ peak.
The remaining branches are separate curves disconnected from each other and
correspond to the plot of the functions, $t\left(  \tau\right)  =\pm\frac
{1}{2}\ln\left\vert y\left(  \tau\right)  \right\vert +c_{\pm}$.

\section{Minkowski free field in extended Rindler basis \label{fieldMR}}

Consider a complex scalar field in $1+1$ dimensional Minkowski spacetime,
$-\infty<x^{0}<\infty$ and $-\infty<x^{1}<\infty,$ that satisfies the massive
or massless Klein Gordon equation
\begin{equation}
\left(  \nabla^{2}-\mu^{2}\right)  \varphi\left(  x^{0},x^{1}\right)  =0.
\end{equation}
The well-known general solution \cite{BjorkenDrell} is a superposition of
relativistic plane waves, $e^{-iEx^{0}+ik^{1}x^{1}}/\sqrt{4\pi E}$, and their
complex conjugates, that form a properly normalized complete set of modes,
where $E=\sqrt{k_{1}^{2}+\mu^{2}}$ is the energy,%
\begin{equation}
\varphi\left(  x^{0},x^{1}\right)  =\int_{-\infty}^{\infty}dk^{1}\left(
A\left(  k^{1}\right)  \frac{e^{-iEx^{0}+ik^{1}x^{1}}}{\sqrt{4\pi E}%
}+B^{\dagger}\left(  k^{1}\right)  \frac{e^{iEx^{0}-ik^{1}x^{1}}}{\sqrt{4\pi
E}}\right)  . \label{PsiM1}%
\end{equation}
For this paper, it will be convenient to split the integral$\ $into two parts,
$\int_{-\infty}^{\infty}dk^{1}=\int_{0}^{\infty}dk+\int_{-\infty}^{0}dk,$
where $k^{1}$ has been renamed as $k.$ Changing $k\rightarrow-k$ in the second
integral and defining, $A_{\pm}\left(  k\right)  \equiv A\left(  \pm k\right)
$ with positive $k$, the field is re-written as
\begin{equation}%
\begin{array}
[c]{l}%
\varphi\left(  x^{0},x^{1}\right)  =\int_{0}^{\infty}dk\left(  A_{+}\left(
k\right)  \frac{e^{-iEx^{0}+ikx^{1}}}{\sqrt{4\pi E}}+A_{-}\left(  k\right)
\frac{e^{-iEx^{0}-ikx^{1}}}{\sqrt{4\pi E}}+hc_{A_{\pm}^{\dagger}\rightarrow
B_{\pm}^{\dagger}}\right)  ,\\
\;\;\;\;=\int_{0}^{\infty}dk\left(  A_{+}\left(  k\right)  \frac
{e^{-i\frac{E-k}{2}u}e^{-i\frac{E+k}{2}v}}{\sqrt{4\pi E}}+A_{-}\left(
k\right)  \frac{e^{-i\frac{E+k}{2}u}e^{-i\frac{E-k}{2}v}}{\sqrt{4\pi E}%
}+hc.\right)
\end{array}
\label{PsiM}%
\end{equation}
where \textquotedblleft$hc_{A^{\dagger}\rightarrow B^{\dagger}}$%
\textquotedblright\ stands for hermitian conjugates of the first two terms but
with $A_{\pm}^{\dagger}\left(  k\right)  $ replaced by $B_{\pm}^{\dagger
}\left(  k\right)  $ for a complex field. From now on we will sometimes
abbreviate this piece simply as \textquotedblleft$hc$\textquotedblright%
\ unless some clarification is needed. For a real field, we simply replace
$B_{\pm}^{\dagger}\left(  k\right)  $ by $A_{\pm}^{\dagger}\left(  k\right)  $ everywhere.

In \textit{classical field theory,} ($A_{\pm}\left(  k\right)  ,B_{\pm
}^{\dagger}\left(  k\right)  $) are complex functions of the positive momentum
$k.$ These ($A_{\pm}\left(  k\right)  ,B_{\pm}^{\dagger}\left(  k\right)  $)
could be fixed by initial/final boundary conditions that correspond to some
wave packets. In \textit{quantum field theory}, the ($A_{\pm}\left(  k\right)
,A_{\pm}^{\dagger}\left(  k\right)  $) and ($B_{\pm}\left(  k\right)  ,B_{\pm
}^{\dagger}\left(  k\right)  $) are pairs of annihilation/creation operators
for particles $\left(  A\right)  $ and antiparticles $\left(  B\right)  $
acting in the Fock space built on the Minkowski vacuum \cite{BjorkenDrell}.
\begin{equation}
\left[  A_{\pm}\left(  k\right)  ,A_{\pm^{\prime}}^{\dagger}\left(  k^{\prime
}\right)  \right]  =\delta_{\pm,\pm^{\prime}}\delta\left(  k-k^{\prime
}\right)  =\left[  B_{\pm}\left(  k\right)  ,B_{\pm^{\prime}}^{\dagger}\left(
k\right)  \right]  ,\;(A_{\pm}\left(  k\right)  \text{ or }B_{\pm}\left(
k\right)  )|0_{M}\rangle=0. \label{Mvac}%
\end{equation}

Now we would like to setup the equivalent general superposition of the same
field in terms of Rindler modes rather than the plane wave Minkowski modes.
Rindler modes $\varphi_{\pm}\left(  t,y\right)  $ are the complete set of
solutions to the Rindler Klein-Gordon equation given in (\ref{Laplace}). The
positive frequency modes, $\varphi_{\pm}\left(  t,y\right)  =e^{-i\omega
t}\varphi_{\mp\omega}\left(  y\right)  ,$ and their complex conjugate negative
energy modes $\varphi_{\pm}^{\ast}\left(  t,y\right)  =e^{i\omega t}%
\varphi_{\mp\omega}^{\ast}\left(  y\right)  $, satisfy the time independent
differential equation
\begin{equation}
\left(  \partial_{y}^{2}+\frac{1}{y}\partial_{y}+\frac{\omega^{2}}{4y^{2}%
}-\frac{\mu^{2}}{2y}\right)  \varphi_{\mp\omega}\left(  y\right)  =0.
\label{SDiffEq}%
\end{equation}
The linearly independent solutions $\varphi_{\mp\omega}\left(  y\right)  $ are
proportional to the Bessel functions $I_{\mp i\omega}\left(  \sqrt{2y\mu^{2}%
}\right)  .$ The \textit{normalized}\footnote{The so called \textquotedblleft
Klein-Gordon\textquotedblright\ dot product between two relativistic
wavefunctions $\varphi_{1},\varphi_{2}$ in curved spacetime is given by an
integral over a spacelike Cauchy surface, $\langle\varphi_{1}|\varphi
_{2}\rangle=-i\int d\Sigma_{\mu}\sqrt{-g}g^{\mu\nu}\left(  \varphi
_{1}^{\dagger}\partial_{\nu}\varphi_{2}-\partial_{\nu}\varphi_{1}^{\dagger
}\varphi_{2}\right)  .$ This relies on the Klein-Gordon current, $J_{1,2}%
^{\mu}=-i\sqrt{-g}g^{\mu\nu}\left(  \varphi_{1}^{\dagger}\partial_{\nu}%
\varphi_{2}-\partial_{\nu}\varphi_{1}^{\dagger}\varphi_{2}\right)  $ which is
conserved $\partial_{\mu}J_{1,2}^{\mu}\left(  x\right)  =0$. This dot product
is \textit{independent of the choice of the Cauchy surface}. In the Minkowski
case one chooses $d\Sigma_{x^{0}}=dx^{1}$ as a fixed-$x^{0}$ surface, while in
the Rindler case in regions I\&III one chooses $d\Sigma_{t}=dy$ as a fixed-$t$
surface since $i\partial_{t}$ is a Killing vector. Using $\sqrt{-g}=1$ and
$g^{tt}=-\left(  2y\right)  ^{-1},$ one finds%
\begin{equation}
\langle\varphi_{1}|\varphi_{2}\rangle=i\int_{0}^{\infty}\frac{dy}{2y}\left(
\varphi_{1}^{\dagger}\partial_{t}\varphi_{2}-\partial_{t}\varphi_{1}^{\dagger
}\varphi_{2}\right)  . \label{normR}%
\end{equation}
Hence, the basis functions $\varphi_{\mp\omega}^{\mp^{\prime}}\equiv
e^{\mp^{\prime}i\omega t}\phi_{\mp\omega}^{\mp^{\prime}}\left(  y\right)  $
are ortho-normalized as follows%
\[
\langle\varphi_{\mp_{1}\omega_{1}}^{\mp_{1}^{\prime}}|\varphi_{\mp_{2}%
\omega_{2}}^{\mp_{2}^{\prime}}\rangle=\delta^{\mp_{1}^{\prime},\mp_{2}%
^{\prime}}\delta_{\mp_{1},\mp_{2}}\left(  \pm_{1}^{\prime}2\omega_{1}\right)
\int_{0}^{\infty}\frac{dy}{2y}\left(  \phi_{\mp,\omega_{1}}^{\mp^{\prime}%
}\left(  y\right)  \right)  ^{\ast}\phi_{\mp,\omega_{2}}^{\mp^{\prime}}\left(
y\right)  =\pm_{1}^{\prime}\delta\left(  \omega_{1}-\omega_{2}\right)
\delta_{\mp_{1},\mp_{2}}\delta^{\mp_{1}^{\prime},\mp_{2}^{\prime}}.
\]
As usual, in relativistic field theory, the Klein-Gordon \textquotedblleft
norm\textquotedblright\ of basis functions is proportional to the
\textquotedblleft charge\textquotedblright\ associated with the conserved
current, while the sign of the frequency term in the exponent of the plane
wave, i.e. $\mp^{\prime},$ is minus the sign of the \textquotedblleft
charge\textquotedblright\ that distinguishes particle/antiparticle. In
accordance with this, note the overall $\pm^{\prime}$signs in front of the
delta functions in the final expression. \label{norm}} positive frequency
solutions in region $I$ are conveniently written in the form
\begin{equation}%
\begin{array}
[c]{l}%
\varphi_{\pm}\left(  t,y\right)  =e^{-i\omega t}\varphi_{\mp\omega}\left(
y\right)  =\frac{e^{-i\omega t}}{\sqrt{4\pi\omega}}\Gamma\left(  1\mp
i\omega\right)  \left(  \frac{\mu}{2}\right)  ^{\pm i\omega}I_{\mp i\omega
}\left(  \sqrt{2y\mu^{2}}\right) \\
\;\;\;\;=\frac{e^{-i\omega t}\left(  2y\right)  ^{\mp i\frac{\omega}{2}}%
}{\sqrt{4\pi\omega}}S_{\mp}\left(  2y\mu^{2}\right)  =\left\{
\begin{array}
[c]{c}%
\frac{u^{-i\omega}}{\sqrt{4\pi\omega}}S_{-}\left(  -\mu^{2}uv\right) \\
\frac{\left(  -v\right)  ^{i\omega}}{\sqrt{4\pi\omega}}S_{+}\left(  -\mu
^{2}uv\right)
\end{array}
\right.  .
\end{array}
\label{normalized}%
\end{equation}
In the last step, the region-I relations, $2y=-uv$ and $e^{2t}=-u/v,$ were
used to re-write $\varphi_{\pm}\left(  t,y\right)  $ in terms of $\left(
u,v\right)  $. The functions $S_{\mp}\left(  z\right)  $ are defined such that
$\lim_{z\rightarrow0}S_{\mp}\left(  z\right)  =1,$ when the argument
$z=2y\mu^{2}=-\mu^{2}uv$ vanishes. $S_{\mp}\left(  z\right)  $ are given by
the hypergeometric function $_{0}F_{1}$
\begin{equation}
S_{\mp}\left(  z\right)  =~_{0}F_{1}\left(  1\mp i\omega,\frac{z}{4}\right)
=\frac{\Gamma\left(  1\mp i\omega\right)  I_{\mp i\omega}\left(  \sqrt
{z}\right)  }{\left(  \frac{1}{2}\sqrt{z}\right)  ^{\mp i\omega}}=\sum
_{n=0}^{\infty}\frac{\left(  \frac{z}{4}\right)  ^{n}\Gamma\left(  1\mp
i\omega\right)  }{n!\Gamma\left(  n+1\mp i\omega\right)  }. \label{S}%
\end{equation}
$S_{\mp}\left(  z\right)  $ are entire analytic functions of $z$ in the finite
complex $z$-plane and have an essential singularity at $z=\infty$
\cite{essentialSing}. For the massless field $S_{\mp}\left(  z\right)  $ are
both replaced by $1$ since $\lim_{\mu\rightarrow0}S_{\mp}\left(  -\mu
^{2}uv\right)  \rightarrow1.$ The analytic properties of the modes
(\ref{normalized}) will play an essential part in our discussion in section
(\ref{analy}) where they will be discussed in detail.

We can now express the general solution $\varphi_{1}\left(  u,v\right)  $ in
region-I as the general superposition of the normalized basis in an analogous
form to the Minkowski case in Eq.(\ref{PsiM}),%
\begin{equation}%
\begin{array}
[c]{l}%
\varphi_{1}\left(  u,v\right)  =\int_{0}^{\infty}d\omega\left[
\begin{array}
[c]{c}%
a_{1-}\left(  \omega\right)  \varphi_{-}\left(  t,y\right)  +b_{1-}^{\dagger
}\left(  \omega\right)  \left(  \varphi_{-}\left(  t,y\right)  \right)
^{\ast}\\
+a_{1+}\left(  \omega\right)  \varphi_{+}\left(  t,y\right)  +b_{1+}^{\dagger
}\left(  \omega\right)  \left(  \varphi_{+}\left(  t,y\right)  \right)
^{\ast}%
\end{array}
\right] \\
\;\;\text{\ \ }=\int_{0}^{\infty}d\omega\left[  a_{1-}\left(  \omega\right)
\frac{u^{-i\omega}S_{-}\left(  -\mu^{2}uv\right)  }{\sqrt{4\pi\omega}}%
+a_{1+}\left(  \omega\right)  \frac{\left(  -v\right)  ^{i\omega}S_{+}\left(
-\mu^{2}uv\right)  }{\sqrt{4\pi\omega}}+hc_{a_{1\pm}^{\dagger}\rightarrow
b_{1\pm}^{\dagger}}\right]  .
\end{array}
\label{f100}%
\end{equation}
Note that both $a_{1\pm}$-coefficients are associated with Rindler wavepackets
of positive frequency $\omega$ (see Eq.(\ref{normalized})), so these represent
Rindler particles, while the $b_{1\pm}^{\dagger}$-coefficients are associated
with the complex conjugate wavepackets that have negative frequency and
represent Rindler antiparticles. In classical field theory (wavefunction in
first quantization) the coefficients $\left(  a_{1\pm},b_{1\pm}^{\dagger
}\right)  $ in region-I serve to specify some Rindler wavepackets that satisfy
some initial/final conditions. In quantum field theory, the pairs $\left(
a_{1\pm},a_{1\pm}^{\dagger}\right)  $ and $\left(  b_{1\pm},b_{1\pm}^{\dagger
}\right)  $ are creation-annihilation operators for Rindler
particles/antiparticles respectively, acting on the Fock space built on the
Rindler vacuum $|0_{R}\rangle$
\begin{equation}
\left[  a_{1\pm}\left(  \omega\right)  ,a_{1\pm^{\prime}}^{\dagger}\left(
\omega^{\prime}\right)  \right]  =\delta_{\pm,\pm^{\prime}}\delta\left(
\omega-\omega^{\prime}\right)  =\left[  b_{1\pm}\left(  \omega\right)
,b_{1\pm^{\prime}}^{\dagger}\left(  \omega\right)  \right]  ,\;(a_{1\pm
}\left(  \omega\right)  \text{ or }b_{1\pm}\left(  \omega\right)
)|0_{R}\rangle=0. \label{Rvac}%
\end{equation}
This defines the quantum field $\varphi_{1}\left(  u,v\right)  $ and its
hermitian conjugate $\varphi_{1}^{\dagger}\left(  u,v\right)  $ in region-I
$\left(  u>0,v<0\right)  .$

Similarly, the quantum field is constructed region by region in every region
of the extended Rindler space. For region-III $\left(  u<0,v>0\right)  $ the
modes of the Laplace equation (\ref{Laplace}) look the same as those in
region-I ($u>0,v<0)$, but one must introduce a new set of coefficients,
$\left(  a_{3\pm},a_{3\pm}^{\dagger}\right)  $ and $\left(  b_{3\pm},b_{3\pm
}^{\dagger}\right)  ,$ to write down the general solution (and its hermitian
conjugate $\varphi_{3}^{\dagger}\left(  u,v\right)  $)
\begin{equation}
\varphi_{3}\left(  u,v\right)  =\int_{0}^{\infty}d\omega\left[  b_{3-}%
^{\dagger}\left(  \omega\right)  \frac{\left(  -u\right)  ^{-i\omega}%
S_{-}\left(  -\mu^{2}uv\right)  }{\sqrt{4\pi\omega}}+b_{3+}^{\dagger}\left(
\omega\right)  \frac{v^{i\omega}S_{+}\left(  -\mu^{2}uv\right)  }{\sqrt
{4\pi\omega}}+hc_{b_{3\pm}\rightarrow a_{3\pm}}\right]  . \label{f300}%
\end{equation}
Note that $\varphi_{3}\left(  u,v\right)  $ is structurally very similar to
$\varphi_{1}\left(  u,v\right)  $ except that the $\left(  u,v\right)  $ in
region $I$ is replaced by the $\left(  u,v\right)  $ in region-III, including
a convenient change of signs $\left(  u,-v\right)  _{I}\rightarrow\left(
-u,v\right)  _{III},$ where the sign change is absorbed in the definition of
the corresponding coefficients. The extra signs do not affect the fact that
the modes in Eq.(\ref{f300}) are ortho-normalized solutions of
Eq.(\ref{Laplace}). In addition, in comparing $\varphi_{3}$ to $\varphi_{1},$
note that the first two terms in $\varphi_{3}$, which are positive frequency
solutions, are associated with Rindler antiparticle $b_{3\mp}^{\dagger}%
$-coefficients by contrast to the Rindler particle $a_{1\mp}$-coefficients in
$\varphi_{1}$, and vice-versa. The reasoning \cite{unruh} for this switch of
particle$\leftrightarrow$antiparticle interpretations of the Rindler waves in
region-I versus region-III, is that the Minkowski time $x^{0}$ increases as
the Rindler time $t$ increases in region-I, but $x^{0}$ decreases as $t$
increases in region-III; this was emphasized in the second paragraph following
Eq.(\ref{Laplace}), and is also evident in the contrast of the directions of
the arrows in the geodesics in Figs.(2,3). For the quantized field
$\varphi_{3}\left(  u,v\right)  ,$ the coefficients turn into pairs of
creation-annihilation operators, $\left(  a_{3\pm},a_{3\pm}^{\dagger}\right)
$ and $\left(  b_{3\pm},b_{3\pm}^{\dagger}\right)  ,$ for Rindler
particles/antiparticles respectively. These act on the Fock space built on the
same Rindler vacuum $|0_{R}\rangle$ as in Eq.(\ref{Rvac}),
\begin{equation}
\left[  a_{3\pm}\left(  \omega\right)  ,a_{3\pm^{\prime}}^{\dagger}\left(
\omega^{\prime}\right)  \right]  =\delta_{\pm,\pm^{\prime}}\delta\left(
\omega-\omega^{\prime}\right)  =\left[  b_{3\pm}\left(  \omega\right)
,b_{3\pm^{\prime}}^{\dagger}\left(  \omega\right)  \right]  ,\;(a_{3\pm
}\left(  \omega\right)  \text{ or }b_{3\pm}\left(  \omega\right)
)|0_{R}\rangle=0. \label{Rvac3}%
\end{equation}
Appropriate boundary conditions discussed in section (\ref{bc}) will provide
certain relations between the four complex functions in region-I, $\left(
a_{1\pm},b_{1\pm}^{\dagger}\right)  $ and those in region-III, $\left(
a_{3\pm},b_{3\pm}^{\dagger}\right)  ,$ consistently with the quantum
commutation relations above. Before boundary conditions are applied $\left(
a_{1\pm},b_{1\pm}^{\dagger}\right)  ,\left(  a_{3\pm},b_{3\pm}^{\dagger
}\right)  $ are treated as if they are unrelated to each other.

Similarly, one obtains the general solutions in regions II and IV of the
extended $\left(  u,v\right)  $ Rindler space, and then must insure that the
wavefunction in the full $\left(  u,v\right)  $ space is continuous across all
horizons. It turns out that the fields $\varphi_{2}\left(  u,v\right)  $ and
$\varphi_{4}\left(  u,v\right)  $ in regions II and IV respectively are fully
determined by analytic continuation of the fields $\varphi_{1}\left(
u,v\right)  $ and $\varphi_{3}\left(  u,v\right)  $ across the horizons. So,
there are no new $a,b$ coefficients beyond those already introduced above. The
full continuous field throughout the extended Rindler space $\left(
u,v\right)  $ is%
\begin{equation}
\varphi\left(  u,v\right)  =\varphi_{0}+\theta\left(  I\right)  \varphi
_{1}\left(  u,v\right)  +\theta\left(  II\right)  \varphi_{2}\left(
u,v\right)  +\theta\left(  III\right)  \varphi_{3}\left(  u,v\right)
+\theta\left(  IV\right)  \varphi_{4}\left(  u,v\right)  . \label{fullF}%
\end{equation}
The theta functions, $\theta\left(  I\right)  \equiv\theta\left(  u\right)
\theta\left(  -v\right)  ,$ etc. enforce the regions I-IV as defined in
Fig.(1) and Eq.(\ref{uv-tyRegions}). $\varphi_{0}$ is a constant zero mode
that is justified in Eq.(\ref{boundaA}). The expressions for $\varphi
_{1}\left(  u,v\right)  $ and $\varphi_{3}\left(  u,v\right)  $ given above,
as well as $\varphi_{2}\left(  u,v\right)  $ and $\varphi_{4}\left(
u,v\right)  $ obtained by analytic continuation are
\begin{equation}%
\begin{array}
[c]{l}%
\varphi_{1}\left(  u,v\right)  =\int_{0}^{\infty}d\omega\left[  a_{1-}\left(
\omega\right)  \frac{u^{-i\omega}S_{-}\left(  -\mu^{2}uv\right)  }{\sqrt
{4\pi\omega}}+a_{1+}\left(  \omega\right)  \frac{\left(  -v\right)  ^{i\omega
}S_{+}\left(  -\mu^{2}uv\right)  }{\sqrt{4\pi\omega}}+hc_{a_{1\pm}^{\dagger
}\rightarrow b_{1\pm}^{\dagger}}\right]  ,\\
\varphi_{2}\left(  u,v\right)  =\int_{0}^{\infty}d\omega\left[  a_{1-}\left(
\omega\right)  \frac{u^{-i\omega}S_{-}\left(  -\mu^{2}uv\right)  }{\sqrt
{4\pi\omega}}+b_{3+}^{\dagger}\left(  \omega\right)  \frac{v^{i\omega}%
S_{+}\left(  -\mu^{2}uv\right)  }{\sqrt{4\pi\omega}}+hc_{a_{1-}^{\dagger
}\rightarrow b_{1-}^{\dagger}}^{b_{3+}\rightarrow a_{3+}}\right]  ,\\
\varphi_{3}\left(  u,v\right)  =\int_{0}^{\infty}d\omega\left[  b_{3-}%
^{\dagger}\left(  \omega\right)  \frac{\left(  -u\right)  ^{-i\omega}%
S_{-}\left(  -\mu^{2}uv\right)  }{\sqrt{4\pi\omega}}+b_{3+}^{\dagger}\left(
\omega\right)  \frac{v^{i\omega}S_{+}\left(  -\mu^{2}uv\right)  }{\sqrt
{4\pi\omega}}+hc_{b_{3\pm}\rightarrow a_{3\pm}}\right]  ,\\
\varphi_{4}\left(  u,v\right)  =\int_{0}^{\infty}d\omega\left[  b_{3-}%
^{\dagger}\left(  \omega\right)  \frac{\left(  -u\right)  ^{-i\omega}%
S_{-}\left(  -\mu^{2}uv\right)  }{\sqrt{4\pi\omega}}+a_{1+}\left(
\omega\right)  \frac{\left(  -v\right)  ^{i\omega}S_{+}\left(  -\mu
^{2}uv\right)  }{\sqrt{4\pi\omega}}+hc_{b_{3-}\rightarrow a_{3-}}%
^{a_{1+}^{\dagger}\rightarrow b_{1+}^{\dagger}}\right]  .
\end{array}
\label{list}%
\end{equation}

The analytic continuation of the field across the horizons needs some
explanation. Compare $\varphi_{1}\left(  u,v\right)  $ to $\varphi_{2}\left(
u,v\right)  $ at the horizon that separates regions I\&II where $v=0$ and
$0<u<\infty.$ For continuity of the field we want to argue that, $\varphi
_{1}\left(  u,0\right)  =\varphi_{2}\left(  u,0\right)  $ for $u>0.$ In the
first half of the field it is clear that $a_{1-}\left(  \omega\right)
u^{-i\omega}S_{-}\left(  -\mu^{2}uv\right)  $ is analytically continued from
one side of the horizon, $v<0,$ to the other, $v>0,$ and noting that $S_{\pm
}\left(  z\right)  ,$ that satisfy $S_{\pm}\left(  0\right)  =1$ (see
Eq.(\ref{S})), are analytic entire functions in the finite complex plane. In
the second half of the field, continuity is not evident because different
coefficients $a_{1+}\left(  \omega\right)  $ and $b_{3+}^{\dagger}\left(
\omega\right)  $ appear in $\varphi_{1}\left(  u,v\right)  $ versus
$\varphi_{2}\left(  u,v\right)  $. However, in these apparently problematic
terms$,$ the factors $\left(  -v\right)  ^{i\omega}$ or $\left(  v\right)
^{i\omega}$ vanish as \textit{distributions} near $v\sim0$ on both sides of
the horizon. This is because $\left\vert v\right\vert ^{i\omega}=e^{i\omega
\ln\left\vert v\right\vert }$ oscillates wildly as $\left\vert v\right\vert
\rightarrow0,$ so the integral $\int_{0}^{\infty}d\omega$ can be computed by
the steepest descent method. Noting that only the neighborhood of $\omega=0$
can contribute to the leading behavior of such an integral, a typical smooth
integrand $F\left(  \omega\right)  $ can be approximated by its value near
$\omega=0$ to give
\begin{equation}
\lim_{v\rightarrow0}\int_{0}^{\infty}d\omega F\left(  \omega\right)
e^{i\omega\ln\left\vert v\right\vert }\simeq F\left(  0\right)  \lim
_{v\rightarrow0}\lim_{\varepsilon\rightarrow0^{+}}\int_{0}^{\infty}d\omega
e^{-\varepsilon\omega}e^{i\omega\ln\left\vert v\right\vert }=\lim
_{v\rightarrow0}\frac{F\left(  0\right)  }{i\ln\left\vert v\right\vert }=0.
\label{epsilon}%
\end{equation}
The $e^{-\varepsilon\omega}$ factor is introduced as a device to enforce the
integration region to remain close to $\omega=0,$ thus producing a convergent
integral in regions far from $\omega=0$. This shows that the terms involving
$\left(  -v\right)  ^{i\omega}$ or $\left(  v\right)  ^{i\omega}$ in
$\varphi_{1,2}\left(  u,v\right)  $ vanish at the $v=0$ horizon. Hence we have
shown that, despite the fact that $a_{1+}\left(  \omega\right)  $ and
$b_{3+}^{\dagger}\left(  \omega\right)  $ are different, the field is
continuous at the horizon and is given by
\begin{equation}
\varphi_{1}\left(  u,0\right)  =\varphi_{2}\left(  u,0\right)  =\int
_{0}^{\infty}d\omega\left[  a_{1-}\left(  \omega\right)  \frac{u^{-i\omega}%
}{\sqrt{4\pi\omega}}+b_{1-}^{\dagger}\left(  \omega\right)  \frac{u^{i\omega}%
}{\sqrt{4\pi\omega}}\right]  . \label{contin}%
\end{equation}

Even though the field is continuous, its derivative is discontinuous at the
$v=0$ horizon because $a_{1+}\left(  \omega\right)  $ and $b_{3+}^{\dagger
}\left(  \omega\right)  $ are different. However continuity of the derivative
is not required to have a solution of the Klein-Gordon equation in the
$\left(  u,v\right)  $ variables because the Laplacian (proportional to
$\partial_{u}\partial_{v}$) is linear rather than quadratic in $\partial_{u}$
and similarly in $\partial_{v}.$ The same argument applies at every horizon,
thus determining all the terms in $\varphi_{2,4}\left(  u,v\right)  $ by
analytic continuation from regions I and III. Therefore, we can state that the
full field $\varphi\left(  u,v\right)  $ as given in Eqs.(\ref{fullF}%
,\ref{list}) has just sufficient amount of continuity throughout the extended
Rindler space $\left(  u,v\right)  $ to be a solution of the Klein-Gordon
equation without any sources.

\section{Analyticity in the extended Rindler space \label{analy}}

We have already required some analyticity properties in $\left(  u,v\right)  $
space in order to establish the continuity of the field. The field discussed
in the previous section has branch points and corresponding branch cuts in the
complex $\left(  u,v\right)  $ planes. This defines an infinite number of
sheets in both the complex $u$-plane and complex $v$-plane. On the real axes
on each sheet, the field $\varphi\left(  u,v\right)  $ of Eqs.(\ref{fullF}%
,\ref{list}) takes on different values that are related to each other
continuously by analytic continuation from sheet to sheet. This presents
itself as a geometry consisting of an infinite stack of different real
Minkowski spaces $\left(  u,v\right)  $ on which the first quantized
wavefunction (i.e. the classical field) as well as the second quantized field
take on values. The branch points at $u=0$ or $v=0$ correspond to the horizons
within each Minkowski plane as shown in Fig.1. So the connection between the
infinite stack of real Minkowski spaces is precisely at the horizons. The
field is analytically continued from one Minkowski plane to another by going
slightly off the real axis at the horizons and winding around a branch cut in
either the $u$ or $v$ complex planes and then back to the real axis on a
different sheet. Hence, information could potentially flow at the horizons
from one Minkowski plane to any other Minkowski plane. This shows that there
are previously missed aspects of extended Rindler spacetime, namely the
natural presence of a multiverse structure consisting of a spacetime with an
infinite stack of Minkowski planes connected to each other at the horizons.
This was not evident from the classical metric or the geodesics discussed
before; it emerged only at the quantum level.

We now expand on these properties and clarify them. The analyticity properties
of the basis functions, $\varphi_{+}\left(  u,v\right)  \sim u^{-i\omega}%
S_{-}\left(  -\mu^{2}uv\right)  $ and $\varphi_{-}\left(  u,v\right)
\sim\left(  -v\right)  ^{i\omega}S_{+}\left(  -\mu^{2}uv\right)  ,$ follow
from their properties in the complex $u$ and complex $v$ planes. Since
$S_{\mp}\left(  z\right)  $ are entire functions (see remarks after
Eq.(\ref{S})), the pre-factors $u^{-i\omega}$ and $\left(  -v\right)
^{i\omega}$ determine the analyticity properties for both the massless and
massive fields. Clearly, these have branch points at $u=0$ and $v=0$
respectively. We choose the branch cuts to be on the positive imaginary axes
in both the complex $u$ and $v$ planes as shown in Figs.(5,6). This defines an
infinite number of sheets in the analytic $u$ and $v$ planes. The $u$-sheets
($v$-sheets) are labeled by an integer $n$ ($m),$ so the stack of universes is
labeled by these integers $\left(  n,m\right)  .$

Some typical points $\left(  u_{I},v_{I}\right)  ,\left(  u_{II}%
,v_{II}\right)  ,\left(  u_{III},v_{III}\right)  ,\left(  u_{IV}%
,v_{IV}\right)  $ in regions I-IV in universe $\left(  0,0\right)  ,$
connected to each other with some arbitrary analyticity path, are shown in
Fig.(7) (this ignores the small excursions into the complex $u$ or $v$ plane
near the horizons). The images of the points on the real axes in the complex
$\left(  u,v\right)  $ planes are shown in Figs.(5,6). The analyticity path
that connects them is also shown, such that, for clarity, the path goes
slightly under the branch points in the complex planes to stay within the
$\left(  0,0\right)  $ universe. Of course, the analyticity path within the
same $\left(  0,0\right)  $ level can be any other curve in the complex planes
that connects the points as long as it does not cross the branch cuts in
Figs.(5,6). This is the path of analyticity on level (0,0) used to establish
the continuity of the field $\varphi\left(  u,v\right)  $ as given in
Eqs.(\ref{fullF},\ref{list}).

\begin{center}%
{\parbox[b]{3.0753in}{\begin{center}
\includegraphics[
height=1.5558in,
width=3.0753in
]%
{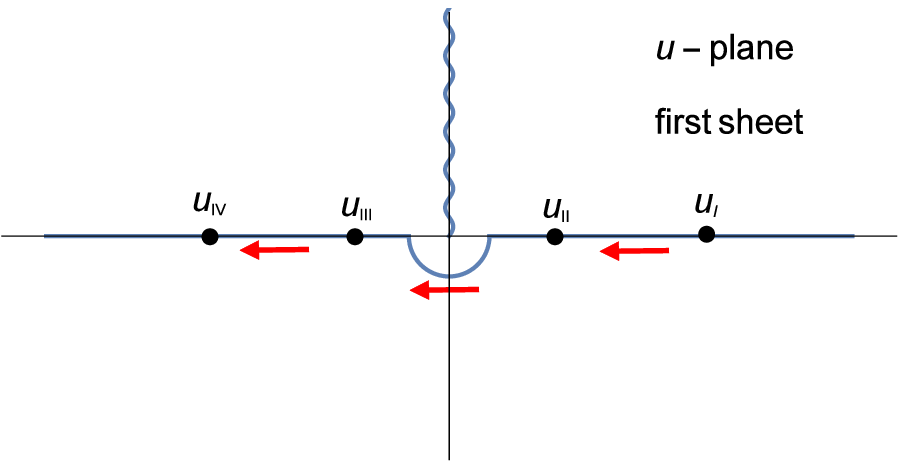}%
\\
{\protect\scriptsize Fig.5- A path on 0}$^{th}${\protect\scriptsize \ sheet in
analytic }$u${\protect\scriptsize -plane.}%
\end{center}}}
\
{\parbox[b]{3.0986in}{\begin{center}
\includegraphics[
height=1.5662in,
width=3.0986in
]%
{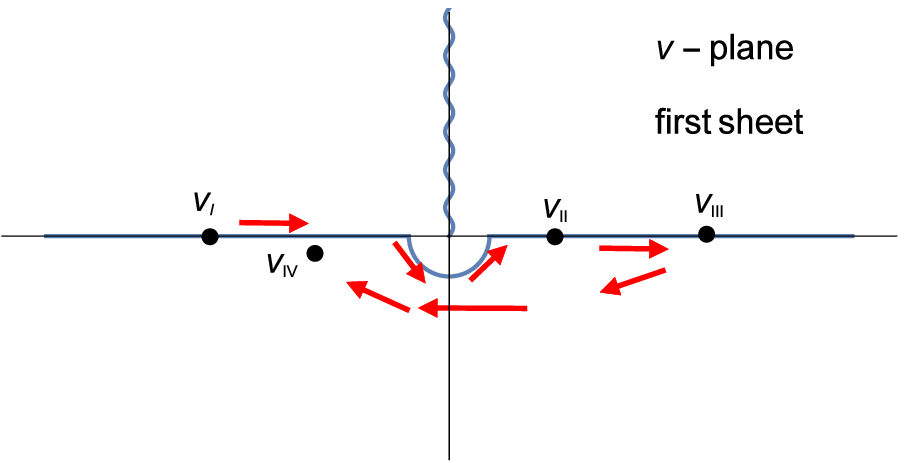}%
\\
{\protect\scriptsize Fig.6- A path on 0}$^{th}${\protect\scriptsize \ sheet in
analytic }$v${\protect\scriptsize -plane.}%
\end{center}}}

\end{center}

%

\begin{center}
\includegraphics[
height=2.2044in,
width=2.2044in
]%
{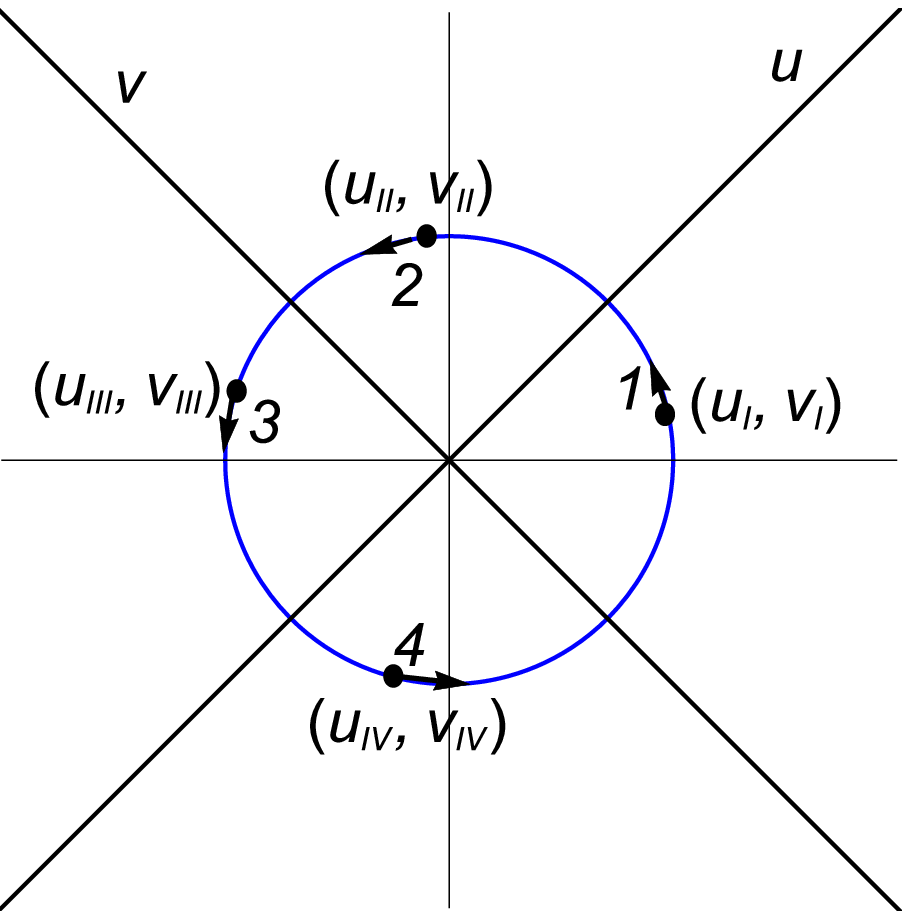}%
\\
{\protect\scriptsize Fig.(7)- A path on the real }$(u,v)$%
{\protect\scriptsize \ plane in the (0,0) universe. }%
\end{center}

Now we can analytically continue from level (0,0) to any other level $\left(
n,m\right)  $ by crossing the branch points and coming back again to the real
axes of $u$ and $v.$ This provides the value of the field in Eqs.(\ref{fullF}%
,\ref{list}) at different levels $\left(  n,m\right)  $ that again look like
the Minkowski plane. The basis $\varphi_{\pm}\left(  u,v\right)  $ on the real
axis of universe $\left(  n,m\right)  $ is related by monodromy to the basis
$\varphi_{\pm}\left(  u,v\right)  $ in Eq.(\ref{normalized}) for universe
$\left(  0,0\right)  $. Recall that $S_{\mp}\left(  -\mu^{2}uv\right)  $ given
in Eq.(\ref{S}) are entire functions of their arguments, so they have no
discontinuities under analytic continuation, so we only need to analytically
continue $u^{-i\omega}$ and $v^{-i\omega}$.

The analytic continuation involves replacing, $u\rightarrow ue^{2\pi in}$ or
$v\rightarrow ve^{2\pi im}$ with integer $n$ or $m,$ to indicate how many
times we wind around the branch points at $u=0$ or $v=0$ in the respective $u$
or $v$ complex planes in Figs.(5,6) when the horizons are crossed in an
analytic path on the real $\left(  u,v\right)  $ plane shown in Fig.(7). The
winding numbers, that may be different at each horizon, will lead to some
sheet in the multiverse. An analytic continuation of the field $\varphi\left(
u,v\right)  $ in Eqs.(\ref{fullF},\ref{list}) from level $\left(  0,0\right)
$ to level $\left(  n,m\right)  ,$ which is consistent with the boundary
conditions\footnote{A crucial consequence of boundary conditions is
Eq.(\ref{a+-relations}). This requires the oscilators of regions I and III to
satisfy
\begin{equation}
\bar{a}_{1-}a_{1-}=\bar{a}_{1+}a_{1+},\;\bar{b}_{1-}b_{1-}=\bar{b}_{1+}%
b_{1+},\;\bar{a}_{3-}a_{3-}=\bar{a}_{3+}a_{3+},\;\bar{b}_{3-}b_{3-}=\bar
{b}_{3+}b_{3+}%
\end{equation}
A physical consequence of these relations is that the fields $\varphi
_{1,3}\left(  u,v\right)  $ vanish at the asymptotic regions $\left\vert
u\right\vert \rightarrow\infty$ or $\left\vert v\right\vert \rightarrow\infty$
in regions I or III. Without these conditions probablity would become infinite
in those asymptotic regions. This is explained in section-\ref{bc}. The same
physical conditions exist also for all layers of the multiverse. To respect
these boundary conditions only certain patterns of winding numbers are allowed
in the definition of the layers of the multiverse as shown in Eq.(\ref{Phinm}%
). The consequence of these patterns leads to the oscillators for level
$\left(  n,m\right)  $ given Eq.(\ref{can1}). It can be observed that these
rescaled oscillators also satisfy the physical boundary conditions at all such
levels, $\bar{a}_{1-}^{\left(  n\right)  }a_{1-}^{\left(  n\right)  }=\bar
{a}_{1+}^{\left(  n\right)  }a_{1+}^{\left(  n\right)  },$ etc. because the
relation is true at $n=0$ as given in Eq.(\ref{a+-relations}). A different
pattern of windings at the horizons violate the physical boundary conditions
discussed above, and this is the reason why they are not consistent.} that are
later explained in section-\ref{bc}, can only have the following
pattern\footnote{This pattern applies only to the extended Rindler space. For
black hole spacetimes, $S_{\mp}\left(  uv\right)  $ have a branch cut that
starts at the black hole singularity $uv=1$, therefore the monodromy in
Eq.(\ref{continuation}) as well as the pattern of analytic continuation is
different for black holes (see \cite{barsArayaBH}).}%
\begin{equation}%
\begin{array}
[c]{l}%
\varphi^{\left(  n,m\right)  }\left(  u,v\right)  =\left(
\begin{array}
[c]{c}%
\varphi_{0}+\theta\left(  I\right)  \varphi_{1}\left(  ue^{2\pi in},e^{-2\pi
in}v\right)  +\theta\left(  II\right)  \varphi_{2}\left(  ue^{2\pi
in},ve^{2\pi im}\right) \\
+\theta\left(  III\right)  \varphi_{3}\left(  ue^{-2\pi im},ve^{2\pi
im}\right)  +\theta\left(  IV\right)  \varphi_{4}\left(  ue^{-2\pi
im},ve^{-2\pi in}\right)
\end{array}
\right) \\
\bar{\varphi}^{\left(  n,m\right)  }\left(  u,v\right)  =\left(
\begin{array}
[c]{c}%
\varphi_{0}+\theta\left(  I\right)  \varphi_{1}^{\dagger}\left(  ue^{2\pi
in},e^{-2\pi in}v\right)  +\theta\left(  II\right)  \varphi_{2}^{\dagger
}\left(  ue^{2\pi in},ve^{2\pi im}\right) \\
+\theta\left(  III\right)  \varphi_{3}^{\dagger}\left(  ue^{-2\pi im},ve^{2\pi
im}\right)  +\theta\left(  IV\right)  \varphi_{4}^{\dagger}\left(  ue^{-2\pi
im},ve^{-2\pi in}\right)
\end{array}
\right)
\end{array}
. \label{Phinm}%
\end{equation}
where $\left(  n,m\right)  $ are integers, and the hermitian conjugates
$\varphi_{1,2,3,4}^{\dagger}$ are defined for real $\left(  u,v\right)  $.
Here $\bar{\varphi}^{\left(  n,m\right)  }\left(  u,v\right)  $ is the
canonical conjugate to $\varphi^{\left(  n,m\right)  }\left(  u,v\right)  .$
The alert reader will note that in these expressions the regional fields
$\varphi_{1,2,3,4}$ and $\varphi_{1,2,3,4}^{\dagger}$ are analytically
continued in a pattern that is different in each region. In each region
$\left(  u,v\right)  $ are real in the respective ranges as seen in Fig.1. The
phases $e^{\pm2\pi in}=e^{\pm2\pi im}=1$ do not change the reality of the
continued $\left(  u,v\right)  $ for each region, but because of the monodromy
properties of the factors, $\left(  u^{\mp i\omega},v^{\pm i\omega}\right)  $
that appear in the expressions for $\varphi\left(  u,v\right)  ,$ such as
\begin{equation}
\left(  ue^{2\pi in}\right)  ^{\mp i\omega}=u^{\mp i\omega}e^{\pm2\pi\omega
n},\;\;\left(  ve^{-2\pi in}\right)  ^{\pm i\omega}=v^{\pm i\omega}e^{\pm
2\pi\omega n},\text{ }S_{\mp}\left(  -\mu^{2}uve^{\pm2\pi\omega k}\right)
=S_{\mp}\left(  -\mu^{2}uv\right)  , \label{continuation}%
\end{equation}
the result for $\varphi^{\left(  n,m\right)  }\left(  u,v\right)  $ in
Eq.(\ref{Phinm}) is different than $\varphi\left(  u,v\right)  $. The
analytically continued fields $\varphi_{1,2,3,4}$ in $\varphi^{\left(
n,m\right)  }\left(  u,v\right)  $ have the same form as the $\varphi
_{1,2,3,4}$ in Eq.(\ref{list}) except for the fact that the oscillators
$\left(  a_{1\mp},b_{1\mp}^{\dagger},b_{3\pm}^{\dagger},a_{3\pm}\right)  $ in
Eq.(\ref{list}) are now replaced by new ones in $\varphi^{\left(  n,m\right)
}\left(  u,v\right)  $ that we label as $(a_{1\mp}^{\left(  n\right)  }%
,\bar{b}_{1\mp}^{\left(  n\right)  },\bar{b}_{3\pm}^{m},a_{3\pm}^{m}).$ The
relation between $(a_{1\mp}^{\left(  n\right)  },\bar{b}_{1\mp}^{\left(
n\right)  },\bar{b}_{3\pm}^{m},a_{3\pm}^{m})$ and $\left(  a_{1\mp},b_{1\mp
}^{\dagger},b_{3\pm}^{\dagger},a_{3\pm}\right)  $ is obtained by inserting
Eq.(\ref{continuation}) into the $\varphi_{1,2,3,4}$ in Eq.(\ref{list}), and
similarly for the hermitian conjugates $\varphi_{1,2,3,4}^{\dagger}$. The
result of the analytic continuation in Eq.(\ref{Phinm}) then yields the
desired relations between levels $\left(  n,m\right)  $ and $\left(
0,0\right)  $
\begin{equation}%
\begin{array}
[c]{c}%
a_{1\mp}^{\left(  n\right)  }=a_{1\mp}e^{2\pi\omega n},\;\bar{a}_{1\mp
}^{\left(  n\right)  }=a_{1\mp}^{\dagger}e^{-2\pi\omega n},\;b_{1\mp}^{\left(
n\right)  }=b_{1\mp}e^{2\pi\omega n},\;\bar{b}_{1\mp}^{\left(  n\right)
}=b_{1\mp}^{\dagger}e^{-2\pi\omega n},\\
a_{3\pm}^{\left(  m\right)  }=a_{3\pm}e^{2\pi\omega m},\;\bar{a}_{3\pm
}^{\left(  m\right)  }=a_{3\pm}^{\dagger}e^{-2\pi\omega m},\;b_{3\pm}^{\left(
m\right)  }=a_{3\pm}e^{2\pi\omega m},\;\bar{b}_{3\pm}^{\left(  m\right)
}=b_{3\pm}^{\dagger}e^{-2\pi\omega m}.
\end{array}
\label{can1}%
\end{equation}
This analyticity-induced map is a canonical transformation since%
\begin{equation}
\left[  a_{1\mp}^{\left(  n\right)  },\bar{a}_{1\mp}^{\left(  n\right)
}\right]  =\left[  a_{1\mp},a_{1\mp}^{\dagger}\right]  ,~\text{etc.}
\label{can2}%
\end{equation}

We explain some notation. We used the overbar symbol in $(\bar{a}_{1\mp
}^{\left(  n\right)  },\bar{b}_{1\mp}^{\left(  n\right)  },\bar{a}_{3\pm
}^{\left(  m\right)  },\bar{b}_{3\pm}^{\left(  m\right)  })$ to indicate the
canonical conjugates of ($a_{1\mp}^{\left(  n\right)  },b_{1\mp}^{\left(
n\right)  },a_{3\pm}^{\left(  m\right)  },b_{3\pm}^{\left(  m\right)  })$
respectively. For $n=m=0$ the overbar is defined to be actually the same the
hermitian conjugate, $(\bar{a}_{1\mp}^{\left(  0\right)  },\bar{b}_{1\mp
}^{\left(  0\right)  },\bar{a}_{3\pm}^{\left(  0\right)  },\bar{b}_{3\pm
}^{\left(  0\right)  })\equiv\left(  a_{1\mp}^{\dagger},b_{1\mp}^{\dagger
},a_{3\pm}^{\dagger},b_{3\pm}^{\dagger}\right)  ,$ but for general $n$,
$\bar{a}_{1\mp}^{\left(  n\right)  }$ is not the hermitian conjugate of
$a_{1\mp}^{\left(  n\right)  },$ although it is its canonical conjugate, and
similarly for the other oscillators.

This shows that the analytic continuation to universe $\left(  n,m\right)  $
defined by Eqs.(\ref{Phinm}-\ref{can2}) amounts to a canonical transformation
of the creation-annihilation operators. The full field $\bar{\varphi}^{\left(
n,m\right)  }\left(  u,v\right)  $ is not the naive hermitian conjugate of
$\varphi^{\left(  n,m\right)  }\left(  u,v\right)  $ but it is its canonical
conjugate$.$ The equal time quantum commutator, $\left[  \varphi^{\left(
n,m\right)  }\left(  t,y\right)  ,\bar{\varphi}^{\left(  n,m\right)  }\left(
t,y^{\prime}\right)  \right]  ,$ produces a delta function $\delta\left(
y-y^{\prime}\right)  $ at every universe $\left(  n,m\right)  $ just as the
$\left(  0,0\right)  $ universe. It should be emphasized that the fields on
different levels $\left(  \varphi^{\left(  n_{1},m_{1}\right)  }\left(
t,y\right)  ,\bar{\varphi}^{\left(  n_{2},m_{2}\right)  }\left(  t,y^{\prime
}\right)  \right)  $ with $\left(  n_{1},m_{1}\right)  $ different from
$\left(  n_{2},m_{2}\right)  $ are not independent of each other since the
corresponding oscillators $(a_{1\mp}^{\left(  n_{1}\right)  },\bar{b}_{1\mp
}^{\left(  n_{1}\right)  },\bar{b}_{3\pm}^{\left(  m_{1}\right)  },a_{3\pm
}^{\left(  m_{1}\right)  })$ and $(a_{1\mp}^{\left(  n_{2}\right)  },\bar
{b}_{1\mp}^{\left(  n_{2}\right)  },\bar{b}_{3\pm}^{\left(  m_{2}\right)
},a_{3\pm}^{\left(  m_{2}\right)  })$ are all related to the same set of basic
oscillators $\left(  a_{1\mp},b_{1\mp}^{\dagger},b_{3\pm}^{\dagger},a_{3\pm
}\right)  $ and their hermitian conjugates that define the field in universe
$\left(  0,0\right)  .$ So the quantum rules for the entire multiverse,
including non-trivial commutators among fields at different levels, such as,
$\left[  \varphi^{\left(  n_{1},m_{1}\right)  }\left(  t,y\right)
,\bar{\varphi}^{\left(  n_{2},m_{2}\right)  }\left(  t,y^{\prime}\right)
\right]  $, depend only on the quantum rules established at level $\left(
0,0\right)  .$

Propagators, various correletors and probabilities of processes computed with
$\varphi^{\left(  n,m\right)  }\left(  u,v\right)  $ may be different for
different levels because of the shift of normalizations of the Rindler
coefficients as given in Eq.(\ref{can1}). Hence one must also specify the
universe on which boundary conditions are imposed. We define level $\left(
0,0\right)  $ as the reference universe at which boundary conditions are
applied as shown in the next section. Then, using $\varphi^{\left(
n,m\right)  }\left(  u,v\right)  $ as given above, probabilities for various
physical processes at universe $\left(  n,m\right)  ,$ that may depend on the
boundary conditions in the $\left(  0,0\right)  $ universe, can be determined.
In particular, we will try to answer the question: is there probability (or
information) flow from one universe labeled by $\left(  n,m\right)  $ to other
universes labeled by $\left(  n^{\prime},m^{\prime}\right)  $?

\section{Horizon boundary conditions \label{bc}}

Boundary conditions imposed in universe $(0,0)$ will automatically fix all
boundary conditions for all $(n,m)$ as determined in the previous section.
Accordingly, given the relationship between Minkowski and Rindler coordinates
as given in Eq.(\ref{uv-tyRegions}), we require the $(0,0)$ Rindler field
$\varphi^{\left(  0,0\right)  }(u,v)=\varphi(u,v)$ given in Eqs.(\ref{fullF}%
,\ref{list}), to be identical to the Minkowski field $\varphi(u,v)$ given in
Eqs$.$(\ref{PsiM1},\ref{PsiM}). For this, it is sufficient to impose boundary
conditions at each horizon, $\varphi^{\left(  0,0\right)  }(u,0)=\varphi(u,0)$
and $\varphi^{\left(  0,0\right)  }(0,v)=\varphi(0,v)$. Other boundary
conditions could be considered for the extended Rindler case, however we
emphasize these horizon boundary conditions since in the case of a black hole
it is also appropriate to impose the same boundary conditions as done in an
upcoming paper \cite{barsArayaBH}. This is because, near the horizons, the
black hole metric and field behave \textit{locally} just like the flat
Minkowski metric and field. So the horizon boundary conditions used here for
the Rindler case will be used also identically for black hole case. Hence, for
either Rindler or a black hole metric, using $S_{\mp}(0)=1$ at horizons in
Eqs.(\ref{fullF},\ref{list}), we have the desired boundary conditions%
\begin{equation}%
\begin{array}
[c]{l}%
\varphi_{0}=\varphi(0,0)=\int_{-\infty}^{\infty}\frac{dk}{\sqrt{4\pi E}%
}\left(  A\left(  k\right)  +B^{\dagger}\left(  k\right)  \right) \\
\varphi_{1\text{ or }2}\left(  u,0\right)  =\int_{0}^{\infty}d\omega\left[
a_{1-}\left(  \omega\right)  \frac{u^{-i\omega}}{\sqrt{4\pi\omega}}%
+b_{1-}^{\dagger}\left(  \omega\right)  \frac{u^{i\omega}}{\sqrt{4\pi\omega}%
}\right]  =\theta\left(  u\right)  \left(  \varphi(u,0)-\varphi_{0}\right) \\
\varphi_{3\text{ or }4}\left(  u,0\right)  =\int_{0}^{\infty}d\omega\left[
b_{3-}^{\dagger}\left(  \omega\right)  \frac{\left(  -u\right)  ^{-i\omega}%
}{\sqrt{4\pi\omega}}+a_{3-}\left(  \omega\right)  \frac{\left(  -u\right)
^{i\omega}}{\sqrt{4\pi\omega}}\right]  =\theta\left(  -u\right)  \left(
\varphi(u,0)-\varphi_{0}\right) \\
\varphi_{2\text{ or }3}\left(  0,v\right)  =\int_{0}^{\infty}d\omega\left[
b_{3+}^{\dagger}\left(  \omega\right)  \frac{v^{i\omega}}{\sqrt{4\pi\omega}%
}+a_{3+}\left(  \omega\right)  \frac{v^{-i\omega}}{\sqrt{4\pi\omega}}\right]
=\theta\left(  v\right)  \left(  \varphi(0,v)-\varphi_{0}\right) \\
\varphi_{1\text{ or }4}\left(  0,v\right)  =\int_{0}^{\infty}d\omega\left[
a_{1+}\left(  \omega\right)  \frac{\left(  -v\right)  ^{i\omega}}{\sqrt
{4\pi\omega}}+b_{1+}^{\dagger}\left(  \omega\right)  \frac{\left(  -v\right)
^{-i\omega}}{\sqrt{4\pi\omega}}\right]  =\theta\left(  -v\right)  \left(
\varphi(0,v)-\varphi_{0}\right)
\end{array}
\label{boundaA}%
\end{equation}
where the $\varphi(u,0)$ or $\varphi(0,v)$ on the right hand side of
Eq.(\ref{boundaA}) is the Minkowski field (\ref{PsiM}) evaluated at the
horizons%
\begin{equation}%
\begin{array}
[c]{c}%
\varphi(u,0)=\int_{0}^{\infty}dk\left(  \frac{A_{+}\left(  k\right)
e^{-i\frac{E-k}{2}u}+B_{+}^{\dagger}\left(  k\right)  e^{i\frac{E-k}{2}u}%
}{\sqrt{4\pi E}}+\frac{A_{-}\left(  k\right)  e^{-i\frac{E+k}{2}u}%
+B_{-}^{\dagger}\left(  k\right)  e^{i\frac{E+k}{2}u}}{\sqrt{4\pi E}}\right)
,\\
\varphi(0,v)=\int_{0}^{\infty}dk\left(  \frac{A_{+}\left(  k\right)
e^{-i\frac{E+k}{2}v}+B_{+}^{\dagger}\left(  k\right)  e^{i\frac{E+k}{2}u}%
}{\sqrt{4\pi E}}+\frac{A_{-}\left(  k\right)  e^{-i\frac{E-k}{2}v}%
+B_{-}^{\dagger}\left(  k\right)  e^{i\frac{E-k}{2}v}}{\sqrt{4\pi E}}\right)
.
\end{array}
\end{equation}
We see in the first line of Eq.(\ref{boundaA}) that a non-trivial zero mode
$\varphi_{0}$ is necessary because $\varphi_{1,2,3,4}\left(  0,0\right)  $ all
vanish according to the arguments in Eq.(\ref{epsilon}). From the second
expression in Eq.(\ref{boundaA}) at the horizon I\&II, $a_{1-}$ can be
extracted by using the orthonormality and completeness of the basis $u^{\mp
i\omega}/\sqrt{4\pi\omega}$ on the half-line $u>0$; namely$^{\ref{norm}}$,
$a_{1-}\left(  \omega\right)  =\int_{0}^{\infty}du\left(  \frac{u^{i\omega}%
}{\sqrt{4\pi\omega}}i\partial_{u}\varphi(u,0)-\varphi(u,0)i\partial_{u}%
\frac{u^{i\omega}}{\sqrt{4\pi\omega}}\right)  ,$ and similarly for
$b_{1-}^{\dagger}$ by replacing $u^{-i\omega}$ instead of $u^{i\omega}$. After
performing the integrals we obtain%
\begin{equation}%
\begin{array}
[c]{c}%
a_{1-}\left(  \omega\right)  =\frac{\Gamma\left(  1+i\omega~\right)  }%
{i\sqrt{2\pi\omega}}\int_{-\infty}^{\infty}\frac{dk}{\sqrt{4\pi E}}\left(
\frac{E-k}{2}\right)  ^{-i\omega}\left(  A\left(  k\right)  e^{\frac{\pi
\omega}{2}}+B^{\dagger}\left(  k\right)  e^{-\frac{\pi\omega}{2}}\right)  ,\\
b_{1-}^{\dagger}\left(  \omega\right)  =\frac{\Gamma\left(  1-i\omega~\right)
}{-i\sqrt{2\pi\omega}}\int_{-\infty}^{\infty}\frac{dk}{\sqrt{4\pi E}}\left(
\frac{E-k}{2}\right)  ^{i\omega}\left(  B^{\dagger}\left(  k\right)
e^{\frac{\pi\omega}{2}}+A\left(  k\right)  e^{-\frac{\pi\omega}{2}}\right)  .
\end{array}
\label{abAB}%
\end{equation}
Note that on the right hand side the integral $\int_{-\infty}^{\infty}dk$
contains $A\left(  k\right)  ,B^{\dagger}\left(  k\right)  $ over the full
momentum range. To make contact with the notation $A_{\pm}\left(  k\right)
,B_{\pm}^{\dagger}\left(  k\right)  ,$ with only $k>0$, the integral can be
split to the positive and negative intervals.

The hermitian conjugate $a_{1-}^{\dagger}$ looks like $b_{1-}^{\dagger}$ above
but with $A\leftrightarrow B$ interchanged on the right hand side of
(\ref{abAB}). As a consistency check, it can then be verified that the
commutation rules (\ref{Rvac}) of the Rindler modes, $\left[  a_{1-}\left(
\omega\right)  ,a_{1-}^{\dagger}\left(  \omega^{\prime}\right)  \right]
=\delta\left(  \omega-\omega^{\prime}\right)  ,$ etc., can be obtained by
using only the commutation rules (\ref{Mvac}) of the Minkowski modes, $\left[
A\left(  k\right)  ,A^{\dagger}\left(  k^{\prime}\right)  \right]
=\delta\left(  k-k^{\prime}\right)  =\left[  B\left(  k\right)  ,B^{\dagger
}\left(  k^{\prime}\right)  \right]  ,$ by using the relations above. Similar
expressions are obtained at the 4 horizons.

This fixes the 8 Rindler complex coefficients in the level $\left(
0,0\right)  $ universe, $a_{1\mp},b_{1\mp}^{\dagger},a_{3\mp},b_{3\mp
}^{\dagger}$, in terms of the 4 Minkowski complex coefficients, $A_{\pm
}\left(  k\right)  ,B_{\pm}^{\dagger}\left(  k\right)  $ (similarly, for the
black hole \cite{barsArayaBH}). It is revealing to re-write the 8 relations in
level $\left(  0,0\right)  $ in the form of Bogoliubov transformations as
follows, where $\frac{\Gamma\left(  1\pm i\omega~\right)  }{\sqrt{2\pi\omega}%
}=\frac{e^{\pm i\theta\left(  \omega\right)  }e^{-\pi\omega/2}}{\sqrt
{1-e^{-2\pi\omega}}}$ is used,
\begin{equation}%
\begin{array}
[c]{c}%
\frac{ie^{-i\theta\left(  \omega\right)  }}{\sqrt{1-e^{-2\pi\omega}}}\left(
\begin{array}
[c]{cc}%
1 & -e^{-\pi\omega}\\
-e^{-\pi\omega} & 1
\end{array}
\right)  \left(
\begin{array}
[c]{c}%
a_{1-}\left(  \omega\right)  \\
b_{3+}^{\dagger}\left(  \omega\right)
\end{array}
\right)  =\int_{-\infty}^{\infty}dk\frac{\left(  \frac{E-k}{2}\right)
^{-i\omega}}{\sqrt{4\pi E}}\left(
\begin{array}
[c]{c}%
A\left(  k\right)  \\
B^{\dagger}\left(  k\right)
\end{array}
\right)  \\
\frac{-ie^{i\theta\left(  \omega\right)  }}{\sqrt{1-e^{-2\pi\omega}}}\left(
\begin{array}
[c]{cc}%
1 & -e^{-\pi\omega}\\
-e^{-\pi\omega} & 1
\end{array}
\right)  \left(
\begin{array}
[c]{c}%
a_{3+}\left(  \omega\right)  \\
b_{1-}^{\dagger}\left(  \omega\right)
\end{array}
\right)  =\int_{-\infty}^{\infty}dk\frac{\left(  \frac{E-k}{2}\right)
^{i\omega}}{\sqrt{4\pi E}}\left(
\begin{array}
[c]{c}%
A\left(  k\right)  \\
B^{\dagger}\left(  k\right)
\end{array}
\right)  \\
\frac{-ie^{-i\theta\left(  \omega\right)  }}{\sqrt{1-e^{-2\pi\omega}}}\left(
\begin{array}
[c]{cc}%
1 & -e^{-\pi\omega}\\
-e^{-\pi\omega} & 1
\end{array}
\right)  \left(
\begin{array}
[c]{c}%
a_{3-}\left(  \omega\right)  \\
b_{1+}^{\dagger}\left(  \omega\right)
\end{array}
\right)  =\int_{-\infty}^{\infty}dk\frac{\left(  \frac{E+k}{2}\right)
^{-i\omega}}{\sqrt{4\pi E}}\left(
\begin{array}
[c]{c}%
A\left(  k\right)  \\
B^{\dagger}\left(  k\right)
\end{array}
\right)  \\
\frac{ie^{i\theta\left(  \omega\right)  }}{\sqrt{1-e^{-2\pi\omega}}}\left(
\begin{array}
[c]{cc}%
1 & -e^{-\pi\omega}\\
-e^{-\pi\omega} & 1
\end{array}
\right)  \left(
\begin{array}
[c]{c}%
a_{1+}\left(  \omega\right)  \\
b_{3-}^{\dagger}\left(  \omega\right)
\end{array}
\right)  =\int_{-\infty}^{\infty}dk\frac{\left(  \frac{E+k}{2}\right)
^{i\omega}}{\sqrt{4\pi E}}\left(
\begin{array}
[c]{c}%
A\left(  k\right)  \\
B^{\dagger}\left(  k\right)
\end{array}
\right)
\end{array}
\label{bogol1}%
\end{equation}
and their hermitian conjugates. These equations can be easily
inverted\footnote{$A\left(  k\right)  $ and $B^{\dagger}\left(  k\right)  $
are isolated from the right hand side of Eq.(\ref{bogol1}) by multiplying with
$\left(  E^{\prime}\mp k^{\prime}\right)  ^{\pm^{\prime}i\omega}$ as
appropriate and adding two terms so that one may use $\int_{0}^{\infty}%
d\omega\left(  \left(  E-k\right)  ^{-i\omega}\left(  E^{\prime}\mp k^{\prime
}\right)  ^{i\omega}+\left(  E-k\right)  ^{i\omega}\left(  E^{\prime}\mp
k^{\prime}\right)  ^{-i\omega}\right)  =\delta\left(  k\mp k^{\prime}\right)
2\pi E$ when $E\left(  k\right)  =\sqrt{k^{2}+\mu^{2}}$. To check that
Eq.(\ref{invert}) satisfies Eq.(\ref{bogol1}) use $\int_{-\infty}^{\infty
}\frac{dk}{2\pi E\left(  k\right)  }\left(  E\left(  k\right)  \pm k\right)
^{i\lambda}=\delta\left(  \lambda\right)  $. The contributions of the zero
modes to the integrals in Eq.(\ref{bogol1}) are proportional to $\delta\left(
\omega\right)  ,$ but these vanish since $\omega>0$. \label{invert1}} to
obtain an explicit expression for the Minkowski oscillators $\left(  A\left(
k\right)  ,B^{\dagger}\left(  k\right)  \right)  $ in terms of the Rindler
oscillators $\left(  a_{1\mp},b_{1\mp}^{\dagger},a_{3\mp},b_{3\mp}^{\dagger
}\right)  .$
\begin{equation}%
\begin{array}
[c]{c}%
A\left(  k\right)  =\frac{\varphi_{0}/c+\sigma_{0}}{2\sqrt{\pi E}}+\int
_{0}^{\infty}d\omega\left[
\begin{array}
[c]{c}%
\frac{\left(  \frac{E\mp k}{2}\right)  ^{\pm i\omega}}{\sqrt{\pi E}}%
\frac{ie^{\mp i\theta\left(  \omega\right)  }}{\sqrt{1-e^{-2\pi\omega}}%
}\left(  a_{1\mp}\left(  \omega\right)  -e^{-\pi\omega}b_{3\pm}^{\dagger
}\left(  \omega\right)  \right)  \\
+\frac{\left(  \frac{E\mp k}{2}\right)  ^{\mp i\omega}}{\sqrt{\pi E}}%
\frac{-ie^{\pm i\theta\left(  \omega\right)  }}{\sqrt{1-e^{-2\pi\omega}}%
}\left(  a_{3\pm}\left(  \omega\right)  -e^{-\pi\omega}b_{1\mp}^{\dagger
}\left(  \omega\right)  \right)
\end{array}
\right]  \\
B^{\dagger}\left(  k\right)  =\frac{\varphi_{0}/c-\sigma_{0}}{2\sqrt{\pi E}%
}+\int_{0}^{\infty}d\omega\left[
\begin{array}
[c]{c}%
\frac{\left(  \frac{E\mp k}{2}\right)  ^{\mp i\omega}}{\sqrt{\pi E}}%
\frac{ie^{\pm i\theta\left(  \omega\right)  }}{\sqrt{1-e^{-2\pi\omega}}%
}\left(  b_{1\mp}^{\dagger}\left(  \omega\right)  -e^{-\pi\omega}a_{3\pm
}\left(  \omega\right)  \right)  \\
+\frac{\left(  \frac{E\mp k}{2}\right)  ^{\pm i\omega}}{\sqrt{\pi E}}%
\frac{-ie^{\mp i\theta\left(  \omega\right)  }}{\sqrt{1-e^{-2\pi\omega}}%
}\left(  b_{3\pm}^{\dagger}\left(  \omega\right)  -e^{-\pi\omega}a_{1\mp
}\left(  \omega\right)  \right)
\end{array}
\right]
\end{array}
\label{invert}%
\end{equation}
where $-\infty<k<\infty.$ Here $\varphi_{0}$ is the zero mode that satisfies
Eq.(\ref{boundaA}) where$^{\ref{invert1}}$ $c\equiv\int_{-\infty}^{\infty
}\frac{dk}{2\pi E\left(  k\right)  }=\delta\left(  0\right)  ,$ while
$\sigma_{0}$ is another Rindler zero mode. Note that the two integrands with
the upper/lower signs \textquotedblleft$\pm$\textquotedblright\ in
Eq.(\ref{invert}) are equal to each other for the massive case on account of
the relations between $a_{1\pm}\left(  \omega\right)  $ etc. explained below
in Eqs.(\ref{E+-k},\ref{a+-relations}).

Based on analyticity properties of the wavefunction, Unruh \cite{unruh} gave a
simple argument to derive the so called Unruh modes. An Unruh mode is the
following combination of the Rindler modes that annihilates the Minkowski
vacuum $|0_{M}\rangle$, such as%
\begin{equation}
\frac{a_{1-}\left(  \omega\right)  -e^{-\pi\omega}b_{3+}^{\dagger}\left(
\omega\right)  }{\sqrt{1-e^{-2\pi\omega}}}|0_{M}\rangle=0.
\end{equation}
This is in agreement with the first line of Eq.(\ref{bogol1}) where the Unruh
mode above is seen in more detail to be equal to, $-ie^{i\theta\left(
\omega\right)  }\int_{-\infty}^{\infty}dk\frac{A\left(  k\right)  }{\sqrt{4\pi
E}}\left(  \frac{E-k}{2}\right)  ^{-i\omega}$, which is clearly a combination
of the Minkowski annihilation operators as defined in Eq.(\ref{Mvac}). Knowing
the additional detail given here, of how to write the Unruh mode in terms of
the Minkowski modes as in Eq.(\ref{bogol1}), is important because this can be
used to compute the action of the Unruh modes, or more generally $a_{1-}$ or
$b_{3+}^{\dagger}$ on their own as in Eq.(\ref{abAB}), on any general
Minkowski state, not only the vacuum state $|0_{M}\rangle$. Our explicit
relations in Eq.(\ref{bogol1}) should be useful for various applications
involving quantum effects in Rindler space. Furthermore, our expressions
(\ref{bogol1}) are more general because they apply also to black holes
\cite{barsArayaBH}.

The relations (\ref{bogol1}) reveal additional important properties. For
example, the expressions for the region-I coefficients $a_{1\pm},b_{1\pm
}^{\dagger}$ extracted from Eq.(\ref{bogol1}) are%
\begin{equation}%
\begin{array}
[c]{c}%
a_{1\mp}\left(  \omega\right)  =\frac{e^{\pm i\theta\left(  \omega\right)  }%
}{\pm i\sqrt{1-e^{-2\pi\omega}}}\int_{-\infty}^{\infty}\frac{dk}{\sqrt{4\pi
E}}\left(  \frac{E\mp k}{2}\right)  ^{\mp i\omega}\left(  A\left(  k\right)
+B^{\dagger}\left(  k\right)  e^{-\pi\omega}\right)  ,\\
b_{1\mp}^{\dagger}\left(  \omega\right)  =\frac{e^{\mp i\theta\left(
\omega\right)  }}{\mp i\sqrt{1-e^{-2\pi\omega}}}\int_{-\infty}^{\infty}%
\frac{dk}{\sqrt{4\pi E}}\left(  \frac{E\mp k}{2}\right)  ^{\pm i\omega}\left(
B^{\dagger}\left(  k\right)  +A\left(  k\right)  e^{-\pi\omega}\right)  .
\end{array}
\label{a+-AB}%
\end{equation}
Now concentrate on the Rindler case because we will next use the fact that
$E=\sqrt{k^{2}+\mu^{2}}$ (for a black hole $E$ and $k$ have a different
relation \cite{barsArayaBH}). Then in Eq.(\ref{a+-AB}) insert
\begin{equation}
\frac{E-k}{2}=\frac{\mu^{2}}{4}\left(  \frac{E+k}{2}\right)  ^{-1}%
,\label{E+-k}%
\end{equation}
and then see that $\left(  a_{1-},b_{1-}^{\dagger}\right)  $ and $\left(
a_{1+},b_{1+}^{\dagger}\right)  $ are proportional to each other with overall
phases. Similar arguments hold also for $\left(  a_{3-},b_{3-}^{\dagger
}\right)  $ and $\left(  a_{3+},b_{3+}^{\dagger}\right)  ,$ so we find%
\begin{equation}%
\begin{array}
[c]{l}%
a_{1+}\left(  \omega\right)  =-\left(  \mu^{2}/4\right)  ^{i\omega
}e^{-2i\theta\left(  \omega\right)  }a_{1-}\left(  \omega\right)
,\;\;a_{3-}\left(  \omega\right)  =-\left(  \mu^{2}/4\right)  ^{i\omega
}e^{-2i\theta\left(  \omega\right)  }a_{3+}\left(  \omega\right)  ,\\
b_{1+}^{\dagger}\left(  \omega\right)  =-\left(  \mu^{2}/4\right)  ^{-i\omega
}e^{2i\theta\left(  \omega\right)  }b_{1-}^{\dagger}\left(  \omega\right)
,\;\;b_{3-}^{\dagger}\left(  \omega\right)  =-\left(  \mu^{2}/4\right)
^{-i\omega}e^{2i\theta\left(  \omega\right)  }b_{3+}^{\dagger}\left(
\omega\right)  .
\end{array}
\label{a+-relations}%
\end{equation}
A significant consequence of these relations is the vanishing of the fields
$\varphi_{1,3}\left(  u,v\right)  $ in (\ref{list}) in the asymptotic
regions-I and III when either $\left\vert u\right\vert $ or $\left\vert
v\right\vert $ goes to infinity at fixed $t$ and large $\left\vert
y\right\vert $, namely%
\begin{equation}
\lim_{u\text{ or (}-v)\rightarrow\infty}\varphi_{1}\left(  u,v\right)
=0=\lim_{\left(  -u\right)  ~\text{or }v\rightarrow\infty}\varphi_{3}\left(
u,v\right)  .\;\label{F13asympt}%
\end{equation}
This can be verified by using the asymptotic behavior of the Bessel functions
$I_{\mp i\omega}\left(  \sqrt{-2\mu^{2}uv}\right)  $ as given in
Eqs.(\ref{normalized},\ref{S}) when $uv<0,$ and using the relations in
Eq.(\ref{a+-relations}).

Quite independently, the vanishing of the field in the asymptotic regions of
I\&III is required on physical grounds because without it the field (or the
quantum wavefunction) would blow up at infinity, implying infinite
probability. It is gratifying that this required physical behavior emerged
from the boundary conditions at the horizons automatically without having to
impose it as an additional boundary condition, thus giving confidence that the
horizon boundary condition is a correct physical approach.

Furthermore, note that the vanishing of the field asymptotically in regions
I\&III is the expected behavior for wavepackets, on the basis of the classical
geodesics in Figs.(2,3), as well as on the basis of the intuitive physical
approach using the effective classical mechanical potential in Fig.(4), and
the effective quantum potential explained below in Eq.(\ref{VeffQ}) and
plotted in Fig.(8).

Turning next to regions II\&IV, the asymptotic behavior of the Rindler field
is oscillatory as seen from analyzing the asymptotic behavior of $I_{\mp
i\omega}\left(  \sqrt{-2\mu^{2}uv}\right)  $ when $uv>0$. Consistent with the
geodesics in Figs.(2,3), this is allowed physical behavior for incoming or
outgoing particles/antiparticles, or oscillatory waves and wavepackets built
from them. Further boundary conditions may be imposed in regions II\&IV to
correspond to physical processes for either incoming or outgoing wavepackets
for particles or antiparticles.

There is another intuitive approach to understand the same general behavior of
the wavefunction, without doing any calculations, which is in agreement with
the results of the horizon boundary conditions given in the preceding
paragraphs. Namely, by defining $\psi\left(  y\right)  \equiv\sqrt{2y}%
\varphi\left(  y\right)  ,$ Eq.(\ref{SDiffEq}) takes the standard form of the
non-relativistic Schr\"{o}dinger equation
\begin{equation}
\left[  -\partial_{y}^{2}+V_{eff}\left(  y\right)  \right]  \psi\left(
y\right)  =0,\;V_{eff}\left(  y\right)  =\frac{\mu^{2}}{2y}-\frac{\omega
^{2}+1}{\left(  2y\right)  ^{2}}, \label{VeffQ}%
\end{equation}
where the effective quantum potential, $V_{eff}\left(  y\right)  ,$ is plotted
in Fig.(8).%
\begin{center}
\includegraphics[
height=1.5982in,
width=2.5754in
]%
{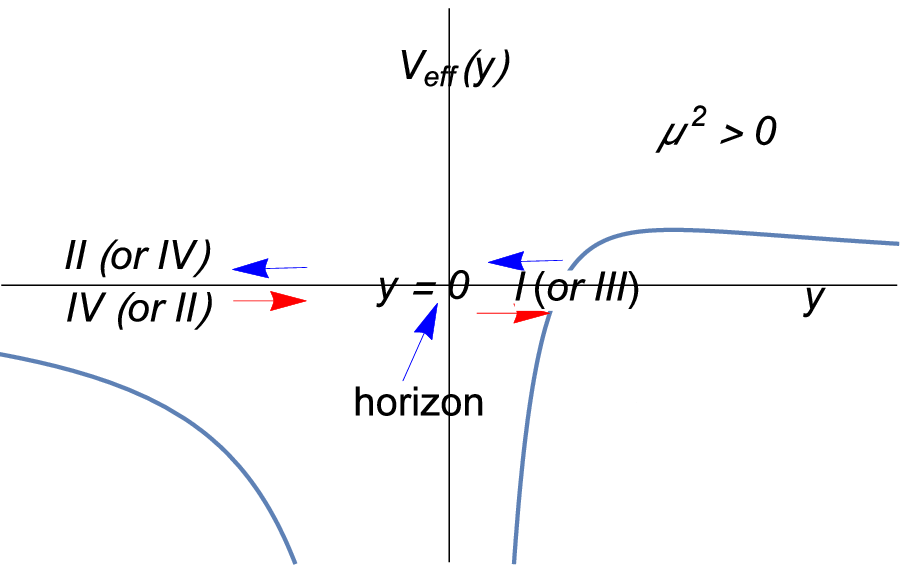}%
\\
{\protect\scriptsize Fig.(8) - Incoming/reflected waves at }$y\rightarrow
-\infty${\protect\scriptsize \ (regions II\&IV); vanishing wavefunction at
}$y\rightarrow+\infty${\protect\scriptsize \ in I \& III.}%
\end{center}

The \textquotedblleft Schr\"{o}dinger energy\ level\textquotedblright\ in
Eq.(\ref{VeffQ}) is zero, which corresponds to the real axis in Fig.(8). The
intuitive physics extracted from this figure is that of scattering of waves
from the barrier presented by the \textquotedblleft hill\textquotedblright.
Thus, oscillating waves approaching from $y\sim-\infty$ in region IV (or II)
pass the horizon at $y=0$ and move into region I (or III), they get scattered
from the barrier and move within region I (or III) toward the horizon at
$y=0,$ then they continue into region II (or IV) and go on to its asymptotic
regions, $y\sim-\infty,$ as oscillating waves. This behavior of the quantum
wavefunction is fully consistent with the geodesics in Figs.(2,3) and the
intuition gained from the mechanical potential for geodesics in Fig.(4). The
effective potential approaches of Figs.(4,8) are very important especially
when explicit solutions are not available (such as the case of general black
holes, see e.g. \cite{barsArayaJames}). A figure of the potential
$V_{eff}\left(  y\right)  $ conveys much of the physical behavior, including
boundary conditions, such as the vanishing of the wavefunction for
$y\rightarrow+\infty$ in Fig.(8), consistent with Eq.(\ref{F13asympt}).

Next, it is worth outlining the behavior of the massless field ( $\mu^{2}$ =
0) in contrast to the massive field. In this case Eq.(\ref{E+-k}) cannot be
used naively because in the massless limit either $\left(  E-k\right)  $ or
$\left(  E+k\right)  $ vanishes. So the consequences of Eqs.(\ref{bogol1}) for
the massless case need to be analyzed separately for $\left(  E-\left\vert
k\right\vert \right)  =0$. For this purpose, in these equations the integral
$\int_{-\infty}^{\infty}dk$ must be split to its positive and negative regions
and the limit $\left(  E-\left\vert k\right\vert \right)  \rightarrow0 $
taken. The integrals that contain the wildly oscillating factors $\left(
E-\left\vert k\right\vert \right)  ^{\pm i\omega}$ vanish in the limit,
leaving behind the correct massless limit of Eq.(\ref{bogol1}) for region-I in
the Rindler case (not black hole case),
\begin{equation}%
\begin{array}
[c]{c}%
a_{1\mp}\left(  \omega\right)  =\frac{e^{\pm i\theta\left(  \omega\right)  }%
}{\pm i\sqrt{1-e^{-2\pi\omega}}}\int_{0}^{\infty}\frac{dk}{\sqrt{4\pi k}%
}k^{\mp i\omega}\left(  A_{\mp}\left(  k\right)  +B_{\mp}^{\dagger}\left(
k\right)  e^{-\pi\omega}\right) \\
b_{1\mp}^{\dagger}\left(  \omega\right)  =\frac{e^{\mp i\theta\left(
\omega\right)  }}{\mp i\sqrt{1-e^{-2\pi\omega}}}\int_{0}^{\infty}\frac
{dk}{\sqrt{4\pi k}}k^{\pm i\omega}\left(  B_{\mp}^{\dagger}\left(  k\right)
+A_{\mp}\left(  k\right)  e^{-\pi\omega}\right)
\end{array}
\label{a+-massless}%
\end{equation}
noting that only half of the $A_{\mp}\left(  k\right)  ,B_{\mp}^{\dagger
}\left(  k\right)  $ survive in each line. A similar set of equations hold for
region III. Together, these may be written as Bogoliubov transformations that
correspond to the massless limit of Eq.(\ref{bogol1})
\begin{equation}%
\begin{array}
[c]{c}%
\frac{ie^{-i\theta\left(  \omega\right)  }}{\sqrt{1-e^{-2\pi\omega}}}\left(
\begin{array}
[c]{cc}%
1 & -e^{-\pi\omega}\\
-e^{-\pi\omega} & 1
\end{array}
\right)  \left(
\begin{array}
[c]{c}%
a_{1-}\left(  \omega\right) \\
b_{3+}^{\dagger}\left(  \omega\right)
\end{array}
\right)  =\int_{0}^{\infty}dk\frac{k^{-i\omega}}{\sqrt{4\pi k}}\left(
\begin{array}
[c]{c}%
A_{-}\left(  k\right) \\
B_{-}^{\dagger}\left(  k\right)
\end{array}
\right) \\
\frac{-ie^{i\theta\left(  \omega\right)  }}{\sqrt{1-e^{-2\pi\omega}}}\left(
\begin{array}
[c]{cc}%
1 & -e^{-\pi\omega}\\
-e^{-\pi\omega} & 1
\end{array}
\right)  \left(
\begin{array}
[c]{c}%
a_{3+}\left(  \omega\right) \\
b_{1-}^{\dagger}\left(  \omega\right)
\end{array}
\right)  =\int_{0}^{\infty}dk\frac{k^{i\omega}}{\sqrt{4\pi k}}\left(
\begin{array}
[c]{c}%
A_{-}\left(  k\right) \\
B_{-}^{\dagger}\left(  k\right)
\end{array}
\right) \\
\frac{-ie^{-i\theta\left(  \omega\right)  }}{\sqrt{1-e^{-2\pi\omega}}}\left(
\begin{array}
[c]{cc}%
1 & -e^{-\pi\omega}\\
-e^{-\pi\omega} & 1
\end{array}
\right)  \left(
\begin{array}
[c]{c}%
a_{3-}\left(  \omega\right) \\
b_{1+}^{\dagger}\left(  \omega\right)
\end{array}
\right)  =\int_{0}^{\infty}dk\frac{k^{-i\omega}}{\sqrt{4\pi k}}\left(
\begin{array}
[c]{c}%
A_{+}\left(  k\right) \\
B_{+}^{\dagger}\left(  k\right)
\end{array}
\right) \\
\frac{ie^{i\theta\left(  \omega\right)  }}{\sqrt{1-e^{-2\pi\omega}}}\left(
\begin{array}
[c]{cc}%
1 & -e^{-\pi\omega}\\
-e^{-\pi\omega} & 1
\end{array}
\right)  \left(
\begin{array}
[c]{c}%
a_{1+}\left(  \omega\right) \\
b_{3-}^{\dagger}\left(  \omega\right)
\end{array}
\right)  =\int_{0}^{\infty}dk\frac{k^{i\omega}}{\sqrt{4\pi k}}\left(
\begin{array}
[c]{c}%
A_{+}\left(  k\right) \\
B_{+}^{\dagger}\left(  k\right)
\end{array}
\right)
\end{array}
\label{bogol2}%
\end{equation}
The contrast with the massive case in Eq.(\ref{bogol1}) is the right hand side
of these equations, noting that only half of the $A_{\mp}\left(  k\right)
,B_{\mp}^{\dagger}\left(  k\right)  $ survive in each line for the massless
case. Furthermore, Eqs.(\ref{a+-massless},\ref{bogol2}) show that, unlike
Eq.(\ref{a+-relations}), the $a_{1\pm}$ are independent of each other for the
massless case. This implies that the massless field as written in
Eqs.(\ref{list}), but with $S\left(  0\right)  =$1, does not vanish at the
asymptotic regions of I or III, but rather has an oscillating behavior. This
is consistent with the behavior of the massless limit of the geodesics in
Figs.(2,3) as discussed in the last paragraph of section (\ref{geodesics}),
which indicate that massless particles/antiparticles do indeed travel to such
asymptotic regions. This is also evident from the intuitive effective
potential approach in Figs.(4,8) after the corresponding effective potentials
are replaced by their $\mu^{2}$ = 0 counterparts. Thus, the mechanical
potential for geodesics becomes a constant (see Eqs.(\ref{constraintR}%
,\ref{yconstraint})) so the maximum position $y_{\ast}$ in the modified
Fig.(4) moves to infinity. Similarly the effective quantum potential in the
modified Fig.(8) no longer has a barrier, so waves can move both ways from
$y=\mp\infty$ to $y=\pm\infty.$

As a check of our expressions we may compute the expectation value of the
Rindler number density operators $a_{1-}^{\dagger}a_{1-}$ etc. in the
Minkowski vacuum, $\langle0_{M}|a_{1-}^{\dagger}\left(  \omega\right)
a_{1-}\left(  \omega^{\prime}\right)  |0_{M}\rangle,$ by using directly the
Bogoliubov relation between $a_{1-}\left(  \omega\right)  $ and $A\left(
k\right)  $ \& $B^{\dagger}\left(  k\right)  $ given in Eq.(\ref{abAB}). Using
the properties of the Minkowski vacuum, $\langle0_{M}|A^{\dagger}\left(
k\right)  =0=B\left(  k\right)  |0_{M}\rangle,$ we obtain,
\begin{equation}
\langle0_{M}|a_{1-}^{\dagger}\left(  \omega\right)  a_{1-}\left(
\omega^{\prime}\right)  |0_{M}\rangle=\frac{e^{-\pi\left(  \omega
+\omega^{\prime}\right)  }\int_{-\infty}^{\infty}\frac{dk}{4\pi E\left(
k\right)  }\left(  \frac{1}{2}E\left(  k\right)  -\frac{1}{2}k\right)
^{i\left(  \omega-\omega^{\prime}\right)  }}{\sqrt{\left(  1-e^{-2\pi\omega
}\right)  \left(  1-e^{-2\pi\omega^{\prime}}\right)  }}=\frac{1}{2}%
\frac{\delta\left(  \omega-\omega^{\prime}\right)  }{e^{2\pi\omega}-1}.
\label{numVac}%
\end{equation}
The integral is given in footnote (\ref{invert1}). The result for other number
operators, $a_{1\pm}^{\dagger}a_{1\pm},~b_{1\pm}^{\dagger}b_{1\pm},$ $a_{3\pm
}^{\dagger}a_{3\pm},$ $b_{3\pm}^{\dagger}b_{3\pm},$ is the same. The factor
$\left(  e^{2\pi\omega}-1\right)  ^{-1}$ is the well known thermal
distribution which, as expected, is in agreement with previous results
\cite{unruh}.

\section{Multiverse levels in Minkowski basis \label{canonMink}}

In this section we display the multiverse directly in the Minkowski basis by
obtaining the relation between the general level $\left(  n,m\right)  $ field
and the level-$\left(  0,0\right)  $ field of Eq.(\ref{PsiM}), both expressed
in terms of Minkowski plane waves. The level $\left(  n,m\right)  $ field
$\varphi^{\left(  n,m\right)  }\left(  u,v\right)  $ can be written in terms
of level $\left(  n,m\right)  $ Minkowski oscillators $A^{\left(  n,m\right)
}\left(  k\right)  ,\bar{B}^{\left(  n,m\right)  }\left(  k\right)  $ as
follows, just like Eq.(\ref{PsiM}),
\begin{equation}
\varphi^{\left(  n,m\right)  }\left(  u,v\right)  =\int_{-\infty}^{\infty
}dk\left(  A^{\left(  n,m\right)  }\left(  k\right)  \frac{e^{-i\frac{E-k}%
{2}u}e^{-i\frac{E+k}{2}v}}{\sqrt{4\pi E}}+\bar{B}^{\left(  n,m\right)
}\left(  k\right)  \frac{e^{i\frac{E-k}{2}u}e^{i\frac{E+k}{2}v}}{\sqrt{4\pi
E}}\right)  . \label{PsinmM}%
\end{equation}
This is equivalent to the same field $\varphi^{\left(  n,m\right)  }\left(
u,v\right)  $ given in Eq.(\ref{Phinm}) in terms of Rindler oscillators. We
will derive a very non-trivial canonical transformation between the
oscillators $\left(  A^{\left(  n,m\right)  }\left(  k\right)  ,\bar
{B}^{\left(  n,m\right)  }\left(  k\right)  \right)  $ and the level $\left(
0,0\right)  $ oscillators $\left(  A\left(  k\right)  ,B^{\dagger}\left(
k\right)  \right)  $ that appear in Eqs.(\ref{PsiM1},\ref{PsiM}). This
relation represents, in the Minkowski basis, the consistent analytic
continuation of the field in the Rindler basis, as given in Eq.(\ref{Phinm}),
and it could not be obtained without going through the Rindler basis$.$

In section-\ref{bc} we related the level $\left(  0,0\right)  $ Rindler
oscillators $\left(  \left(  a_{1\mp},b_{3\pm}^{\dagger}\right)  ,\left(
a_{3\pm},b_{1\mp}^{\dagger}\right)  \right)  $ to the Minkowski oscillators
$\left(  A,\bar{B}\right)  $ and vice versa via Bogoliubov transformations in
Eqs.(\ref{bogol1},\ref{invert})$.$ The same arguments can be given for level
$\left(  n,m\right)  $ to claim the analogous forward and inverse Bogoliubov
transformations that relate $\left(  A^{\left(  n,m\right)  }\left(  k\right)
,\bar{B}^{\left(  n,m\right)  }\left(  k\right)  \right)  $ $\leftrightarrow$
$\left(  \left(  a_{1\mp}^{\left(  n\right)  },\bar{b}_{3\pm}^{\left(
m\right)  }\right)  ,\left(  a_{3\pm}^{\left(  m\right)  },\bar{b}_{1\mp
}^{\left(  n\right)  }\right)  \right)  $. These have the same formal
appearance as Eqs.(\ref{bogol1},\ref{invert}) except for inserting the level
$\left(  n,m\right)  $ oscillators instead of the level $\left(  0,0\right)  $
ones. Now, consider the pair $\left(  A^{\left(  n,m\right)  }\left(
k\right)  ,\bar{B}^{\left(  n,m\right)  }\left(  k\right)  \right)  $ in the
Bogoliubov relation analogous to (\ref{invert}), and on the right hand side
insert the level $\left(  n,m\right)  $ doublets in the following form (using
Eq.(\ref{can1}))
\begin{equation}%
\begin{array}
[c]{l}%
\left(
\begin{array}
[c]{c}%
a_{1\mp}^{\left(  n\right)  }\left(  \omega\right) \\
\bar{b}_{3\pm}^{\left(  m\right)  }\left(  \omega\right)
\end{array}
\right)  =\left(
\begin{array}
[c]{cc}%
e^{2\pi\omega n} & 0\\
0 & e^{-2\pi\omega m}%
\end{array}
\right)  \left(
\begin{array}
[c]{c}%
a_{1\mp}\left(  \omega\right) \\
b_{3\pm}^{\dagger}\left(  \omega\right)
\end{array}
\right)  ,\text{ }\\
\left(
\begin{array}
[c]{c}%
a_{3\pm}^{\left(  m\right)  }\left(  \omega\right) \\
\bar{b}_{1\mp}^{\left(  n\right)  }\left(  \omega\right)
\end{array}
\right)  =\left(
\begin{array}
[c]{cc}%
e^{2\pi\omega m} & 0\\
0 & e^{-2\pi\omega n}%
\end{array}
\right)  \left(
\begin{array}
[c]{c}%
a_{3\pm}\left(  \omega\right) \\
b_{1\mp}^{\dagger}\left(  \omega\right)
\end{array}
\right)  .
\end{array}
\label{doublets}%
\end{equation}
Moreover, replace the $\left(  0,0\right)  $ doublets $\left(  \left(
a_{1\mp},b_{3\pm}^{\dagger}\right)  ,\left(  a_{3\pm},b_{1\mp}^{\dagger
}\right)  \right)  $ that appear in (\ref{doublets}) in terms of the $\left(
0,0\right)  $ level Minkowski doublets $\left(  A\left(  k\right)  ,\bar
{B}\left(  k\right)  \right)  $ by using Eq.(\ref{bogol1}). This gives the
relation between $\left(  A^{\left(  n,m\right)  }\left(  k\right)  ,\bar
{B}^{\left(  n,m\right)  }\left(  k\right)  \right)  $ and $\left(  A\left(
k\right)  ,\bar{B}\left(  k\right)  \right)  .$ The result takes the following
form%
\begin{equation}
\left(
\begin{array}
[c]{c}%
A^{\left(  n,m\right)  }\left(  k\right) \\
\bar{B}^{\left(  n,m\right)  }\left(  k\right)
\end{array}
\right)  =\int_{-\infty}^{\infty}dk^{\prime}~M^{\left(  n,m\right)  }\left(
k,k^{\prime}\right)  \left(
\begin{array}
[c]{c}%
A\left(  k^{\prime}\right) \\
B^{\dagger}\left(  k^{\prime}\right)
\end{array}
\right)  ,\; \label{AnmA}%
\end{equation}
where the 2$\times2$ matrix $M^{\left(  n,m\right)  }\left(  k,k^{\prime
}\right)  $ in infinite momentum space, $-\infty<k,k^{\prime}<\infty,$ is
given by,
\begin{equation}
M^{\left(  n,m\right)  }\left(  k,k^{\prime}\right)  =\frac{1}{2\pi
\sqrt{E\left(  k\right)  E\left(  k^{\prime}\right)  }}\int_{-\infty}^{\infty
}d\omega~\left(  \frac{E\left(  k\right)  +k}{E\left(  k^{\prime}\right)
+k^{\prime}}\right)  ^{-i\omega}M^{\left(  n,m\right)  }\left(  \omega\right)
, \label{Mnmk}%
\end{equation}
with $E\left(  k\right)  =\sqrt{k^{2}+\mu^{2}},$ while the 2$\times2$ matrices
$M^{\left(  n,m\right)  }\left(  \omega\right)  $ in Rindler frequency space
$-\infty<\omega<\infty$ are given by
\begin{equation}%
\begin{array}
[c]{l}%
M^{\left(  n,m\right)  }\left(  \omega\right)  \equiv\frac{1}{1-e^{-2\pi
\omega}}\left(
\genfrac{}{}{0pt}{}{1}{-e^{-\omega}}%
\genfrac{}{}{0pt}{}{-e^{-\omega}}{1}%
\right)  \left(
\genfrac{}{}{0pt}{}{e^{2\pi\omega n}}{0}%
\genfrac{}{}{0pt}{}{0}{e^{-2\pi\omega m}}%
\right)  \left(
\genfrac{}{}{0pt}{}{1}{e^{-\omega}}%
\genfrac{}{}{0pt}{}{e^{-\omega}}{1}%
\right) \\
\;\;\;\;\;\;=\left(
\begin{array}
[c]{cc}%
\frac{e^{\pi\omega\left(  2n+1\right)  }-e^{-\pi\omega\left(  2m+1\right)  }%
}{e^{\pi\omega}-e^{-\pi\omega}} & \frac{e^{\pi\omega\left(  2n\right)
}-e^{-\pi\omega\left(  2m\right)  }}{e^{\pi\omega}-e^{-\pi\omega}}\\
-\frac{e^{\pi\omega\left(  2n\right)  }-e^{-\pi\omega\left(  2m\right)  }%
}{e^{\pi\omega}-e^{-\pi\omega}} & -\frac{e^{\pi\omega\left(  2n-1\right)
}-e^{-\pi\omega\left(  2m-1\right)  }}{e^{\pi\omega}-e^{-\pi\omega}}%
\end{array}
\right)  .
\end{array}
\label{Mnm}%
\end{equation}
For example, for $\left(  n,m\right)  =\left(  0,1\right)  $ or $\left(
1,0\right)  $, they are%
\begin{equation}
M^{\left(  0,1\right)  }\left(  \omega\right)  =\left(
\begin{array}
[c]{cc}%
1+e^{-2\pi\omega} & e^{-\pi\omega}\\
-e^{-\pi\omega} & 0
\end{array}
\right)  ,\;M^{\left(  1,0\right)  }\left(  \omega\right)  =\left(
\begin{array}
[c]{cc}%
1+e^{2\pi\omega} & e^{\pi\omega}\\
-e^{\pi\omega} & 0
\end{array}
\right)  . \label{Mexample}%
\end{equation}
Note the property of $M^{\left(  n,m\right)  }\left(  \omega\right)  ,$ that
when the integers are interchanged $\left(  n,m\right)  \rightarrow\left(
m,n\right)  $ and $\omega\rightarrow-\omega$, we obtain the same matrix
$M^{\left(  n,m\right)  }\left(  \omega\right)  $
\begin{equation}
M^{\left(  m,n\right)  }\left(  -\omega\right)  =M^{\left(  n,m\right)
}\left(  \omega\right)  .
\end{equation}
The diagonal matrices in Eq.(\ref{doublets}) are interchanged under the same
transformation. This explains how we end up with an integral $\int_{-\infty
}^{\infty}d\omega$ over positive and negative Rindler frequency in
Eq.(\ref{Mnmk}) even though the integrals in the Bogoliubov transformation
(\ref{bogol1}) are only over positive Rindler frequency. Thus, when the
integral $\int_{-\infty}^{\infty}d\omega$ is split into its positive and
negative pieces, and in the negative piece we replace $\omega\rightarrow
-\omega$ and interchange $\left(  n,m\right)  \rightarrow\left(  m,n\right)
,$ we see that the positive (negative) piece comes from the contribution of
the first (second) doublet in Eq.(\ref{doublets}).

The canonical conjugates that appear in the field $\bar{\varphi}^{\left(
n,m\right)  }\left(  u,v\right)  $, written in a row matrix form, are%
\begin{equation}
\left(
\begin{array}
[c]{ll}%
\bar{A}^{\left(  n,m\right)  }\left(  k\right)  & B^{\left(  n,m\right)
}\left(  k\right)
\end{array}
\right)  =\int_{-\infty}^{\infty}dk^{\prime}~\left(
\begin{array}
[c]{ll}%
A^{\dagger}\left(  k^{\prime}\right)  & B\left(  k^{\prime}\right)
\end{array}
\right)  \bar{M}^{\left(  n,m\right)  }\left(  k^{\prime},k\right)  .
\label{AnmAbar}%
\end{equation}
This is obtained by taking the hermitian conjugate of Eq.(\ref{AnmA}) and
replacing $\left(  n,m\right)  $ by $\left(  -n,-m\right)  $ on the right
side$.$ This gives the matrix $\bar{M}^{\left(  n,m\right)  }\left(
k^{\prime},k\right)  $ as follows
\begin{equation}
\bar{M}^{\left(  n,m\right)  }\left(  k^{\prime},k\right)  =\frac{1}{2\pi
\sqrt{E\left(  k\right)  E\left(  k^{\prime}\right)  }}\int_{-\infty}^{\infty
}d\omega~\left(  \frac{E\left(  k\right)  +k}{E\left(  k^{\prime}\right)
+k^{\prime}}\right)  ^{i\omega}\left(  M^{\left(  n,m\right)  }\left(
\omega\right)  \right)  ^{-1T}, \label{Mnmkbar}%
\end{equation}
where the exponent $\left(  -1T\right)  $ in $\left(  M^{\left(  n,m\right)
}\right)  ^{-1T}$ means inverse and transpose of the 2$\times2$ matrix
$M^{\left(  n,m\right)  }$, noting that the matrices in Eq.(\ref{Mnm})
satisfy, $M^{\left(  -n,-m\right)  }\left(  \omega\right)  =\left(  M^{\left(
n,m\right)  }\left(  \omega\right)  \right)  ^{-1}$ and $M^{\dagger\left(
-n,-m\right)  }\left(  \omega\right)  =\left(  M^{\left(  n,m\right)  }\left(
\omega\right)  \right)  ^{-1T}.$

These matrices satisfy the following remarkable properties in Rindler
frequency space,
\begin{equation}
M^{\left(  n,m\right)  }\left(  \omega\right)  \left(
\genfrac{}{}{0pt}{}{1}{0}%
\genfrac{}{}{0pt}{}{0}{-1}%
\right)  \left(  M^{\left(  n,m\right)  }\left(  \omega\right)  \right)
^{-1T}=\left(
\genfrac{}{}{0pt}{}{1}{0}%
\genfrac{}{}{0pt}{}{0}{-1}%
\right)  ,
\end{equation}
and in Minkowski momentum space\footnote{To prove this property we use the
following integrals, $\int_{-\infty}^{\infty}\frac{dk}{2\pi E\left(  k\right)
}\left(  E\left(  k\right)  +k\right)  ^{i\left(  \omega_{1}-\omega
_{2}\right)  }=\delta\left(  \omega_{1}-\omega_{2}\right)  ,$ and
$\int_{-\infty}^{\infty}d\omega\left(  \frac{E\left(  k_{1}\right)  +k_{1}%
}{E\left(  k_{2}\right)  +k_{2}}\right)  ^{-i\omega}=2\pi\sqrt{E_{1}E_{2}%
}\delta\left(  k_{1}-k_{2}\right)  .$}%
\begin{equation}
\int_{-\infty}^{\infty}dk~M^{\left(  n,m\right)  }\left(  k_{1},k\right)
\left(
\genfrac{}{}{0pt}{}{1}{0}%
\genfrac{}{}{0pt}{}{0}{-1}%
\right)  \bar{M}^{\left(  n,m\right)  }\left(  k,k_{2}\right)  =\left(
\genfrac{}{}{0pt}{}{1}{0}%
\genfrac{}{}{0pt}{}{0}{-1}%
\right)  \delta\left(  k_{1}-k_{2}\right)  . \label{MnmkSimplectic}%
\end{equation}
These matrix properties indicate that the transformations in Eqs.(\ref{AnmA}%
,\ref{AnmAbar}) are canonical transformations since it can be verified that
they satisfy the standard oscillator commutation rules in momentum space for
all $\left(  n,m\right)  $
\begin{equation}
\left[  \left(
\begin{array}
[c]{c}%
A^{\left(  n,m\right)  }\left(  k_{1}\right) \\
\bar{B}^{\left(  n,m\right)  }\left(  k_{2}\right)
\end{array}
\right)  ,\left(
\begin{array}
[c]{ll}%
\bar{A}^{\left(  n,m\right)  }\left(  k_{2}\right)  & B^{\left(  n,m\right)
}\left(  k_{2}\right)
\end{array}
\right)  \right]  =\left(
\genfrac{}{}{0pt}{}{1}{0}%
\genfrac{}{}{0pt}{}{0}{-1}%
\right)  \delta\left(  k_{1}-k_{2}\right)  .
\end{equation}
This includes the original Minkowski commutators given in Eq.(\ref{Mvac}) that
are reproduced in the case of $n=m=0$. Clearly, by the construction of
Eq.(\ref{AnmA}), the level $\left(  n,m\right)  $ commutator follows directly
from the level $\left(  0,0\right)  $ commutator and the remarkable properties
of the matrix $M^{\left(  n,m\right)  }\left(  k,k^{\prime}\right)  $ that
relates the $\left(  0,0\right)  $ and $\left(  n,m\right)  $ levels to each
other as a canonical transformation.

This establishes the quantum properties of the multiverse in the Minkowski
basis for all levels $\left(  n,m\right)  $. It is evident that the
Minkowski-basis field $\varphi^{\left(  n,m\right)  }\left(  u,v\right)  $ at
all levels given in Eqs.(\ref{PsinmM},\ref{AnmA},\ref{Mnmk},\ref{Mnm}),
including $n=m=0,$ inherits its properties only from the analyticity
properties of the $\left(  0,0\right)  $ level field $\varphi\left(
u,v\right)  $ in the extended Rindler basis.

\section{Charge conservation and information flow \label{prob}}

In this section we address the question on whether information flows from one
level of the Rindler multiverse to other levels. To do this we consider the
probability associated with a wavepacket. As expected, a wavepacket will on
the average follow the path of a geodesic as it develops as a function of
time. Earlier in the paper we discussed the geodesics at the classical level,
and of course at the classical level, since there is no multiverse, the
geodesics cannot give information on our question. However, a wavepacket may
leak to other levels of the multiverse when it crosses the horizons. The
question is whether it does or not.

For the Klein-Gordon field in curved spacetime, that is normalized according
to the Klein-Gordon dot product in footnote (\ref{norm}), probablity is
directly related to the conserved charge current up to the sign of the charge.
While the probability density is always positive, the charge density is
positive/negative for particles/antiparticles respectively (i.e. $a$ versus
$b$ symbols in the wavepacket). Therefore, to understand probability (or
information) flow we study the flow of the charge current with all $a$ and $b$
coefficients included in order to undestand the flow based on the most general
wavepacket including particles and antiparticles.

The conserved current for the Klein-Gordon equation, $\left(  \nabla^{2}%
-\mu^{2}\right)  \varphi=0,$ in curved spacetime is
\begin{equation}
J^{\mu}\left(  x\right)  =-i\sqrt{-g}g^{\mu\nu}\left(  \varphi^{\dagger
}\partial_{\nu}\varphi-\partial_{\nu}\varphi^{\dagger}\varphi\right)
,\;\;\partial_{\mu}J^{\mu}\left(  x\right)  =0. \label{current}%
\end{equation}
The conservation $\partial_{\mu}J^{\mu}\left(  x\right)  =0$ is verified by
using the Klein-Gordon equation $\left(  \sqrt{-g}\right)  ^{-1}\partial_{\mu
}\left(  \sqrt{-g}g^{\mu\nu}\partial_{\nu}\varphi\right)  =\mu^{2}\varphi$.
The conserved charge associated with this current is computed as an integral
over a spacelike Cauchy surface $\Sigma$
\begin{equation}
Q=\int_{\Sigma}J^{\mu}d\Sigma_{\mu}%
\end{equation}
The conserved charge is independent of the surface $\Sigma.$

The Cauchy surface can be specified differently in the Minkowski versus
Rindler bases. In the Minkowski version the surface is defined by taking a
slice of constant $x^{0}$ and evaluating the integral as given in textbooks%
\begin{equation}
Q_{M}\left(  x^{0}\right)  =\int_{-\infty}^{\infty}dx^{1}J^{0}\left(
x^{0},x^{1}\right)  =\int_{-\infty}^{\infty}dk^{1}\left(  A^{\dagger}\left(
k^{1}\right)  A\left(  k^{1}\right)  -B^{\dagger}\left(  k^{1}\right)
B\left(  k^{1}\right)  \right)  , \label{QMi}%
\end{equation}
where the computation is performed by using the Minkowski version of the field
in Eq.(\ref{PsiM}) at constant finite values $x^{0}.$ The time derivative
$\partial_{x^{0}}Q_{M}\left(  x^{0}\right)  $ is
\begin{equation}
\partial_{0}Q_{M}\left(  x^{0}\right)  =\int_{-\infty}^{\infty}dx^{1}%
\partial_{0}J^{0}=\int_{-\infty}^{\infty}dx^{1}\left(  \partial_{\mu}J^{\mu
}-\partial_{1}J^{1}\right)  =-J^{1}\left(  x^{0},\infty\right)  +J^{1}\left(
x^{0},-\infty\right)  .
\end{equation}
where the Klein-Gordon equation is used to set $\partial_{\mu}J^{\mu}=0$, and
then Stoke's theorem is applied to write the result in terms of the current
$J^{1}\left(  x^{0},x^{1}\right)  $ evaluated at the asymptotic boundaries. In
general the current at the boundaries,
\begin{equation}
J^{1}\left(  x^{0},\pm\infty\right)  =\lim_{x^{1}\rightarrow\pm\infty}\left(
-i\left(  \varphi^{\dagger}\left(  x^{0},x^{1}\right)  \partial_{1}%
\varphi\left(  x^{0},x^{1}\right)  -\partial_{1}\varphi^{\dagger}\left(
x^{0},x^{1}\right)  \varphi\left(  x^{0},x^{1}\right)  \right)  \right)  ,
\end{equation}
does not vanish as this represents the charge flux of incoming/outgoing
particles, so in such physical processes $\partial_{0}Q_{M}\left(
x^{0}\right)  $ cannot vanish at asymptotic boundaries. On the other hand
Eq.(\ref{QMi}) shows that $Q_{M}\left(  x^{0}\right)  $ is time independent at
finite $x^{0}.$ These observations are reconciled by noting that the support
of $J^{1}\left(  x^{0},\pm\infty\right)  $ is not only at space infinity
$x^{1}=\pm\infty,$ but also at time infinity $x^{0}=\pm\infty,$ such as
$J^{1}\left(  x^{0},\pm\infty\right)  =\pm\delta\left(  x^{0}\pm\infty\right)
J,$ where $J$ is a constant determined in terms of $\left(  A\left(  k\right)
,B^{\dagger}\left(  k\right)  \right)  $ as shown in Eq.(\ref{J}) below. Then
the charge conservation equation takes the form
\begin{equation}
\partial_{x^{0}}Q_{M}\left(  x^{0}\right)  =\left(  \delta\left(  x^{0}%
+\infty\right)  -\delta\left(  x^{0}-\infty\right)  \right)  J. \label{dQM}%
\end{equation}
This result implies that $Q_{M}\left(  x^{0}\right)  $ is not in general a
constant at the asymptotic past and future boundaries of Minkowski space.
Furthermore, the conservation of charge $Q_{M}$ in Minkowski space at finite
$x^{0}$ is explained by the fact that the flux of charge $J$ into Minkowski
space at $x^{0}=-\infty$ is exactly equal to the flow of charge $J$ out of the
space at $x^{0}=\infty.$ This is the statement of conservation of charge and
it implies conservation of information within the Minkowski spacetime. It also
leads to unitarity in the complete Hilbert space in the quantum field theory.

In the Rindler case the \textit{spacelike} Cauchy surface needs to be
specified differently in each region because the roles of $\left(  t,y\right)
$ alternate between time and space in regions I\&III versus regions II\&IV.
For example, in region I, a spacelike surface correspond to a fixed value of
the Rindler time $t$ (any ray in Fig.1) so that the charge of a field
configuration is given by integrating over $d\Sigma_{t}=dy$ at fixed $t,$
\begin{equation}
q_{1}\left(  t\right)  =\int_{\Sigma}J^{\mu}\left(  x\right)  d\Sigma_{\mu
}=\int_{y_{1}}^{y_{2}}dyJ^{t}\left(  t,y\right)  ,\text{ with }J^{t}=\frac
{i}{2y}\left(  \varphi_{1}^{\dagger}\partial_{t}\varphi_{1}-\partial
_{t}\varphi_{1}^{\dagger}\varphi_{1}\right)  , \label{rho}%
\end{equation}
where, in $J^{t}$ we used $\sqrt{-g}=1,$ $g^{tt}=-\left(  2y\right)  ^{-1},$
and $\varphi_{1}\left(  t,y\right)  $ given in Eq.(\ref{list}). Here
$J^{t}\left(  t,y\right)  $ is the charge density, so $\int_{y_{1}}^{y_{2}%
}dyJ^{t}$ is the total charge contained in the interval $y_{1}<y<y_{2}$ at
time\ $t.$ Changing the value of $t$ in the range, $-\infty<t<\infty,$ covers
the spacetime bounded by the hyperbolas shown in Fig(1) within region-I.
Sending $y_{1}\rightarrow0$ and $y_{2}\rightarrow\infty$ covers the entire
region-I. Then $q_{1}\left(  t\right)  ,$ with $y_{1}=0$ and $y_{2}=\infty,$
is the total regional charge within region I at an arbitrary time $t.$

By contrast to Eq.(\ref{rho}), in region II a \textit{spacelike} Cauchy
surface\footnote{One may be tempted to ignore the spacelike requirement of the
Cauchy surface, and based on the fact that $\partial_{t}$ is the conserved
Killing vector in all regions, including region II, one may take the surface
of integration in region II to be again $d\Sigma_{t}=dy$ just like region I.
Applying this reasoning uniformly to every region, one may wish to define
$q_{1,2,3,4}$ as integrals over $y$ at fixed $t,$ just as in Eq.(\ref{rho}).
This turns out to give the wrong set of sign patterns for $q_{1,2,3,4}$
contrary to the correct patterns displayed in our results in Eqs.(\ref{rho1}%
-\ref{drhoRin}): i.e. $+a^{\dagger}a$ for the charges associated to particles
and the opposite signs $-b^{\dagger}b$ for antiparticles. The wrong set of
signs that occur differently in different regions fail the self consistency
check involving the Bogoliubov transformations as given in Eqs.(\ref{Bo}%
,\ref{J}). \label{wrong}} corresponds to a fixed value of the
\textquotedblleft time\textquotedblright\ $y$ (a fixed hyperbola in region II
in Fig.1, not shown) so that the charge of a field configuration is given by
integrating over $d\Sigma_{y}=dt$ at fixed $y,$
\begin{equation}
q_{2}\left(  y\right)  =\int_{\Sigma}J^{\mu}\left(  x\right)  d\Sigma_{\mu
}=-\int_{-\infty}^{\infty}dtJ^{y}\left(  t,y\right)  ,\text{ with }%
J^{y}=\left(  -2yi\right)  \left(  \varphi_{2}^{\dagger}\partial_{y}%
\varphi_{2}-\partial_{y}\varphi_{2}^{\dagger}\varphi_{2}\right)  , \label{Jyy}%
\end{equation}
where, in $J^{y}$ we used$\sqrt{-g}=1,$ $g^{yy}=2y,$ and $\varphi_{2}\left(
t,y\right)  $ given in Eq.(\ref{list}). The reason for the extra overall sign
in the integral $-\int_{-\infty}^{\infty}dt$ will be explained below after
Eq.(\ref{drhoRin}).

The Rindler version of the total charge $Q_{R}$ for the full extended Rindler
space in the $\left(  0,0\right)  $ universe (the equivalent of $Q_{M}$ for
the Minkowski space in Eq.(\ref{QMi})) is given by integrals at
\textit{spacelike} Cauchy surfaces of constant $t$ in regions I \& III and
constant $y$ in regions II \& IV. This is because in regions I \& III $t$ is
the timelike coordinate because sign$\left(  y\right)  =+1$ for the spacetime
geometry described by the line element in Eq.(\ref{Laplace}), while in regions
II \& IV $y$ is the timelike coordinate because sign$\left(  y\right)  =-1$.
The computation of the total charge $Q_{R}$ is to be performed by using the
regional Rindler fields $\varphi_{1,2,3,4}$ given in Eq.(\ref{list}). We
define the total charge, $Q_{R}=\int_{\Sigma}J^{\mu}\left(  x\right)
d\Sigma_{\mu},$ as an integral over the union of spacelike Cauchy surfaces
that were used in the definition of $q_{1,2,3,4}.$ Then we find
\begin{equation}
Q_{R}=q_{1}+q_{2}+q_{3}+q_{4}, \label{QRi}%
\end{equation}
where
\begin{equation}%
\begin{array}
[c]{l}%
q_{1}\left(  t\right)  =\int_{0}^{\infty}dyJ_{1}^{t}=\int_{0}^{\infty}%
dy\frac{i}{2y}\left(  \varphi_{1}^{\dagger}\partial_{t}\varphi_{1}%
-\partial_{t}\varphi_{1}^{\dagger}\varphi_{1}\right) \\
q_{2}\left(  y\right)  =-\int_{-\infty}^{\infty}dtJ_{2}^{y}=-\int_{-\infty
}^{\infty}dt~\left(  -2yi\right)  \left(  \varphi_{2}^{\dagger}\partial
_{y}\varphi_{2}-\partial_{y}\varphi_{2}^{\dagger}\varphi_{2}\right) \\
q_{3}\left(  t\right)  =-\int_{0}^{\infty}dyJ_{3}^{t}=-\int_{0}^{\infty
}dy\frac{i}{2y}\left(  \varphi_{3}^{\dagger}\partial_{t}\varphi_{3}%
-\partial_{t}\varphi_{3}^{\dagger}\varphi_{3}\right) \\
q_{4}\left(  y\right)  =\int_{-\infty}^{\infty}dtJ_{4}^{y}=\int_{-\infty
}^{\infty}dt~\left(  -2yi\right)  \left(  \varphi_{4}^{\dagger}\partial
_{y}\varphi_{4}-\partial_{y}\varphi_{4}^{\dagger}\varphi_{4}\right)
\end{array}
\label{rhoRin}%
\end{equation}
Furthermore, the rate of change of these charges with respect to
\textquotedblleft time\textquotedblright\ is given by the \textquotedblleft
time\textquotedblright\ derivatives in the respective regions
\begin{equation}
\partial_{t}q_{1}\left(  t\right)  ,\;-\partial_{y}q_{2}\left(  y\right)
,\;-\partial_{t}q_{3}\left(  t\right)  ,\;\;\partial_{y}q_{4}\left(  y\right)
. \label{drhoRin}%
\end{equation}

Note the extra overall minus signs in the definitions of the charges
$q_{2}\left(  t\right)  $ and $q_{3}\left(  y\right)  $ as well as the extra
signs in taking their \textquotedblleft time\textquotedblright\ derivatives.
The justification for such extra signs is the comparison of the time
increments for the Minkowski sign$\left(  dx^{0}\right)  $ to the Rindler
sign$\left(  dT\right)  $ where $T$ is the monotonically \textit{increasing}
\textquotedblleft time\textquotedblright\ in the corresponding regions. Thus,
in region III we have $T\equiv t$ and sign$\left(  dx^{0}\right)
=-$sign$\left(  dt\right)  $ because in region III as $t$ decreases as $x^{0}$
increases. This explains why $q_{3}\left(  t\right)  $ and $-\partial_{t}%
q_{3}\left(  t\right)  $ have an extra sign: it is because the extra sign is
absorbed into $\partial/\partial\left(  -t\right)  $ both in the definition of
the current $J^{t}$ in Eq.(\ref{rho}) and in the rate of change, so that
$\partial/\partial\left(  -t\right)  $ implies an increment of time with the
same sign as $\partial/\partial x^{0}.$ The same explanation works for region
II, where $T\equiv-y$ since $y$ is negative, and noting that in region II
sign$\left(  dx^{0}\right)  =-$sign$\left(  dy\right)  =$sign$\left(
d\left\vert y\right\vert \right)  ;$ consequently the overall sign is absorbed
into $\partial/\partial\left(  -y\right)  =\partial/\partial\left\vert
y\right\vert .$ By contrast, in region IV where $y$ is negative, we have
$T=-y$ but sign$\left(  dx^{0}\right)  =$sign$\left(  dy\right)
=-$sign$\left(  d\left\vert y\right\vert \right)  ,$ therefore no extra signs
are needed in region IV. When these extra signs are combined with the signs
produced when the currents are integrated on \textit{spacelike} Cauchy
surfaces (see footnote (\ref{wrong})) one obtains the correct sign patterns
for particle/antiparticle charges and charge fluxes in our results given below.

The explicit computation of $q_{1,2,3,4}$ shows that they are constants within
each region, but $\left(  \partial_{t}q_{1}\left(  t\right)  ,\partial
_{-y}q_{2}\left(  y\right)  ,\partial_{-t}q_{3}\left(  t\right)  ,\partial
_{y}q_{4}\left(  y\right)  \right)  $ receive non-trivial contributions at the
horizons and the asymptotic boundaries of each region (analogous to constant
$Q_{M}$ but nontrivial $\partial Q_{M}$ at boundaries as in Eqs.(\ref{QMi}%
,\ref{dQM})). The results are as follows.

For region I, according to the computations shown in Appendix (\ref{appendix})
we have
\begin{equation}%
\begin{array}
[c]{l}%
q_{1}=\int_{0}^{\infty}d\omega\left(  \left(  a_{1-}^{\dagger}a_{1-}%
-b_{1-}^{\dagger}b_{1-}\right)  +\left(  a_{1+}^{\dagger}a_{1+}-b_{1+}%
^{\dagger}b_{1+}\right)  \right)  ,\\
\left.  \partial_{t}q_{1}\left(  t\right)  \right\vert _{\mu^{2}=0}=\int
_{0}^{\infty}d\omega\left[
\begin{array}
[c]{l}%
\left(  \lim_{v\rightarrow-\infty}-\lim_{v\rightarrow0}\right)  \delta
_{\varepsilon}\left(  \ln u\right)  \left(  a_{1-}^{\dagger}a_{1-}%
-b_{1-}^{\dagger}b_{1-}\right) \\
+\left(  \lim_{u\rightarrow0}-\lim_{u\rightarrow\infty}\right)  \delta
_{\varepsilon}\left(  \ln\left\vert v\right\vert \right)  \left(
a_{1+}^{\dagger}a_{1+}-b_{1+}^{\dagger}b_{1+}\right)
\end{array}
\right] \\
\left.  \partial_{t}q_{1}\left(  t\right)  \right\vert _{\mu^{2}\neq0}%
=\int_{0}^{\infty}d\omega\left[
\begin{array}
[c]{c}%
-\lim_{v\rightarrow0}\delta_{\varepsilon}\left(  \ln u\right)  \left(
a_{1-}^{\dagger}a_{1-}-b_{1-}^{\dagger}b_{1-}\right) \\
+\lim_{u\rightarrow0}\delta_{\varepsilon}\left(  \ln\left\vert v\right\vert
\right)  \left(  a_{1+}^{\dagger}a_{1+}-b_{1+}^{\dagger}b_{1+}\right)
\end{array}
\right]
\end{array}
\label{rho1}%
\end{equation}
Here the symbol $\delta_{\varepsilon}\left(  z\right)  $ is a smeared delta
function defined in Eqs.(\ref{smeared}-\ref{smeared3}). We discuss briefly the
meaning of these equations. First note that the charge $q_{1}$ is conserved
\textit{within} region I by itself, $\partial_{t}q_{1}\left(  t\right)  =0$ at
finite $t,$ since $q_{1}$ is explicitly time independent. However, the
expression for massless particles $\left.  \partial_{t}q_{1}\left(  t\right)
\right\vert _{\mu^{2}=0}$ shows that charge is not conserved locally at both
horizons $u=0$ or $v=0$ and at both asymptotic boundaries $u=\infty$ or
$v=\infty$; similarly for massive massive particles $\left.  \partial_{t}%
q_{1}\left(  t\right)  \right\vert _{\mu^{2}\neq0}$ shows that charge is not
conserved locally at both horizons (the wavefunction and current in regions
I\&III\ vanish asymptotically for massive particles, so $\left.  \partial
_{t}q_{1}\left(  t\right)  \right\vert _{\mu^{2}\neq0}$ at $y\rightarrow
\pm\infty$).%
\begin{center}
\includegraphics[
height=2.444in,
width=2.444in
]%
{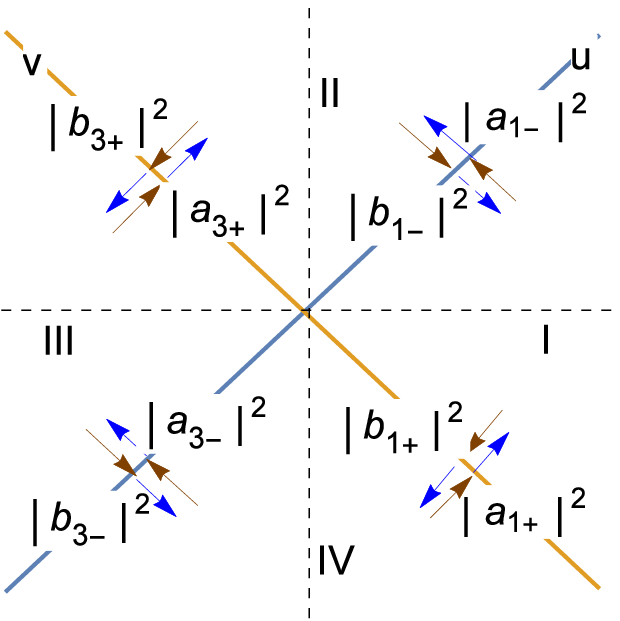}%
\\
Fig.(9) - {\protect\scriptsize Magnitudes of incoming and outgoing fluxes at
the horizons of each region. Blue arrows = }$+${\protect\scriptsize \ sign,
and red arrows = }$-${\protect\scriptsize \ sign. For example, for region I,
at the future horizon }$\left(  v=0\right)  ,${\protect\scriptsize \ the
outgoing particle current is proportional to }$\left\vert a_{1-}^{2}\left(
\omega\right)  \right\vert ${\protect\scriptsize \ (red) and the incoming
antiparticle current is proportional to }$\left\vert b_{1-}\left(
\omega\right)  \right\vert ^{2}${\protect\scriptsize \ (blue). Similarly at
the past horizon of region I }$\left(  u=0\right)  ,$%
{\protect\scriptsize \ the incoming particle current is proportional to
}$\left\vert a_{1+}\left(  \omega\right)  \right\vert ^{2}$%
{\protect\scriptsize \ (blue) and the outgoing antiparticle current is
}$\left\vert b_{1+}^{\dagger}\left(  \omega\right)  \right\vert ^{2}%
${\protect\scriptsize \ (red). Similar in and out currents are indicated for
each region.}%
\end{center}

That we should expect non-trivial charge flow at the horizons for the massive
particle, $\left.  \partial_{t}q_{1}\left(  t\right)  \right\vert _{\mu
^{2}\neq0}\neq0$ at $t\rightarrow\pm\infty$ and $y=0,$ was evident in
Figs.(2,3) that depict the classical geodesics for massive particles that show
geodesics crossing the horizons $\left(  y=0\right)  $ at $t=\pm\infty.$ In
the quantum computation, using general wavepackets with particles and
antiparticles, we see in Eq.(\ref{rho1}) and Fig.(9) that at each frequency
$\omega$ there is a charge flux ($-\left\vert a_{1-}\left(  \omega\right)
\right\vert ^{2}+\left\vert b_{1-}^{\dagger}\left(  \omega\right)  \right\vert
^{2}$) due to outgoing particles $\left(  \text{overall }-\text{sign}\right)
$ and incoming antiparticles $\left(  +\text{sign}\right)  $ at the future
horizon in region I $\left(  v=0\right)  ,$ and another charge flux
($+\left\vert a_{1+}\left(  \omega\right)  \right\vert ^{2}-\left\vert
b_{1+}^{\dagger}\left(  \omega\right)  \right\vert ^{2}$) due to incoming
particles $\left(  +\text{sign}\right)  $ and outgoing antiparticles $\left(
-\text{sign}\right)  $ at the past horizon in region I $\left(  u=0\right)  .$
These incoming and outgoing fluxes sum up to zero because, for the massive
particle, the boundary conditions are $\left\vert a_{1-}\left(  \omega\right)
\right\vert =\left\vert a_{1+}\left(  \omega\right)  \right\vert $ and
$\left\vert b_{1-}^{\dagger}\left(  \omega\right)  \right\vert =\left\vert
b_{1+}^{\dagger}\left(  \omega\right)  \right\vert $ as seen in Eq.$\left(
\text{\ref{a+-relations}}\right)  .$ For massless particles $\left\vert
a_{1\pm}\left(  \omega\right)  \right\vert $ and similarly $\left\vert
b_{1\pm}^{\dagger}\left(  \omega\right)  \right\vert $ are unrelated to each
other, but the result for $\left.  \partial_{t}q_{1}\left(  t\right)
\right\vert _{\mu^{2}=0}$ given above indicates that again the total incoming
flux of charge into region I is equal to the total outgoing flux of charge.

We emphasize the fact that the incoming and outgoing particle/antiparticle
fluxes sum up to zero \textit{separately} for every species of particle or
antiparticle, for either massless or massive particles of every frequency
$\omega$. This indicates that what comes into region I goes out fully in the
same form (species of particle or antiparticle) at each frequency $\omega$.
This is why the total charge \textit{within} region I remains a constant
$\partial_{t}q_{1}=0$ at finite $t,$ for every species \textit{separately},
and not by cancelletion among the different species (i.e. $a_{1\pm},a_{3\pm
},b_{1\pm},b_{3\pm}$).

For region II, the result of the computations shown in Appendix
(\ref{appendix}) is%
\begin{equation}%
\begin{array}
[c]{l}%
q_{2}=\int_{0}^{\infty}d\omega\left(  \left(  a_{1-}^{\dagger}a_{1-}%
-b_{1-}^{\dagger}b_{1-}\right)  +\left(  -b_{3+}^{\dagger}b_{3+}%
+a_{3+}^{\dagger}a_{3+}\right)  \right)  ,\\
\left.  \partial_{-y}q_{2}\left(  y\right)  \right\vert _{\mu^{2}=0}=\int
_{0}^{\infty}d\omega\left[
\begin{array}
[c]{l}%
\left(  \lim_{v\rightarrow0}-\lim_{v\rightarrow\infty}\right)  \delta
_{\varepsilon}\left(  \ln u\right)  \left(  a_{1-}^{\dagger}a_{1-}%
-b_{1-}^{\dagger}b_{1-}\right) \\
+\left(  \lim_{u\rightarrow0}-\lim_{u\rightarrow\infty}\right)  \delta
_{\varepsilon}\left(  \ln v\right)  \left(  -b_{3+}^{\dagger}b_{3+}%
+a_{3+}^{\dagger}a_{3+}\right)
\end{array}
\right] \\
\left.  \partial_{-y}q_{2}\left(  y\right)  \right\vert _{\mu^{2}\neq0}%
=\int_{0}^{\infty}d\omega\left[
\begin{array}
[c]{l}%
+\lim_{v\rightarrow0}\delta_{\varepsilon}\left(  \ln u\right)  \left(
a_{1-}^{\dagger}a_{1-}-b_{1-}^{\dagger}b_{1-}\right) \\
+\lim_{u\rightarrow0}\delta_{\varepsilon}\left(  \ln v\right)  \left(
-b_{3+}^{\dagger}b_{3+}+a_{3+}^{\dagger}a_{3+}\right) \\
-\lim_{v\rightarrow\infty}\delta_{\varepsilon}\left(  \ln u\right)
\frac{\left\vert a_{1-}^{\dagger}+e^{-\pi\omega}\left(  \mu^{2}\right)
^{-i\omega}e^{i2\theta}b_{3+}^{\dagger}\right\vert ^{2}-\left\vert
b_{1-}+e^{-\pi\omega}\left(  \mu^{2}\right)  ^{i\omega}e^{-i2\theta}%
a_{3+}\right\vert ^{2}}{1-e^{-2\pi\omega}}\\
-\lim_{u\rightarrow\infty}\delta_{\varepsilon}\left(  \ln v\right)
\frac{-\left\vert b_{3+}^{\dagger}+e^{-\pi\omega}\left(  \mu^{2}\right)
^{i\omega}e^{-i2\theta}a_{1-}\right\vert ^{2}+\left\vert a_{3+}+e^{-\pi\omega
}\left(  \mu^{2}\right)  ^{-i\omega}e^{i2\theta}b_{1-}^{\dagger}\right\vert
^{2}}{1-e^{-2\pi\omega}}%
\end{array}
\right]
\end{array}
\label{rho2}%
\end{equation}
The interpretation of these expressions for region II is similar to the one
above for region I. The total charge $q_{2}$ is explicitly time independent
\textit{within} region II, but its rate of change locally at each horizon
$u=0$ or $v=0,$ or asymptotic boundaries $u\rightarrow\infty$ or
$v\rightarrow\infty$, is generally nonzero. However, again what comes into
region II goes out of region II in the same total form (either particle or
antiparticle) at each frequency $\omega.$ The vanishing of the sum of incoming
and outgoing fluxes for all boundaries of region II is evident without any
computation for the massless particle. For the massive particle, simple
algebra such as
\begin{equation}
\frac{-\left\vert a_{1-}^{\dagger}+e^{-\pi\omega}\left(  \mu^{2}\right)
^{-i\omega}e^{i2\theta}b_{3+}^{\dagger}\right\vert ^{2}+\left\vert
b_{3+}^{\dagger}+e^{-\pi\omega}\left(  \mu^{2}\right)  ^{i\omega}e^{-i2\theta
}a_{1-}\right\vert ^{2}}{1-e^{-2\pi\omega}}=-\left\vert a_{1-}\right\vert
^{2}+\left\vert b_{3+}^{\dagger}\right\vert ^{2}, \label{simple}%
\end{equation}
shows that the \textit{sum} of asymptotic fluxes at $u\rightarrow\infty$ and
$v\rightarrow\infty,$ namely [($-\left\vert a_{1-}\right\vert ^{2}+\left\vert
b_{3+}^{\dagger}\right\vert ^{2})+(-\left\vert a_{3+}\right\vert
^{2}+\left\vert b_{1-}^{\dagger}\right\vert ^{2})],$ matches the \textit{sum}
of the fluxes at the horizons $u=0$ and $v=0$ except for an overall sign.
Hence, the total sum over all boundaries vanishes. We conclude (similar to
region I) that charge, information or probability, are conserved within region
II by itself.

Observe that the incoming (outgoing) particle (antiparticle) charge at the
$v=0$ horizon of region II, is identical to the particle (antiparticle) flux
that leaves (enters) region I. This shows that the flux of particles and
antiparticles is continuous across the horizon at the boundary of regions
I\&II as indicated in Fig.(9).

For region III, the computations are parallel to those for region I. The
result is obtained from Eq.(\ref{rho1}) simply by replacing $\left(
u,-v\right)  \rightarrow\left(  -u,v\right)  $ (see Fig.1) and $\left(
a_{1-},a_{1+}\right)  \rightarrow\left(  b_{3-}^{\dagger},b_{3+}^{\dagger
}\right)  $ (see Eq.(\ref{list})), and multiplying by an overall minus sign
for $q$ but not for $\partial q$ (see Eq.(\ref{rhoRin},\ref{drhoRin})). The
result is%
\begin{equation}%
\begin{array}
[c]{l}%
q_{3}=\int_{0}^{\infty}d\omega\left(  \left(  -b_{3-}^{\dagger}b_{3-}%
+a_{3-}^{\dagger}a_{3-}\right)  +\left(  -b_{3+}^{\dagger}b_{3+}%
+a_{3+}^{\dagger}a_{3+}\right)  \right) \\
\left.  \partial_{-t}q_{3}\left(  t\right)  \right\vert _{\mu^{2}=0}=\int
_{0}^{\infty}d\omega\left[
\begin{array}
[c]{l}%
\left(  \lim_{v\rightarrow\infty}-\lim_{v\rightarrow0}\right)  \delta
_{\varepsilon}\left(  \ln\left\vert u\right\vert \right)  \left(
b_{3-}^{\dagger}b_{3-}-a_{3-}^{\dagger}a_{3-}\right) \\
+\left(  \lim_{u\rightarrow0}-\lim_{u\rightarrow-\infty}\right)
\delta_{\varepsilon}\left(  \ln v\right)  \left(  b_{3+}^{\dagger}%
b_{3+}-a_{3+}^{\dagger}a_{3+}\right)
\end{array}
\right] \\
\left.  \partial_{-t}q_{3}\left(  t\right)  \right\vert _{\mu^{2}\neq0}%
=\int_{0}^{\infty}d\omega\left[
\begin{array}
[c]{c}%
-\lim_{v\rightarrow0}\delta_{\varepsilon}\left(  \ln\left\vert u\right\vert
\right)  \left(  b_{3-}^{\dagger}b_{3-}-a_{3-}^{\dagger}a_{3-}\right) \\
+\lim_{u\rightarrow0}\delta_{\varepsilon}\left(  \ln v\right)  \left(
b_{3+}^{\dagger}b_{3+}-a_{3+}^{\dagger}a_{3+}\right)
\end{array}
\right]
\end{array}
\label{rho3}%
\end{equation}
Recall that for the massive particle $\left\vert a_{3+}\right\vert =\left\vert
a_{3-}\right\vert $ and $\left\vert b_{3+}\right\vert =\left\vert
b_{3-}\right\vert $ according to the boundary conditions obtained in
Eq.(\ref{a+-relations}). For the massless particle there are no such
relations. The interpretation is parallel to the discussion above for region
I. Furthermore, at the horizon at the common boundary for regions II\&III we
see that what leaves (enters) region II fully enters (leaves) region III. From
this we conclude (similar to region I or II) that charge, information or
probability, are conserved \textit{within} region III by itself, independent
of what goes on in other regions of the extended Rindler space. This leads
also to unitarity of the scattering matrix in the quantum Hilbert space of
region III by itself.

For region IV, the computations are parallel to those for region II. The
result is obtained from Eq.(\ref{rho2}) simply by replacing $\left(
u,v\right)  \rightarrow\left(  -u,-v\right)  $ (see Fig.1) and $\left(
a_{1-},b_{3+}^{\dagger}\right)  \rightarrow\left(  b_{3-}^{\dagger}%
,a_{1+}\right)  $ (see Eq.(\ref{list})), and multiplying by an overall minus
sign for $q$ but not for $\partial q$ (see Eq.(\ref{rhoRin},\ref{drhoRin})).
The result is%
\begin{equation}%
\begin{array}
[c]{l}%
q_{4}=\int_{0}^{\infty}d\omega\left(  \left(  -b_{3-}^{\dagger}b_{3-}%
+a_{3-}^{\dagger}a_{3-}\right)  +\left(  a_{1+}^{\dagger}a_{1+}-b_{1+}%
^{\dagger}b_{1+}\right)  \right) \\
\left.  \partial_{y}q_{4}\left(  y\right)  \right\vert _{\mu^{2}=0}=\int
_{0}^{\infty}d\omega\left[
\begin{array}
[c]{l}%
\left(  \lim_{v\rightarrow0}-\lim_{v\rightarrow-\infty}\right)  \delta
_{\varepsilon}\left(  \ln\left\vert u\right\vert \right)  \left(
b_{3-}^{\dagger}b_{3-}-a_{3-}^{\dagger}a_{3-}\right) \\
+\left(  \lim_{u\rightarrow-\infty}-\lim_{u\rightarrow0}\right)
\delta_{\varepsilon}\left(  \ln\left\vert v\right\vert \right)  \left(
a_{1+}^{\dagger}a_{1+}-b_{1+}^{\dagger}b_{1+}\right)
\end{array}
\right] \\
\left.  \partial_{y}q_{4}\left(  y\right)  \right\vert _{\mu^{2}\neq0}%
=\int_{0}^{\infty}d\omega\left[
\begin{array}
[c]{l}%
+\lim_{v\rightarrow0}\delta_{\varepsilon}\left(  \ln\left\vert u\right\vert
\right)  \left(  b_{3-}^{\dagger}b_{3-}-a_{3-}^{\dagger}a_{3-}\right) \\
-\lim_{u\rightarrow0}\delta_{\varepsilon}\left(  \ln\left\vert v\right\vert
\right)  \left(  a_{1+}^{\dagger}a_{1+}-b_{1+}^{\dagger}b_{1+}\right) \\
-\lim_{v\rightarrow-\infty}\delta_{\varepsilon}\left(  \ln\left\vert
u\right\vert \right)  \frac{\left\vert b_{3-}^{\dagger}+e^{-\pi\omega}\left(
\mu^{2}\right)  ^{-i\omega}e^{i2\theta}a_{1+}\right\vert ^{2}-\left\vert
a_{3-}+e^{-\pi\omega}\left(  \mu^{2}\right)  ^{i\omega}e^{-i2\theta}%
b_{1+}^{\dagger}\right\vert ^{2}}{1-e^{-2\pi\omega}}\\
+\lim_{u\rightarrow-\infty}\delta_{\varepsilon}\left(  \ln\left\vert
v\right\vert \right)  \frac{\left\vert a_{1+}+e^{-\pi\omega}\left(  \mu
^{2}\right)  ^{i\omega}e^{-i2\theta}b_{3-}^{\dagger}\right\vert ^{2}%
-\left\vert b_{1+}^{\dagger}+e^{-\pi\omega}\left(  \mu^{2}\right)  ^{-i\omega
}e^{i2\theta}a_{3-}\right\vert ^{2}}{1-e^{-2\pi\omega}}%
\end{array}
\right]
\end{array}
\label{rho4}%
\end{equation}
The interpretation is similar to those of regions I or II or III, and again we
conclude that charge, information or probability, are conserved within region
IV by itself, independent of what goes on in other regions of the extended
Rindler space.

We may now compute the sum of the charges $Q_{R},$ Eq.(\ref{QRi}), in all the
regions I-IV for universe $\left(  0,0\right)  $, and find%
\begin{equation}%
\begin{array}
[c]{l}%
Q_{R}=2\sum_{\pm}\int_{0}^{\infty}d\omega\left(  \left(  a_{1\pm}^{\dagger
}a_{1\pm}-b_{1\pm}^{\dagger}b_{1\pm}\right)  -\left(  b_{3\pm}^{\dagger
}b_{3\pm}-a_{3\pm}^{\dagger}a_{3\pm}\right)  \right)  ,\text{ }\\
\left.  \partial Q_{R}\right\vert _{\mu^{2}=0}=\int_{0}^{\infty}d\omega\left(
\begin{array}
[c]{c}%
+\left(  \lim_{v\rightarrow-\infty}-\lim_{v\rightarrow\infty}\right)
\delta_{\varepsilon}\left(  \ln\left\vert u\right\vert \right)  \left[
\begin{array}
[c]{c}%
\left(  a_{1-}^{\dagger}a_{1-}-b_{1-}^{\dagger}b_{1-}\right) \\
-\left(  b_{3-}^{\dagger}b_{3-}-a_{3-}^{\dagger}a_{3-}\right)
\end{array}
\right] \\
+\left(  \lim_{u\rightarrow-\infty}-\lim_{u\rightarrow\infty}\right)
\delta_{\varepsilon}\left(  \ln\left\vert v\right\vert \right)  \left[
\begin{array}
[c]{c}%
\left(  a_{1+}^{\dagger}a_{1+}-b_{1+}^{\dagger}b_{1+}\right) \\
-\left(  b_{3+}^{\dagger}b_{3+}-a_{3+}^{\dagger}a_{3+}\right)
\end{array}
\right]
\end{array}
\right) \\
\left.  \partial Q_{R}\right\vert _{\mu^{2}\neq0}=\int_{0}^{\infty}%
\frac{d\omega}{1-e^{-2\pi\omega}}\left[
\begin{array}
[c]{c}%
\lim_{u\rightarrow-\infty}\delta_{\varepsilon}\left(  \ln\left\vert
v\right\vert \right)  \left(
\begin{array}
[c]{c}%
\left\vert a_{1+}+e^{-\pi\omega}\left(  \mu^{2}\right)  ^{i\omega}%
e^{-i2\theta}b_{3-}^{\dagger}\right\vert ^{2}\\
-\left\vert b_{1+}^{\dagger}+e^{-\pi\omega}\left(  \mu^{2}\right)  ^{-i\omega
}e^{i2\theta}a_{3-}\right\vert ^{2}%
\end{array}
\right) \\
+\lim_{v\rightarrow-\infty}\delta_{\varepsilon}\left(  \ln\left\vert
u\right\vert \right)  \left(
\begin{array}
[c]{c}%
\left\vert a_{3-}+e^{-\pi\omega}\left(  \mu^{2}\right)  ^{i\omega}%
e^{-i2\theta}b_{1+}^{\dagger}\right\vert ^{2}\\
-\left\vert b_{3-}^{\dagger}+e^{-\pi\omega}\left(  \mu^{2}\right)  ^{-i\omega
}e^{i2\theta}a_{1+}\right\vert ^{2}%
\end{array}
\right) \\
+\lim_{u\rightarrow\infty}\delta_{\varepsilon}\left(  \ln v\right)  \left(
\begin{array}
[c]{c}%
\left\vert b_{3+}^{\dagger}+e^{-\pi\omega}\left(  \mu^{2}\right)  ^{i\omega
}e^{-i2\theta}a_{1-}\right\vert ^{2}\\
-\left\vert a_{3+}+e^{-\pi\omega}\left(  \mu^{2}\right)  ^{-i\omega
}e^{i2\theta}b_{1-}^{\dagger}\right\vert ^{2}%
\end{array}
\right) \\
+\lim_{v\rightarrow\infty}\delta_{\varepsilon}\left(  \ln u\right)  \left(
\begin{array}
[c]{c}%
\left\vert b_{1-}+e^{-\pi\omega}\left(  \mu^{2}\right)  ^{i\omega}%
e^{-i2\theta}a_{3+}\right\vert ^{2}\\
-\left\vert a_{1-}^{\dagger}+e^{-\pi\omega}\left(  \mu^{2}\right)  ^{-i\omega
}e^{i2\theta}b_{3+}^{\dagger}\right\vert ^{2}%
\end{array}
\right)
\end{array}
\right]
\end{array}
\label{QRtot}%
\end{equation}
The last two equations for $\partial Q_{R}$ are the sums of all the fluxes
$\partial q_{1,2,3,4}$ given in Eqs.(\ref{rho1}-\ref{rho4}); in this sum the
fluxes at each horizon cancel out and only the asymptotic fluxes in each
region remain as shown in Eq.(\ref{QRtot}). Note that the sum of all incoming
terms in $\sum\left(  \partial Q_{R}\right)  _{in}$ is exactly equal to the
sum of all outgoing terms in $\sum\left(  \partial Q_{R}\right)  _{out}.$ This
is easy to see for $\left.  \partial Q_{R}\right\vert _{\mu^{2}=0}.$ Simple
algebra, like Eq.(\ref{simple}), shows that it is also true for $\left.
\partial Q_{R}\right\vert _{\mu^{2}\neq0}$ when we take into account the
results of the boundary conditions given in Eq.(\ref{a+-relations}), namely
\begin{equation}
\mu^{2}\neq0:\left\vert a_{1-}\right\vert =\left\vert a_{1+}\right\vert
,\;\left\vert a_{3-}\right\vert =\left\vert a_{3+}\right\vert ,\;\left\vert
b_{1-}\right\vert =\left\vert b_{1+}\right\vert ,\;\left\vert b_{3-}%
\right\vert =\left\vert b_{3+}\right\vert .
\end{equation}
Eq.(\ref{QRtot}) is the statement of charge conservation for the entire
$\left(  0,0\right)  $ universe: $Q_{R}$ is conserved within the $\left(
0,0\right)  $ universe by itself because the charges that flow in and out its
asymptotic regions balance each other exactly such that the sum of all
influxes is equal to the sum of all outflows. Moreover, each \textit{type} of
charge ($a_{1\pm},a_{3\pm},b_{1\pm},b_{3\pm}$) and corresponding total flux is
\textit{separately conserved}. This amounts to conservation of probability and
information for the overall $\left(  0,0\right)  $ level.

This result for the Rindler total charge, $Q_{R}$ in the $\left(  0,0\right)
$ universe, may be compared to the total charge $Q_{M}$ defined in Minkowski
space as given above in Eq.(\ref{QMi}). We expect the total charge and total
boundary in or out fluxes to be the same in either computation,
\begin{equation}
Q_{R}=Q_{M}\text{ and }\sum\left(  \partial Q_{R}\right)  _{in/out}%
=\sum\left(  \partial Q_{M}\right)  _{in/out}. \label{QRtot2}%
\end{equation}
To relate the Rindler/Minkowski results to each other we use the Bogoliubov
transformations in Eqs.(\ref{bogol1},\ref{bogol2}) and find
\begin{equation}
Q=\left\{
\begin{array}
[c]{l}%
=2\sum_{\pm}\int_{0}^{\infty}d\omega\left[  \left(  a_{1\pm}^{\dagger}a_{1\pm
}-b_{1\pm}^{\dagger}b_{1\pm}\right)  -\left(  b_{3\pm}^{\dagger}b_{3\pm
}-a_{3\pm}^{\dagger}a_{3\pm}\right)  \right] \\
=\int_{-\infty}^{\infty}dk^{1}\left(  A^{\dagger}\left(  k^{1}\right)
A\left(  k^{1}\right)  -B^{\dagger}\left(  k^{1}\right)  B\left(
k^{1}\right)  \right)  ,
\end{array}
\right.  \label{Bo}%
\end{equation}
showing that indeed $Q_{R}=Q_{M}$ according to Eqs.(\ref{QMi},\ref{QRtot}).
Similarly, the identity for the sum of the in or out fluxes can also be proven
by using the Bogoliubov transformations to find%
\begin{equation}
\sum\left(  \partial Q\right)  _{in/out}=\left\{
\begin{array}
[c]{l}%
=\int_{0}^{\infty}d\omega\left[  \left(  a_{1\pm}^{\dagger}a_{1\pm}-b_{1\pm
}^{\dagger}b_{1\pm}\right)  -\left(  b_{3\pm}^{\dagger}b_{3\pm}-a_{3\pm
}^{\dagger}a_{3\pm}\right)  \right] \\
=\frac{1}{2}\int_{0}^{\infty}dk\left(  A_{\pm}^{\dagger}\left(  k\right)
A_{\pm}\left(  k\right)  -B_{\pm}^{\dagger}\left(  k\right)  B_{\pm}\left(
k\right)  \right)
\end{array}
\right.  . \label{J}%
\end{equation}
These checks verify that our approach is self consistent according to
Eqs.(\ref{dQM},\ref{QRtot},\ref{QRtot2}).

This result implies that charge, information or probability, is conserved in
the $\left(  0,0\right)  $ universe by itself and furthermore that the
$\left(  0,0\right)  $ universe formulated in the extended Rindler space is
equivalent to a full Minkowski universe on one sheet. From this we may also
conclude that in the absence of interactions or perturbations, Rindler
information does not leak from the $\left(  0,0\right)  $ universe to any
other $\left(  n,m\right)  $ universe$.$

The same arguments can now be applied at each level by using the Rindler or
Minkowski forms of the same field $\varphi^{\left(  n,m\right)  }\left(
u,v\right)  $ that we have discussed in the previous sections. A little
thought is sufficient to go over the same computations by simply changing the
symbols for the oscillators, and be convinced that charge or information is
again conserved separately within every level $\left(  n,m\right)  .$

Thus, it seems the first quantized level-$\left(  0,0\right)  $ wavefunction
or the quantum field $\varphi\left(  u,v\right)  ,$ analytically continued to
all levels in the extended Rindler spacetime, describes parallel Minkowski
universes. Since all levels are predictably related to each other by
analiticity, one should not think of phenomena in these parallel universes as
being independent from each other, at least not in the present context of free
fields. This is because there is only one set of oscillators to construct
wavepackets, namely those of level-$\left(  0,0\right)  ,$ and as we have
shown, all oscillators at other levels are dependent on the level-$\left(
0,0\right)  $ oscillators.

\section{Discussion \label{discuss}}

In summary, we have shown that, although information does flow between
neighboring regions of the $\left(  0,0\right)  $ universe, regional
information remains constant for each species of particles/antiparticles
$\left(  a_{1\pm},a_{3\pm},b_{1\pm},b_{3\pm}\right)  $ due to the balance of
in/out fluxes for each region \textit{separately}. The conserved regional
charges, $q_{1},q_{1},q_{3},q_{4}$, are generally different in each region and
they are determined by the wavepacket coefficients of the fields in
Eq.(\ref{list}) for each region in universe $\left(  0,0\right)  .$ Note that
the constant $q_{1,2,3,4}$ as well as the fluxes at boundaries depend on the
wavepacket coefficients only in the combinations, $a_{1\pm}^{\dagger}\left(
\omega\right)  a_{1\pm}\left(  \omega\right)  ,b_{1\pm}^{\dagger}\left(
\omega\right)  b_{1\pm}\left(  \omega\right)  ,a_{3\pm}^{\dagger}\left(
\omega\right)  a_{3\pm}\left(  \omega\right)  ,b_{3\pm}^{\dagger}\left(
\omega\right)  b_{3\pm}\left(  \omega\right)  ,$ which turn into number
operators in the second quantized field theory.

This argument is repeated for each $\left(  n,m\right)  $ universe for which
the corresponding fields are fully determined by analyticity. Recall that the
field in the $\left(  n,m\right)  $ universe differs from the field in the
$\left(  0,0\right)  $ universe by the canonical transformations in
Eqs.(\ref{can1},\ref{can2}) or Eqs.(\ref{AnmA},\ref{AnmAbar}). We find that
the regional constant charges $q_{1},q_{2},q_{3},q_{4},$ and the fluxes at the
boundaries, of the Rindler regions in the $\left(  n,m\right)  $ universe, are
identical to those of the $\left(  0,0\right)  $ universe, because, according
to Eqs.(\ref{can1},\ref{can2}), the number operators, $a_{1\pm}^{\dagger
}a_{1\pm}$ etc., in any $\left(  n,m\right)  $ universe are the same as in the
$\left(  0,0\right)  $ universe since these number operators are invariant
under the canonical transformations. This is true despite the fact that the
wavepacket coefficients $a_{1\pm}^{\left(  n,m\right)  }$ etc. in the $\left(
n,m\right)  $ universe are different than the $\left(  n^{\prime},m^{\prime
}\right)  $ universe by real factors (not just phases). Therefore, as far as
information flow and conservation is concerned, the Rindler multiverse seems
to consist of parallel universes that may not communicate with each other.

This conclusion emerged because of information conservation
\textit{separately} in each Rindler quadrant of Minkowski space, at all levels
of the multiverse, which holds as long as the Rindler multiverse system is not
disturbed by interactions that may alter the current $J^{\mu}$ or induce
inter-universe transitions.

Note however that there are non-trivial inter-universe propagators or more
general multi-point correlators with one leg in the $\left(  n,m\right)  $
universe and the other(s) in a different $\left(  n^{\prime},m^{\prime
}\right)  $ universe(s), such as
\begin{equation}
G_{\left(  n,m\right)  }^{\left(  n^{\prime},m^{\prime}\right)  }\left(
u_{i},v_{i};u_{j},v_{j}\right)  \equiv\langle0_{M}|\varphi^{\left(
n,m\right)  }\left(  u,v\right)  \varphi_{j}^{\dagger\left(  n^{\prime
},m^{\prime}\right)  }\left(  u^{\prime},v^{\prime}\right)  |0_{M}\rangle,
\label{interlevel}%
\end{equation}
where $i,j=1,2,3,4,$ indicate the regions I-IV. The creation/annihilation
operators in the analytically continued fields $\varphi_{i}^{\left(
n,m\right)  },\varphi_{j}^{\dagger\left(  n^{\prime},m^{\prime}\right)  }$ are
related to each other but have different real factors that depend on $\left(
n,m\right)  $ or $\left(  n^{\prime},m^{\prime}\right)  $ as given in
Eqs.(\ref{can1},\ref{can2}). When $\left(  n,m\right)  =\left(  n^{\prime
},m^{\prime}\right)  =\left(  0,0\right)  $ these propagators or more general
n-point functions are guarateed to be identical to the well known propagators
or n-point functions of a Klein-Gordon complex scalar field in Minkowski
space. However, in general they will differ because of the $n,m,n^{\prime
},m^{\prime}$ dependent factors that modify computations of the $\left(
0,0\right)  $ universe, such as the modification of the example in
Eq.(\ref{numVac}) by the additional factor as seen below
\begin{equation}
\langle0_{M}|a_{1-}^{\dagger\left(  n,m\right)  }\left(  \omega\right)
a_{1-}^{\left(  n^{\prime},m^{\prime}\right)  }\left(  \omega^{\prime}\right)
|0_{M}\rangle=\frac{1}{2}\frac{\delta\left(  \omega-\omega^{\prime}\right)
}{e^{2\pi\omega}-1}e^{-2\pi\omega\left(  n-n^{\prime}\right)  }.
\label{extrafactor}%
\end{equation}
The propagator $G_{\left(  n,m\right)  }^{\left(  n^{\prime},m^{\prime
}\right)  }\left(  u,v;u^{\prime},v^{\prime}\right)  $ is easily computed by
using such relations that include the extra factor $e^{-2\pi\omega\left(
n-n^{\prime}\right)  }$. The physical meaning of $G_{\left(  n,m\right)
}^{\left(  n^{\prime},m^{\prime}\right)  }\left(  u,v;u^{\prime},v\right)  $
is unclear at the moment when there are no interactions. In any case, these
propagators will surely play a role if there are interactions that cause
inter-universe transitions.

As examples of disturbances of the Rindler parallel universes, we may consider
the geometry of an eternal black hole or the cosmological geometry of the
mini-superspace described in Appendix (\ref{mini}). Either spacetime may be
considered as introducing some gravitational interaction that deforms the
extended Rindler spacetime non-perturbatively. The approach of this paper may
be applied similarly to cosmology as in \cite{barsMiniSuper} or black holes,
as in \cite{barsArayaBH}. We find that, although information conservation as
discussed above holds for the non-interacting Rindler multiverse, it fails for
cases like these. In particular, for black holes it is found that there is
leakage of information precisely at the black hole singularity through which
the current flows between different levels of the multiverse. The information
loss for black holes \cite{hawking},\cite{hawking+Hartle} may be redefined as
a loss of information for the $\left(  0,0\right)  $ universe, but still
conserved in the full eternal black hole multiverse. The flow of information
away from the $\left(  0,0\right)  $ universe can be tracked quantitatively by
computing the amount of information that leaks to specific regions in other
universes in the extended black hole multiverse \cite{barsArayaBH}. The
question remains as to what happens to information if the black hole can fully evaporate.

We have shown that even something as simple as the extended Rindler space is
far richer at the quantum level than the Minkowski geometry specified by the
metric or the geodesics at the classical level. New phenomena of physical
interest may occur due to the natural multiverse predicted by the quantum
field. Even for the Rindler multiverse, it would be interesting to explore
which types of perturbative or non-perturbative interactions (such as black
holes, big bang, and others) may induce communication among the otherwise
apparently non-interacting parallel Rindler universes.

In this paper we discussed a new multiverse concept in an idealized setting
and established certain technical properties of the first quantized
wavefunction or classical field and its second quantization, in the extended
Rindler spacetime. Although this spacetime is related to flat Minkowski
spacetime by a simple coordinate transformation at the classical level, we
showed that the presence of horizons in the Rindler coordinate system led to
subtleties at the quantum level due to cuts in analytic $\left(  u,v\right)  $
spacetime, and that this naturally implied the presence of a multiverse in the
first and second quantized treatment of the field in such a spacetime.
Analyticity of the field in the $\left(  u,v\right)  $ coordinates guaratees
that unavoidably $\varphi\left(  u,v\right)  $ takes unique values throughout
the multiverse. We claim that similar multiverse properties are also shared by
any spacetime that has horizons and/or singularities, such as the full
spacetime of an eternal black hole \cite{barsArayaBH} as well as the
cosmological mini-superspace geometry (in field space) described in
Appendix(\ref{mini}) and in more detail in \cite{barsMiniSuper}. The presence
of the multiverse structure does not seem to be directly detectable by an
observer in Rindler region I, or the analogous region-I observer outside of a
black hole, because, as we have already emphasized such an observer is
incapable of directly detecting anything beyond the horizons of region I.
Possible observable physical effects, that even observers in region I may
notice as indirect consequences of a multiverse, could arise in cosmological
or black hole phenomena. The possibility of transitions through gravitational
singularities (see e.g. \cite{crunchBang},\cite{BST-Higgs}\cite{sailing}%
,\cite{barsArayaJames},\cite{barsJames}) may also include transitions in the
multiverse. How such new mathematical properties of the field are relevant for
some new physical phenomena is under investigation.

It may be worthwhile to emphasize how our multiverse for extended Rindler
spacetime differs from ordinary Minkowski spacetime. Clearly they are quite
different. A field in ordinary Minkowski spacetime has only the level-$\left(
0,0\right)  $ field of our multiverse. Analytic continuation of the ordinary
Minkowski plane-wave basis as in Eq.(\ref{PsiM}), $e^{-i\frac{E-k}{2}%
u}e^{-i\frac{E+k}{2}v}$, by $u\rightarrow ue^{\pm i2\pi}$ or $v\rightarrow
ve^{\pm i2\pi},$ does not lead to any new analyticity results. This is because
the Minkowski coordinate basis is adequate to describe the multiverse one
level at a time and lacks the analyticity information that is available in the
extended Rindler coordinate basis. An analogy to this is the Schwarzchild
coordinate basis for a black hole, that describes only the region outside of
the horizon, versus the Kruskal-Szekeres coordinate basis that provides the
extension to the full eternal black hole spacetime. In a similar way, the
extended Rindler coordinate basis captures the entire multiverse through its
analyticity behavior. What could not be captured directly in the Minkowski
basis is clarified in section-VI. Namely, the level-$\left(  n,m\right)  $
field in the Minkowski basis in Eq.(\ref{PsinmM}) is related by a very
non-trivial canonical transformation to the level-$\left(  0,0\right)  $
field. This canonical transformation is just the result of the non-trivial
analytic continuation in the extended Rindler basis, resulting from
$u\rightarrow ue^{i2\pi n}$ or $v\rightarrow ve^{i2\pi m}$ with integers $n,m$
with the patterns given in detail in Eqs.(\ref{Phinm},\ref{can1}).
Furthermore, as seen via the inter-level correlators that appear in
Eqs.(\ref{interlevel},\ref{extrafactor}), there is a wealth of information in
our multiverse that is absent in ordinary Minkowski spacetime.

The notion and description of a multiverse that emerged in this paper is new
and different than other multiverse notions that originated in the past from
other considerations, such as the multiverse of the many worlds of quantum
mechanics, the multiverse that arises from eternal inflation, or the
multiverse that arises in the landscape of string theory. In particular, our
multiverse contains many levels that are \textit{predictably} connected to
each other by the analyticity properties of the wavefunction. This predictable
aspect is unlike other concepts of a multiverse in the literature. However, in
a complete theory perhaps the different concepts of a multiverse could be
connected to each other; see e.g. \cite{namura}\cite{boussoSusskind} for some
possible relations, which however does not address our new brand of
multiverse. Note that in our case, analyticity connects the different
universes and makes predictions of relations among them. In future
investigations we will consider the physical significance of the ideas
expressed in this paper in a complete realistic theory of fundamental physics
(possibly in cosmology and/or black holes), including models that address the
effects of quantum gravity, such as string theory. The analog of the quantum
wavefunction of a particle is the string field. So, in a deeper investigation
of the multiverse in the sense of the current paper may be possible in string
field theory in which non-trivial backgrounds \cite{nontrivialString} and
string-string interactions are included. This may be a context in which
various notions of a multiverse, including our new one, may be connected to
each other.

We have shown that the multiverse, in the quantum version of certain
spacetimes, is an immutable structure of the wave function - there is no
choice here because it directly follows from quantum mechanics. Our result,
that was not known before, cannot be captured by any amount of analysis of
\textit{classical} general relativity. It is conceivable that indirect
observational consequences of our findings could be analysed through
gravitational waves, since the fluctuations in such gravitational backgrounds,
that are emitted as waves, may encapsulate the predicted multiverse structure
already embedded in the quantum field.

\begin{acknowledgments}
We acknowledge conversations with Albin James on this topic. This research was
supported in part by Perimeter Institute for Theoretical Physics. Research at
Perimeter Institute is supported by the Government of Canada through the
Department of Innovation, Science and Economic Development Canada and by the
Province of Ontario through the Ministry of Research, Innovation and Science.
I.J.A. was supported by CONICYT (Comisi\'{o}n Nacional de Investigaci\'{o}n
Cient\'{\i}fica y Tecnol\'{o}gica - Chilean Government) and by the Fulbright
commission through a joint fellowship.
\end{acknowledgments}

\bigskip

\appendix\bigskip

\section{Computation of charge and boundary fluxes \label{appendix}}

In this appendix we show the computation of $q_{1},\partial_{t}q_{1}$ and
$q_{2},\partial_{-y}q_{2}$ whose results appear in Eqs.(\ref{rho1},\ref{rho2})
respectively. The remaining $q_{3},\partial_{-t}q_{3}$ and $q_{4},\partial
_{y}q_{4}$ are obtained by simple substitution of variables as given just
before Eq.(\ref{rho3}) and Eq.(\ref{rho4}) respectively.

For region I the definitions of $q_{1},\partial_{t}q_{1}$ are given in
Eq.(\ref{rhoRin}),%
\begin{equation}%
\begin{array}
[c]{l}%
q_{1}\left(  t\right)  =\int_{0}^{\infty}dyJ_{1}^{t}=\int_{0}^{\infty}%
dy\frac{i}{2y}\left(  \varphi_{1}^{\dagger}\partial_{t}\varphi_{1}%
-\partial_{t}\varphi_{1}^{\dagger}\varphi_{1}\right) \\
\partial_{t}q_{1}=\int_{0}^{\infty}dy\partial_{t}J_{1}^{t}=\int_{0}^{\infty
}dy\left(  \partial_{\mu}J_{1}^{\mu}-\partial_{y}J_{1}^{y}\right)  =-J_{2}%
^{y}\left(  t,\infty\right)  +J_{2}^{y}\left(  t,0\right)
\end{array}
\label{rho1Int}%
\end{equation}
where the Klein-Gordon equation is used to set $\partial_{\mu}J_{1}^{\mu}=0$,
and then Stoke's theorem is applied to write the result in terms of the
current $J^{y}\left(  t,y\right)  $ evaluated at the asymptotic boundaries.
Here $\varphi_{1}$ that is given in Eq.(\ref{list}) is written in terms of
$\left(  t,y\right)  ,$ and the $y$-component of the current at the boundaries
is given by the following limits%
\begin{equation}%
\begin{array}
[c]{l}%
\varphi_{1}\left(  t,y\right)  =\int_{0}^{\infty}d\omega\left[  e^{-i\omega
t}\left(  a_{1-}\left(  \omega\right)  \frac{\left(  2y\right)  ^{-i\frac
{\omega}{2}}S_{-}\left(  2\mu^{2}y\right)  }{\sqrt{4\pi\omega}}+a_{1+}\left(
\omega\right)  \frac{\left(  2y\right)  ^{+i\frac{\omega}{2}}S_{+}\left(
2\mu^{2}y\right)  }{\sqrt{4\pi\omega}}\right)  +h.c.\right] \\
J_{1}^{y}\left(  t,\infty\text{ or }0\right)  =\lim_{y\rightarrow\infty\text{
or }0}\left(  -i2y\left(  \varphi_{1}^{\dagger}\partial_{y}\varphi
_{1}-\partial_{y}\varphi_{1}^{\dagger}\varphi_{1}\right)  \right)  .
\end{array}
\label{phi1}%
\end{equation}
where the expression for $J^{y}\left(  t,y\right)  $ follows from $J^{\mu}$ in
Eq.(\ref{current}) after using $\sqrt{-g}=1$ and $g^{yy}=2y$ for the Rindler spacetime.

To compute $q_{1}\left(  t\right)  $ one uses the orthonormality of the
positive and negative frequency modes described in footnote (\ref{norm}). Then
the integral in Eq.(\ref{rho1Int}) yields%
\begin{equation}
q_{1}=\int_{0}^{\infty}d\omega\left(  \left(  a_{1-}^{\dagger}a_{1-}%
-b_{1-}^{\dagger}b_{1-}\right)  +\left(  a_{1+}^{\dagger}a_{1+}-b_{1+}%
^{\dagger}b_{1+}\right)  \right)  ,
\end{equation}
as given in Eq.(\ref{rho1}). This shows that $q_{1}\left(  t\right)  $ is time
independent, so the charge is conserved $\partial_{t}q_{1}=0$ within region I
at finite $t.$ We will see that in general it is not conserved, $\partial
_{t}q_{1}\left(  t\right)  \neq0,$ at the $t\rightarrow\pm\infty$ boundaries.

Next we compute the non-trivial fluxes $J_{1}^{y}\left(  t,\infty/0\right)  $
at the $y$-boundaries of region I. Consider at first $J_{1}^{y}\left(
t,\infty\right)  $ for the massive field $\mu^{2}>0$. We had argued in
Eqs.(\ref{a+-relations},\ref{F13asympt}) that the horizon boundary conditions
in section (\ref{bc}) relate $a_{1\pm}\left(  \omega\right)  $ to each other
by a definite phase, and that this implies also the correct physical
asymptotic behavior, $\varphi_{1}\left(  t,y\sim\infty\right)  \rightarrow0.$
In this case the boundary current $J_{I}^{y}\left(  t,\infty\right)  $
vanishes asymptotically, and therefore the charge flow at the asymptotic
boundary of region I vanishes for the massive field, i.e.
\begin{equation}
\mu^{2}>0:J_{1}^{y}\left(  t,\infty\right)  =0,\text{ at all }%
t,~\text{including }t=\pm\infty. \label{Jinf}%
\end{equation}

This result is different for the massless field, $\mu^{2}=0,$ since the
asymptotic $\varphi_{1}\left(  t,\infty\right)  $ does not vanish in that
case. However, due to masslessness, we have $\lim_{\mu\rightarrow0}S_{\mp
}\left(  2\mu^{2}y\right)  =1,$ so the field $\varphi_{1}\left(  t,y\right)  $
in Eq.(\ref{phi1}) simplifies. The $2iy\partial_{y}$ derivatives that occur in
$J_{1}^{y}\left(  t,y\right)  $ in Eq.(\ref{phi1}) are then easily computed by
using $-2iy\partial_{y}\left(  2y\right)  ^{\mp i\frac{\omega}{2}}=\mp
\omega\left(  2y\right)  ^{\mp i\frac{\omega}{2}},$ and we obtain the
following double integral for $J_{1}^{y}\left(  t,y\right)  $
\begin{equation}
\int_{0}^{\infty}\frac{d\omega^{\prime}d\omega e^{i\left(  \omega^{\prime
}-\omega\right)  t}}{\sqrt{4\pi\omega^{\prime}}\sqrt{4\pi\omega}}\left[
\begin{array}
[c]{c}%
\left(  \omega+\omega^{\prime}\right)  \left(  -a_{1-}^{\dagger}\left(
\omega^{\prime}\right)  a_{1-}\left(  \omega\right)  \left(  2y\right)
^{i\frac{\omega^{\prime}-\omega}{2}}+a_{1+}^{\dagger}\left(  \omega^{\prime
}\right)  a_{1+}\left(  \omega\right)  \left(  2y\right)  ^{-i\frac
{\omega^{\prime}-\omega}{2}}\right)  +\cdots\\
+\left(  \omega-\omega^{\prime}\right)  \left(  -a_{1+}^{\dagger}\left(
\omega^{\prime}\right)  a_{1-}\left(  \omega\right)  \left(  2y\right)
^{-i\frac{\omega^{\prime}+\omega}{2}}+a_{1-}^{\dagger}\left(  \omega^{\prime
}\right)  a_{1+}\left(  \omega\right)  \left(  2y\right)  ^{i\frac
{\omega^{\prime}+\omega}{2}}\right)  +\cdots
\end{array}
\right]  \label{double}%
\end{equation}
where \textquotedblleft$\cdots$\textquotedblright\ represent the hermitian
conjugate and mixed terms that are not shown. As $2y\rightarrow\infty$ these
integrals are evaluated by using the steepest descent method because of the
fast oscillating exponentials $\left(  2y\right)  ^{\mp^{\prime}i\left(
\omega\pm\omega^{\prime}\right)  /2}.$ The leading contribution comes only
from the neighborhood $\omega^{\prime}\simeq\omega$ in the first line of
(\ref{double}); then the double integral is approximated by
\begin{equation}
\lim_{y\rightarrow\infty}\int_{0}^{\infty}\frac{d\omega~2\omega}{4\pi\omega
}\left[
\begin{array}
[c]{c}%
-\left\vert a_{1-}\left(  \omega\right)  \right\vert ^{2}\int_{-\infty
}^{\infty}d\zeta e^{-\varepsilon\left\vert \zeta\right\vert }\left(
e^{t}\sqrt{2y}\right)  ^{i\xi}+\cdots\\
+\left\vert a_{1+}\left(  \omega\right)  \right\vert ^{2}\int_{-\infty
}^{\infty}d\zeta e^{-\varepsilon\left\vert \zeta\right\vert }\left(
e^{-t}\sqrt{2y}\right)  ^{-i\xi}+\cdots
\end{array}
\right]
\end{equation}
where the factor $e^{-\varepsilon\left\vert \zeta\right\vert }$ is inserted to
insure the $\int_{-\infty}^{\infty}d\xi$ integrations are limited to the
neighborhood of $\zeta=\omega^{\prime}-\omega\simeq0.$ The $\zeta$ integrals
produce smeared delta functions $\delta_{\varepsilon}\left(  \ln z\right)  $,
\begin{equation}
\int_{-\infty}^{\infty}d\zeta e^{-\varepsilon\left\vert \zeta\right\vert
}z^{\pm i\xi}=2\pi\frac{\varepsilon/\pi}{\left(  \ln z\right)  ^{2}%
+\varepsilon^{2}}\equiv2\pi\delta_{\varepsilon}\left(  \ln z\right)  .
\label{smeared}%
\end{equation}
The result is
\begin{equation}
\mu^{2}=0:-J_{1}^{y}\left(  t,y\sim\infty\right)  =\left(
\begin{array}
[c]{c}%
+\left(  \lim_{v\rightarrow-\infty}\delta_{\varepsilon}\left(  \ln u\right)
\right)  \int_{0}^{\infty}d\omega~\left\vert a_{1-}\left(  \omega\right)
\right\vert ^{2}+\cdots\\
-\left(  \lim_{u\rightarrow\infty}\delta_{\varepsilon}\left(  \ln\left\vert
v\right\vert \right)  \right)  \int_{0}^{\infty}d\omega~\left\vert
a_{1+}\left(  \omega\right)  \right\vert ^{2}+\cdots
\end{array}
\right)  . \label{Jinf0}%
\end{equation}
where we have used,
\begin{equation}
\lim_{y\rightarrow\infty}\delta_{\varepsilon}\left(  \ln\left(  \sqrt{2y}%
e^{t}\right)  \right)  =\lim_{y\rightarrow\infty}\delta_{\varepsilon}\left(
t+\infty\right)  =\lim_{y\rightarrow\infty}\delta_{\varepsilon}\left(
\ln\left(  u\right)  \right)  =\lim_{v\rightarrow-\infty}\delta_{\varepsilon
}\left(  \ln u\right)  , \label{smeared2}%
\end{equation}
and similarly for the second term. This shows that there are non-vanishing
asymptotic contributions proportional to $\left\vert a_{1-}\left(
\omega\right)  \right\vert ^{2}$ when $v\rightarrow-\infty$ and $u$ is finite,
as well as $\left\vert a_{1+}\left(  \omega\right)  \right\vert ^{2}$ when
$u\rightarrow\infty$ and $v$ is finite. These contributions are at the
$\mathcal{I}^{\mp}$ boundaries in a Penrose diagram for region I.

To compute $J_{1}^{y}\left(  t,0\right)  $ for the massive or massless field,
only the $y=0$ neighborhood of the field $\varphi_{1}\left(  t,y\sim0\right)
$ is sufficient, which means $S_{\mp}\left(  2\mu^{2}y\right)  $ in
(\ref{phi1}) may be approximated by $S_{\mp}\left(  0\right)  =1.$ Then
$J_{1}^{y}\left(  t,y\sim0\right)  $ takes the same form as Eq.(\ref{double})
except for setting $2y\sim0.$ The fast oscillations argument is valid again,
and the integral is evaluated as
\begin{equation}
\mu^{2}\geq0:J_{1}^{y}\left(  t,y\sim0\right)  =\left[
\begin{array}
[c]{c}%
-\left(  \lim_{v\rightarrow0}\delta_{\varepsilon}\left(  \ln u\right)
\right)  \int_{0}^{\infty}d\omega~\left\vert a_{1-}\left(  \omega\right)
\right\vert ^{2}+\cdots\\
+\left(  \lim_{u\rightarrow0}\delta_{\varepsilon}\left(  \ln\left\vert
v\right\vert \right)  \right)  \int_{0}^{\infty}d\omega~\left\vert
a_{1+}\left(  \omega\right)  \right\vert ^{2}+\cdots
\end{array}
\right]  . \label{Jzero}%
\end{equation}
where we have used,
\begin{equation}
\lim_{y\rightarrow0}\delta_{\varepsilon}\left(  \ln\left(  \sqrt{2y}%
e^{t}\right)  \right)  =\lim_{y\rightarrow0}\delta_{\varepsilon}\left(
t-\infty\right)  =\lim_{v\rightarrow0}\delta_{\varepsilon}\left(  \ln
u\right)  ,\;etc.. \label{smeared3}%
\end{equation}
This shows that there are non-vanishing contributions when $v\rightarrow0$ and
$u$ is finite as well as when $u\rightarrow0$ and $v$ is finite. These are the
future and past horizons in region I.

Altogether, from Eqs.(\ref{drhoRin},\ref{Jinf},\ref{Jinf0},\ref{Jzero}) we
have
\begin{equation}%
\begin{array}
[c]{l}%
\left.  \partial_{t}q_{1}\left(  t\right)  \right\vert _{\mu^{2}=0}=\int
_{0}^{\infty}d\omega\left[  -\lim_{v\rightarrow0}\delta_{\varepsilon}\left(
\ln u\right)  \left\vert a_{1-}\left(  \omega\right)  \right\vert ^{2}%
+\lim_{u\rightarrow0}\delta_{\varepsilon}\left(  \ln\left\vert v\right\vert
\right)  \left\vert a_{1+}\left(  \omega\right)  \right\vert ^{2}%
+\cdots\right] \\
\left.  \partial_{t}q_{1}\left(  t\right)  \right\vert _{\mu^{2}\neq0}%
=\int_{0}^{\infty}d\omega\left[
\begin{array}
[c]{l}%
-\lim_{v\rightarrow0}\delta_{\varepsilon}\left(  \ln u\right)  \left\vert
a_{1-}\left(  \omega\right)  \right\vert ^{2}+\lim_{u\rightarrow0}%
\delta_{\varepsilon}\left(  \ln\left\vert v\right\vert \right)  \left\vert
a_{1+}\left(  \omega\right)  \right\vert ^{2}+\cdots\\
+\lim_{v\rightarrow-\infty}\delta_{\varepsilon}\left(  \ln\left\vert
u\right\vert \right)  \left\vert a_{1-}\left(  \omega\right)  \right\vert
^{2}-\lim_{u\rightarrow\infty}\delta_{\varepsilon}\left(  \ln v\right)
\left\vert a_{1+}\left(  \omega\right)  \right\vert ^{2}+\cdots
\end{array}
\right]
\end{array}
\label{flowsI}%
\end{equation}
After including the contributions \textquotedblleft$\cdots$\textquotedblright%
\ from the hermitian conjugate terms in $\varphi_{1}$, the results are given
in Eq.(\ref{rho1}).

We now turn to regions II and IV. Since space/time are interchaged in regions
II and IV, we define the conserved charge and its derivative as an integral
over $t$ at fixed $y$ as explained after Eqs.(\ref{rhoRin},\ref{drhoRin})%

\begin{equation}%
\begin{array}
[c]{l}%
q_{2}\left(  y\right)  =-\int_{-\infty}^{\infty}dtJ_{2}^{y}=-\int_{-\infty
}^{\infty}dt~\left(  -2yi\right)  \left(  \varphi_{2}^{\dagger}\partial
_{y}\varphi_{2}-\partial_{y}\varphi_{2}^{\dagger}\varphi_{2}\right) \\
\partial_{-y}q_{2}\left(  y\right)  =\int_{-\infty}^{\infty}dt\partial
_{-y}\left(  -J_{2}^{y}\right)  =\int_{-\infty}^{\infty}dt\left(
\partial_{\mu}J_{2}^{\mu}-\partial_{t}J_{2}^{t}\right)  =-J_{2}^{t}\left(
\infty,y\right)  +J_{2}^{t}\left(  -\infty,y\right)
\end{array}
\label{rho2int}%
\end{equation}
The field $\varphi_{2}\left(  u,v\right)  $ in Eq.(\ref{list}) is now
rewritten in the $\left(  t,y\right)  $ coordinates%
\begin{equation}
\varphi_{2}\left(  t,y\right)  =\int_{0}^{\infty}d\omega\left[  e^{-i\omega
t}\left(  a_{1-}\left(  \omega\right)  \frac{\left(  2y\right)  ^{-i\frac
{\omega}{2}}S_{-}\left(  2\mu^{2}y\right)  }{\sqrt{4\pi\omega}}+b_{3+}%
^{\dagger}\left(  \omega\right)  \frac{\left(  2y\right)  ^{+i\frac{\omega}%
{2}}S_{+}\left(  2\mu^{2}y\right)  }{\sqrt{4\pi\omega}}\right)  +h.c.\right]
. \label{phi2}%
\end{equation}
Apply this first to the massless case to compute $q_{2}\left(  y\right)  $
when $S_{\pm}\left(  0\right)  =1.$ Then, using $i2y\partial_{y}\left(
2y\right)  ^{-i\frac{\omega}{2}}=\pm\omega\left(  2y\right)  ^{-i\frac{\omega
}{2}},$ gives%
\begin{equation}
q_{2}\left(  y\right)  =\int_{0}^{\infty}d\omega\left(  \left(  a_{1-}%
^{\dagger}\left(  \omega\right)  a_{1-}\left(  \omega\right)  -b_{3+}%
^{\dagger}\left(  \omega\right)  b_{3+}\left(  \omega\right)  \right)
+\cdots\right)  \label{rho2y}%
\end{equation}
where \textquotedblleft$\cdots$\textquotedblright\ represents the contribution
from the h.c. part of the field $\varphi_{2}$ above. Note that the signs of
the charges are consistent with the definition of particle/antiparticle as
represented by $a/b$ symbols respectively.

Now compute the non-trivial fluxes $J_{2}^{t}\left(  \pm\infty,y\right)  $ at
the $t\rightarrow\pm\infty$ boundaries of region II. For the massless case we
have
\begin{equation}%
\begin{array}
[c]{l}%
J_{2}^{t}\left(  \infty,y\right)  =\lim_{t\rightarrow\infty}\left(  \frac
{i}{2y}\left(  \varphi_{2}^{\dagger}\partial_{t}\varphi_{2}-\partial
_{t}\varphi_{2}^{\dagger}\varphi_{2}\right)  \right) \\
=\lim_{t\rightarrow\infty}\int_{0}^{\infty}\int_{0}^{\infty}\frac{d\omega
_{1}d\omega_{2}~\left(  \omega_{1}+\omega_{2}\right)  }{2y\sqrt{4\pi\omega
_{1}}\sqrt{4\pi\omega_{2}}}\left(
\begin{array}
[c]{c}%
e^{i\left(  \omega_{1}-\omega_{2}\right)  }\left[  a_{1-}^{\dagger}\left(
\omega_{1}\right)  \left(  -2y\right)  ^{i\frac{\omega_{1}}{2}}+b_{3+}\left(
\omega_{1}\right)  \left(  -2y\right)  ^{-i\frac{\omega_{1}}{2}}\right] \\
\times\left[  a_{1-}\left(  \omega_{2}\right)  \left(  -2y\right)
^{-i\frac{\omega_{2}}{2}}+b_{3+}^{\dagger}\left(  \omega_{2}\right)  \left(
-2y\right)  ^{+i\frac{\omega_{2}}{2}}\right]
\end{array}
\right)  +\cdots
\end{array}
\end{equation}
The \textquotedblleft$+\cdots$\textquotedblright\ represents the hermitian
conjugate and mixed terms that are not shown. Due to wild oscillations, at
large $t$ only the neighborhood of $\omega_{1}\sim\omega_{2}$ can contribute
to this integral. Furthermore, because the support of $\varphi_{2}\left(
t,y\right)  $ at large $t$ is either at large $\left\vert 2y\right\vert
\rightarrow\infty,$ or small $\left\vert 2y\right\vert \rightarrow0,$ terms
with $\left(  2y\right)  ^{\pm i\left(  \omega_{1}+\omega_{2}\right)  /2}$ in
this integral are also negligible since they too vanish at either limit
$\left\vert 2y\right\vert \rightarrow(0$ or $\infty)$ due to wild
oscillations. Therefore the expression above is simplified by keeping the
leading terms and using the same arguments that followed Eq.(\ref{double})
\begin{equation}
J_{2}^{t}\left(  \infty,y\right)  =\int_{0}^{\infty}\frac{d\omega~2\omega
}{4\pi\omega\times\left(  -\left\vert 2y\right\vert \right)  }\left[
\begin{array}
[c]{c}%
\left\vert a_{1-}\left(  \omega\right)  \right\vert ^{2}\lim_{t\rightarrow
\infty}\int_{-\infty}^{\infty}d\zeta e^{-\varepsilon\left\vert \zeta
\right\vert }\left(  e^{t}\sqrt{\left\vert 2y\right\vert }\right)  ^{i\xi
}+\cdots\\
+\left\vert b_{3+}\left(  \omega\right)  \right\vert ^{2}\lim_{t\rightarrow
\infty}\int_{-\infty}^{\infty}d\zeta e^{-\varepsilon\left\vert \zeta
\right\vert }\left(  e^{-t}\sqrt{\left\vert 2y\right\vert }\right)  ^{-i\xi
}+\cdots
\end{array}
\right]  ,
\end{equation}
where we recall that $y$ is negative in region II to rewrite everything in
terms of $\left\vert 2y\right\vert .$ Using the definition of the smeared
delta function in Eqs.(\ref{smeared}-\ref{smeared3}) we evaluate the result as
follows%
\begin{equation}%
\begin{array}
[c]{l}%
\lim_{t\rightarrow\infty}\int_{-\infty}^{\infty}d\zeta e^{-\varepsilon
\left\vert \zeta\right\vert }\frac{\left(  e^{t}\sqrt{\left\vert 2y\right\vert
}\right)  ^{i\xi}}{-\left\vert 2y\right\vert }=\lim_{t\rightarrow\infty}%
\frac{\delta_{\varepsilon}\ln\left(  e^{t}\sqrt{\left\vert 2y\right\vert
}\right)  }{-\left\vert 2y\right\vert }\simeq\lim_{t\rightarrow\infty}%
\frac{\delta_{\varepsilon}\left(  \left\vert y\right\vert \right)  }{-1}%
=-\lim_{v\rightarrow0}\delta_{\varepsilon}\left(  \ln u\right)  ,\\
\lim_{t\rightarrow\infty}\int_{-\infty}^{\infty}d\zeta e^{-\varepsilon
\left\vert \zeta\right\vert }\frac{\left(  e^{-t}\sqrt{\left\vert
2y\right\vert }\right)  ^{i\xi}}{-\left\vert 2y\right\vert }=\lim
_{t\rightarrow\infty}\frac{\delta_{\varepsilon}\ln\left(  e^{-t}%
\sqrt{\left\vert 2y\right\vert }\right)  }{-\left\vert 2y\right\vert }%
\simeq\lim_{t\rightarrow\infty}\frac{\delta_{\varepsilon}\left(  \left\vert
y\right\vert -\infty\right)  }{-1}=-\lim_{u\rightarrow\infty}\delta
_{\varepsilon}\left(  \ln v\right)  .
\end{array}
\end{equation}
Hence we obtain%
\begin{equation}
J_{2}^{t}\left(  \infty,y\right)  =-\lim_{v\rightarrow0}\delta_{\varepsilon
}\left(  \ln u\right)  \int_{0}^{\infty}d\omega\left\vert a_{1-}\left(
\omega\right)  \right\vert ^{2}-\lim_{u\rightarrow\infty}\delta_{\varepsilon
}\left(  \ln v\right)  \int_{0}^{\infty}d\omega\left\vert b_{3+}\left(
\omega\right)  \right\vert ^{2}+\cdots
\end{equation}
The evaluation of $J_{2}^{t}\left(  -\infty,y\right)  $ at $t\rightarrow
-\infty$ proceeds in the same way, leading to%
\begin{equation}
J_{2}^{t}\left(  -\infty,y\right)  =-\lim_{v\rightarrow\infty}\delta
_{\varepsilon}\left(  \ln u\right)  \int_{0}^{\infty}d\omega\left\vert
a_{1-}\left(  \omega\right)  \right\vert ^{2}-\lim_{u\rightarrow0}%
\delta_{\varepsilon}\left(  \ln v\right)  \int_{0}^{\infty}d\omega\left\vert
b_{3+}\left(  \omega\right)  \right\vert ^{2}+\cdots
\end{equation}
The combined result gives the rate of change of the charge at the boundaries
of region II for the massless particle
\begin{equation}%
\begin{array}
[c]{l}%
\left.  \partial_{-y}q_{2}\left(  y\right)  \right\vert _{\mu^{2}=0}%
=-J_{2}^{t}\left(  \infty,y\right)  +J_{2}^{t}\left(  -\infty,y\right) \\
=\int_{0}^{\infty}d\omega\left[
\begin{array}
[c]{c}%
\left(  \lim_{v\rightarrow0}\delta_{\varepsilon}\left(  \ln u\right)
-\lim_{v\rightarrow\infty}\delta_{\varepsilon}\left(  \ln u\right)  \right)
\left(  \left\vert a_{1-}\left(  \omega\right)  \right\vert ^{2}+\cdots\right)
\\
\left(  \lim_{u\rightarrow0}\delta_{\varepsilon}\left(  \ln v\right)
-\lim_{u\rightarrow\infty}\delta_{\varepsilon}\left(  \ln v\right)  \right)
\left(  -\left\vert b_{3+}\left(  \omega\right)  \right\vert ^{2}%
+\cdots\right)
\end{array}
\right]
\end{array}
\end{equation}

For the massive particle, the presence of $S_{\pm}\left(  2\mu^{2}y\right)  $
in $\varphi_{2}\left(  t,y\right)  $ in Eq.(\ref{phi2}) complicates the
calculation somewhat. The integral for $q_{2}\left(  y\right)  $ in
Eq.(\ref{rho2int}) is performed by using the properties of Bessel functions
and the result is just like the massless case given in Eq.(\ref{rho2y}). The
computation of $J_{2}^{t}\left(  \pm\infty,y\right)  $ is more complicated
because as $\left\vert 2y\right\vert \rightarrow\infty$ the nontrivial
asymptotic behavior of $S_{\pm}\left(  2\mu^{2}y\right)  $ must be taken into
account, although for $\left\vert 2y\right\vert \rightarrow0$ one still has
$S_{\pm}\left(  0\right)  =1,$ as in the massless case. Hence compared to the
massless case only the terms involving the $u\rightarrow\infty$ or
$v\rightarrow\infty$ boundaries are altered while the terms at the horizons
are the same. The result is
\begin{equation}%
\begin{array}
[c]{l}%
\left.  \partial_{-y}q_{2}\left(  y\right)  \right\vert _{\mu^{2}\neq0}%
=-J_{2}^{t}\left(  \infty,y\right)  +J_{2}^{t}\left(  -\infty,y\right) \\
=\int_{0}^{\infty}d\omega\left[
\begin{array}
[c]{c}%
\lim_{v\rightarrow0}\delta_{\varepsilon}\left(  \ln u\right)  \left(
\left\vert a_{1-}\left(  \omega\right)  \right\vert ^{2}+\cdots\right) \\
+\lim_{u\rightarrow0}\delta_{\varepsilon}\left(  \ln v\right)  \left(
-\left\vert b_{3+}\left(  \omega\right)  \right\vert ^{2}+\cdots\right) \\
-\lim_{v\rightarrow\infty}\delta_{\varepsilon}\left(  \ln u\right)
\frac{\left\vert a_{1-}^{\dagger}+e^{-\pi\omega}\left(  \mu^{2}\right)
^{-i\omega}e^{i2\theta}b_{3+}^{\dagger}\right\vert ^{2}+\cdots}{1-e^{-2\pi
\omega}}\\
-\lim_{u\rightarrow\infty}\delta_{\varepsilon}\left(  \ln v\right)
\frac{-\left\vert b_{3+}^{\dagger}+e^{-\pi\omega}\left(  \mu^{2}\right)
^{i\omega}e^{-i2\theta}a_{1-}\right\vert ^{2}+\cdots}{1-e^{-2\pi\omega}}%
\end{array}
\right]
\end{array}
\end{equation}
After including the contributions \textquotedblleft$\cdots$\textquotedblright%
\ from the hermitian conjugate terms in $\varphi_{2}$, the results are given
in Eq.(\ref{rho2}).

\section{Mini-superspace and cosmological multiverse \label{mini}}

The Lagrangian for the geodesically complete version of the Standard Model
(SM) coupled to General Relativity (GR) is given in \cite{BST},
\begin{equation}
\mathcal{L}\left(  x\right)  =\sqrt{-g}\left(
\begin{array}
[c]{c}%
L_{\text{SM}}\left(  A_{\mu}^{\gamma,W,Z,g},\;\psi_{q,l},\;\nu_{R}%
,\;\chi\right) \\
+g^{\mu\nu}\left(  \frac{1}{2}\partial_{\mu}\phi\partial_{\nu}\phi-D_{\mu
}H^{\dagger}D_{\nu}H\right) \\
-\left(  \frac{\lambda}{4}\left(  H^{\dagger}H-\omega^{2}\phi^{2}\right)
^{2}+\frac{\lambda^{\prime}}{4}\phi^{4}\right) \\
+\frac{1}{12}\left(  \phi^{2}-2H^{\dagger}H\right)  R\left(  g\right)
\end{array}
\right)  . \label{action}%
\end{equation}
This action is invariant under local scale transformations (Weyl symmetry) and
has a noteworthy unique coupling of conformal scalars to gravity of the form
that appears in the last line above. The relative minus sign in $\left(
\phi^{2}-2H^{\dagger}H\right)  R\left(  g\right)  $ is mandatory so that a
positive gravitational constant $G_{N}$ can be generated by Weyl gauge fixing,
$\frac{1}{12}\left(  \phi^{2}-2H^{\dagger}H\right)  \left(  x^{\mu}\right)
\rightarrow\left(  16\pi G_{N}\right)  ^{-1}$ at least in some patch of
spacetime $x^{\mu},$ but the relative sign is also essential for geodesic
completeness as outlined below. An attractive feature of the Weyl invariant
formulation is that the universe-filling dimensionful constants $G_{N},$ dark
energy $\Lambda$ and the electroweak scale $v_{EW},$ are also generated
simultaneously with $G_{N}$ from the same source \cite{BST}. The uniqueness
and completeness of this form for a Weyl invariant and geodesically complete
approach to the SM+GR was discussed in \cite{BST}, where its emergence from a
deeper gauge symmetry perspective of 2T-physics \cite{barsGrav2T} is also
outlined (for a summary see \cite{IB-cosmoSummary}). See also \cite{IB-sugra}
\cite{ferraraKallosh} \cite{BST} for the occurrence of the same structure in a
supergravity setting.

In this appendix, and with more detail in \cite{barsMiniSuper}, we re-examine
the mini-superspace derived from this theory for cosmological applications.
This was discussed in a series of papers during 2009-2014 in collaborations
between one of the authors of the current paper and C.H. Chen, Paul Steinhardt
and Neil Turok, as summarized in \cite{IB-cosmoSummary}. The
mini-superspace\ consists of the cosmologically most relevant homogeneous
(only time dependent) degrees of freedom, including scalar fields $\left(
\phi\left(  x^{0}\right)  ,h\left(  x^{0}\right)  \right)  ,$ where $h$
represents\footnote{In addition to the Higgs boson there may be more scalar
fields \cite{BST}. In that case the $h$ in mini-superspace represents a
combination of all the scalars. The most economical cosmological scenario is
to have just the Higgs, as this seems to be not impossible \cite{BST-Higgs}.}
the Higgs doublet in a unitary gauge, $H=\left(  0,~h/\sqrt{2}\right)  ,$ and
the cosmological metric, $ds^{2}=a^{2}\left(  x^{0}\right)  \left(  -\left(
dx^{0}\right)  ^{2}e^{2}\left(  x^{0}\right)  +\gamma_{ij}\left(  x^{0}%
,\vec{x}\right)  dx^{i}dx^{j}\right)  ,$ where $a$ is the cosmological scale
factor, $e $ is the lapse function (redefined up the factor $a,$ i.e. $N=ae$)
and $\gamma_{ij}\left(  x^{0},\vec{x}\right)  $ may include spacial curvature
and anisotropies. Moreover, the matter energy-momentum tensor $T_{00}$
includes the radiation density, $\rho_{r}\left(  x^{0}\right)  /a^{4}\left(
x^{0}\right)  ,~$to represent an average \textquotedblleft
fluid\textquotedblright\ behavior of all conformally invariant relativistic
matter (photons, gluons, quarks, leptons, neutrinos, etc.)$.$

The Weyl invariant form of the mini-superspace action was given in
\cite{BCSTcomplete} and \cite{IB-cosmoSummary}. Here we are interested in its
Weyl-fixed form in the so-called $\gamma$-gauge
\begin{equation}
S_{\text{mini}}=\int d\tau\left\{
\begin{array}
[c]{c}%
\frac{1}{2e}\left[  -\left(  \partial_{\tau}\phi_{\gamma}\right)  ^{2}+\left(
\partial_{\tau}h_{\gamma}\right)  ^{2}+\left(  \phi_{\gamma}^{2}-h_{\gamma
}^{2}\right)  \left(  \left(  \partial_{\tau}\alpha_{1}\right)  ^{2}+\left(
\partial_{\tau}\alpha_{2}\right)  ^{2}\right)  \right] \\
-e\left[  \phi_{\gamma}^{4}f\left(  \frac{h_{\gamma}}{\phi_{\gamma}}\right)
-\frac{1}{2}\left(  \phi_{\gamma}^{2}-h_{\gamma}^{2}\right)  v\left(
\alpha_{1},\alpha_{2}\right)  +\rho_{r}\right]
\end{array}
\right\}  , \label{minis}%
\end{equation}
where $\alpha_{1.2}\left(  \tau\right)  $ are anisotropy degres of freedom
with the anisotropy potential $v\left(  \alpha_{1},\alpha_{2}\right)  $ given
in \cite{BCSTcomplete}\cite{IB-cosmoSummary}; $\tau\equiv x^{0}$ is called the
\textquotedblleft conformal time\textquotedblright\ when the $e\left(
\tau\right)  =1$ gauge is chosen by fixing the $\tau$-reparametrization
symmetry of $S_{\text{mini}}$. The Weyl-symmetric version of $S_{\text{mini}}$
starts out with three Weyl-dependent degrees of freedom, namely $\left(
a,\phi,h\right)  ,$ that transform according to $\phi\rightarrow\Omega\phi,$
$h\rightarrow\Omega h,$ $a\rightarrow\Omega^{-1}a.$ The $e,\alpha_{1,2}$ and
$\rho_{r}$ degrees of freedom are Weyl invariant. Furthermore, $a\phi,ah$,
$h/\phi$ and arbitrary functions of these, are also Weyl invariant.

Physics depends only on Weyl invariants, but Weyl gauges that simplify
computations or clarify the physics are welcome. There is an interesting
interplay of four Weyl gauge choices: E-gauge, c-gauge, $\gamma$-gauge
\cite{BCSTcomplete}\cite{IB-cosmoSummary} and string gauge or s-gauge
\cite{nontrivialString}. The action above is in the $\gamma$-gauge which is
defined by freezing the scale factor for all conformal times $\tau$, and
labelling the gauge dependent quantities with $\gamma$ when they are in the
$\gamma$-gauge, namely $a_{\gamma}\left(  \tau\right)  =1,$ and dynamical
$\phi_{\gamma}\left(  \tau\right)  ,h_{\gamma}\left(  \tau\right)  .$ So,
$\left(  \phi_{\gamma},h_{\gamma}\right)  $ are gauge invariant since they can
be written as, $\phi_{\gamma}=a_{\gamma}\phi_{\gamma}=a\phi$ and $h_{\gamma
}=a_{\gamma}h_{\gamma}=ah,$ where $\left(  a\phi,ah\right)  $ may be evaluated
in any other gauge (see below for the case of the E-gauge). The $\gamma$-gauge
is most useful to grasp the geodesic completeness and transitions through
singularities (see e.g.\cite{IB-cosmoSummary} and \cite{crunchBang}%
,\cite{BST-Higgs}\cite{sailing}). Note the light-cone-type structure in
$\left(  \phi_{\gamma},h_{\gamma}\right)  $ field space in Fig.(10) where, in
accordance with the signatures in Eq.(\ref{minis}), the fields $\phi_{\gamma
}\left(  \tau\right)  $ ($h_{\gamma}\left(  \tau\right)  $) play the role of
timelike (spacelike) coordinates (just like $x^{\mu}\left(  \tau\right)  $ in
Eq.(\ref{Sg})).
\begin{center}
\includegraphics[
height=1.9077in,
width=1.9077in
]%
{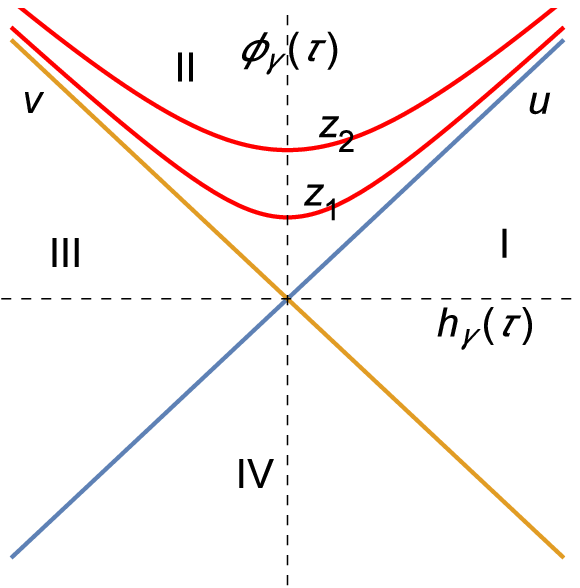}%
\\
Fig.(10) - The $\left(  \phi_{\gamma}\left(  \tau\right)  ,h_{\gamma}\left(
\tau\right)  \right)  $ field space.
\end{center}
We may define $u\equiv\phi_{\gamma}+h_{\gamma}$ and $v\equiv\phi_{\gamma
}-h_{\gamma} $ analogous to lightcone coordinates. The quantity $z\left(
\tau\right)  =\left(  \phi_{\gamma}^{2}\left(  \tau\right)  -h_{\gamma}%
^{2}\left(  \tau\right)  \right)  =u\left(  \tau\right)  v\left(  \tau\right)
$ is positive in regions II\&IV and negative in regions I\&III, while the blue
and orange solid lines, where either $u$ or $v$ vanish, indicate where
$z\left(  \tau\right)  $ vanishes. The hyperbolas in region II labeled by
$0<z_{1}<z_{2}<\infty$ correspond to the curves $\left(  \phi_{\gamma
},h_{\gamma}\right)  |_{z~\text{fixed}}$ for two fixed values of the field
$z\left(  \tau\right)  $; imagine similar hyperbolas in all regions I-IV. The
analogy to the extended Rindler space in Fig.(1) is already apparent. We will
soon explain more precisely the physical relation of the $\left(  \phi
_{\gamma},h_{\gamma}\right)  $ \textit{field-space} to the mathematical
structure of the extended Rindler spacetime discussed in the main body of the paper.

The E-gauge, which puts the full action (\ref{action}) directly in the
Einstein frame, is useful for interpreting the physics because traditionally
physics is discussed in the E-frame. It is defined by freezing the Weyl
invariant, $\int\frac{1}{12}\sqrt{-g}\left(  \phi^{2}-2H^{\dagger}H\right)
R\left(  g\right)  $, to the Einstein-Hilbert form, $\int\left(  \pm16\pi
G_{N}\right)  ^{-1}\sqrt{-g_{E}}R\left(  g_{E}\right)  ,$ where the Weyl-fixed
fields are labeled with an extra letter \textquotedblleft E\textquotedblright,
such as $g_{\mu\nu}^{E},\phi_{E},H_{E}$ to indicate that they are in the
E-gauge. The overall sign, $\pm1=$sign$\left(  \phi^{2}\left(  x\right)
-2H^{\dagger}\left(  x\right)  H\left(  x\right)  \right)  ,$ implies that
there are patches of field space $\left(  \phi,h\right)  ^{\pm}, $ and
corresponding regions of spacetimes $x^{\mu},$ where the E-gauge condition is
satisfied \cite{IB-cosmoSummary}. The $\pm$ signs, which imply a passage
through zero or infinity, are Weyl-invariant because the sign of $\left(
\phi^{2}\left(  x\right)  -2H^{\dagger}\left(  x\right)  H\left(  x\right)
\right)  $ cannot be changed by Weyl transformations. One may ask if a
universe can be complete in a patch with only the + sign. The answer is no,
because the sign$\left(  \phi^{2}-2H^{\dagger}H\right)  $ does flip
dynamically multiple times very generically as a function of $x^{\mu},$ as was
established with an extensive study of analytic solutions in
\cite{BCSTcomplete}\cite{IB-cosmoSummary}. The dynamics show that the field
solutions, and similarly the geodesics, are stopped artificially if only one
sign of $\left(  \phi^{2}-2H^{\dagger}H\right)  $ is imposed by hand. Hence,
quite clearly the traditional SM+GR, that artificially keeps only the positive
sign, is a geodesically incomplete theory. When both signs are kept to
complete the E-gauge field solutions and geodesics, the suddenness of the sign
flip, is just an artifact of the E-gauge. By contrast, the sign change occurs
smoothly in other gauges, such as the $\gamma$-gauge or the c-gauge. We see
that, as compared to the traditional GR+SM, the Weyl invariant GR+SM in
(\ref{action}) describes a larger field space for the same degrees of freedom,
as well as a corresponding larger spacetime. This is how geodesic completeness
is achieved.

Accordingly, in the geodesically complete E-gauge, that freezes $\frac{1}%
{12}\left(  \phi_{E}^{2}\left(  \tau\right)  -h_{E}^{2}\left(  \tau\right)
\right)  =\left(  \pm16\pi G_{N}\right)  ^{-1}$, the mini superspace degrees
of freedom include the two fields $\left(  a_{E}\left(  \tau\right)
,\sigma_{E}\left(  \tau\right)  \right)  $ instead of the three fields
$\left(  a,\phi,h\right)  .$ Here $a_{E}\left(  \tau\right)  $ is the scale
factor and the scalar $\sigma_{E}\left(  \tau\right)  $ is basically a
rewriting of the Higgs in the E-gauge. Naturally, $\left(  a_{E},\sigma
_{E}\right)  $ are related to the $\gamma$-gauge dynamical degrees of freedom
$\left(  \phi_{\gamma},h_{\gamma}\right)  $ by Weyl transformations as given
in \cite{BCSTcomplete}\cite{IB-cosmoSummary}. Consider the Weyl invariants
$a^{2}\left(  \phi^{2}-h^{2}\right)  $ and $\ln\left(  \frac{\phi-h}{\phi
+h}\right)  ;$ by evaluating them in the E-gauge and $\gamma$-gauge and
equating them to each other we find
\begin{equation}
\frac{12~a_{E}^{2}\left(  \tau\right)  }{16\pi G_{N}}=\left\vert \phi_{\gamma
}^{2}\left(  \tau\right)  -h_{\gamma}^{2}\left(  \tau\right)  \right\vert
=\left\vert z\left(  \tau\right)  \right\vert ,\;\sqrt{\frac{12}{16\pi G_{N}}%
}\sigma_{E}\left(  \tau\right)  =\frac{1}{2}\ln\left\vert \frac{\phi_{\gamma
}\left(  \tau\right)  +h_{\gamma}\left(  \tau\right)  }{\phi_{\gamma}\left(
\tau\right)  -h_{\gamma}\left(  \tau\right)  }\right\vert . \label{transforma}%
\end{equation}
This relation is the exact analog of the Rindler-Minkowski relation in
Eq.(\ref{tyRindleruv}); it shows that $\left(  \phi_{\gamma}\pm h_{\gamma
}\right)  $ or $\left(  u,v\right)  $ are Minkowski-like global coordinates in
Fig.(10), while $\left(  \sigma_{E},a_{E}^{2}\right)  $ are non-global
Rindler-like coordinates similar to $\left(  t,y\right)  $ that reparametrize
the four different patches I-IV. Indeed there is a precise correspondence to
the Minkowski and Rindler coordinates used in the rest of this paper; the
translation dictionary is%
\begin{equation}%
\begin{array}
[c]{ll}%
\frac{12~a_{E}^{2}\text{sign}\left(  \phi^{2}-h^{2}\right)  }{16\pi G_{N}%
}=z\leftrightarrow-2y,\;\; & \sqrt{\frac{12}{16\pi G_{N}}}\sigma
_{E}\leftrightarrow t,\\
\left(  \phi_{\gamma}+h_{\gamma}\right)  =u\leftrightarrow\left(  x^{0}%
+x^{1}\right)  ,\; & \left(  \phi_{\gamma}-h_{\gamma}\right)
=v\leftrightarrow\left(  x^{0}-x^{1}\right)  .
\end{array}
\label{correspondence}%
\end{equation}
Then we can insert this information in Eq.(\ref{uv-tyRegions}) to establish
the E-gauge to $\gamma$-gauge relations for every region I-IV in Fig.(10) in
exact correspondence to Fig.(1). With this, we now have a precise
Rindler$\leftrightarrow$Minkowski type map for our cosmological degrees of
freedom $\left(  u,v\right)  $ versus $\left(  \sigma,z\right)  $. This shows
that the cosmological geometry in field space has the same properties as
ordinary extended Rindler spacetime discussed in this paper, but now there are
also interactions that make it much more interesting.

The E-gauge to/from $\gamma$-gauge map described above is helpful to transform
the smooth $\gamma$-gauge solutions \cite{BCSTcomplete}\cite{crunchBang}%
\cite{IB-cosmoSummary} to the geodesically complete but singular E-gauge
solutions and vice-versa. It is then understood that at the instant $z\left(
\tau\right)  =\left(  \phi_{\gamma}^{2}\left(  \tau\right)  -h_{\gamma}%
^{2}\left(  \tau\right)  \right)  =u\left(  \tau\right)  v\left(  \tau\right)
$ vanishes in the $\gamma$-gauge, there is a scalar-curvature singularity in
the E-gauge where $a_{E}^{2}\left(  \tau\right)  =0$ at the same $\tau$
(although not so in $\gamma$-gauge where $a_{\gamma}\left(  \tau\right)  =1$
for all $\tau$). Hence in Fig.(10) the \textquotedblleft
horizons\textquotedblright\ at $u=0$ or $v=0$ correspond to big-crunch or
big-bang instants as interpreted in the E-frame. Also during the periods of
$\tau$ when the quantity $z\left(  \tau\right)  =\left(  \phi_{\gamma}%
^{2}\left(  \tau\right)  -h_{\gamma}^{2}\left(  \tau\right)  \right)
=u\left(  \tau\right)  v\left(  \tau\right)  $ is positive (negative) in the
$\gamma$-gauge, the sign$\left(  \phi^{2}\left(  \tau\right)  -h^{2}\left(
\tau\right)  \right)  $ in any Weyl gauge, including in the E-gauge $\left(
\phi_{E}^{2}\left(  \tau\right)  -h_{E}^{2}\left(  \tau\right)  \right)
=\left(  \pm16\pi G_{N}\right)  ^{-1}$, must be the same sign as sign$\left(
\phi_{\gamma}^{2}\left(  \tau\right)  -h_{\gamma}^{2}\left(  \tau\right)
\right)  $, since Weyl transformations cannot change it. Therefore, in regions
II\&IV (versus I\&III) in Fig.(10), gravity is an attractive (repulsive) force
as interpreted in the E-frame ($+G_{N}$ versus $-G_{N}$). The constants
$z_{1,2}$ that label the hyperbolas in region II correspond to two fixed
values of the scale factor at two instances $z_{1,2}\sim a_{E}^{2}\left(
\tau_{1,2}\right)  .$ So the successive hyperbolas in region II describe the
expanding universe as $\tau$ changes, while similar hyperbolas in region IV
describe a contracting universe in a region of ordinary gravity (+ sign in
E-gauge). By contrast, regions I\&III are antigravity regions that are
unavoidably probed by geodesically complete generic cosmological solutions as
shown in \cite{BCSTcomplete}\cite{crunchBang}\cite{IB-cosmoSummary}, as well
as by the quantum wavefunction of mini-superspace. Therefore, all four regions
are required in a \textit{geodesically complete theory} of SM+GR.

We are now ready for the connection of the mini-superspace in $S_{\text{mini}%
}$ with the multiverse ideas discussed in the current paper. The dynamics of
the cosmological fields in S$_{\text{mini }}$ in Eq.(\ref{minis}) may be
compared to the dynamics of a \textquotedblleft particle\textquotedblright\ on
the worldline parametrized by $\tau$ (like Eq.(\ref{Sg})). The target
spacetime is four dimensional, $X^{\mu}\sim\left(  \phi_{\gamma},h_{\gamma
},\alpha_{1},\alpha_{2}\right)  ;$ the \textquotedblleft
particle\textquotedblright\ (i.e. the universe) moves in a background
gravitational field with metric
\begin{equation}%
\begin{array}
[c]{l}%
ds^{2}=-d\phi_{\gamma}^{2}+dh_{\gamma}^{2}+\left(  \phi_{\gamma}^{2}%
-h_{\gamma}^{2}\right)  \left(  d\alpha_{1}^{2}+d\alpha_{2}^{2}\right)
=-dudv+uv\left(  d\alpha_{1}^{2}+d\alpha_{2}^{2}\right) \\
\;\;=-\frac{1}{4z}dz^{2}+z\left(  d\sigma^{2}+d\alpha_{1}^{2}+d\alpha_{2}%
^{2}\right)
\end{array}
\end{equation}
Note this is a conformally flat metric in field space. The scalar curvature is
$R=6\left(  \phi_{\gamma}^{2}-h_{\gamma}^{2}\right)  ^{-1}.$ There is also a
potential energy,
\begin{equation}
\tilde{V}=[\phi_{\gamma}^{4}f\left(  h_{\gamma}/\phi_{\gamma}\right)
-\frac{1}{2}\left(  \phi_{\gamma}^{2}-h_{\gamma}^{2}\right)  v\left(
\alpha_{1},\alpha_{2}\right)  +\rho_{r}]=[z^{2}v\left(  \sigma\right)
-\frac{1}{2}zv\left(  \alpha_{1},\alpha_{2}\right)  +\rho_{r}],
\end{equation}
where a constant $\rho_{r}>0$ plays the role of \textquotedblleft mass$^{2}%
$\textquotedblright, thus generalizing Eq.(\ref{Sg}) with additional
interactions. Note that $z$ (equivalently the scale factor $a_{E}^{2}$) plays
the role of Rindler \textquotedblleft time\textquotedblright\ in the gravity
regions II\&IV\footnote{Compare a similar timelike role of $2y<0$ in Rindler
regions II\&IV that was explained following Eq.(\ref{rhoRin}). This played a
crucial role in the treatment and interpretation of Eqs.(\ref{rhoRin}%
,\ref{drhoRin}).} where $z>0.$ In the antigravity regions I\&III, where $z<0,$
the overall sign of the metric seems to be wrong, but this is simply
equivalent to replacing $G_{N}$ by $-G_{N}$ in the Einstein-Hilbert
Lagrangian, so the meaning of the overall sign is physically interpreted as
being in the gravity versus antigravity patches of the E-gauge. See
\cite{barsArayaJames}\cite{barsJames} for further applications and
interpretations of this overall sign switch of the metric in the E-gauge.

The quantum wavefunction satisfies the Wheeler-deWitt equation (WdWe) that is
derived from S$_{\text{mini }}$ in Eq.(\ref{minis}) just like
Eq.(\ref{laplace})\footnote{The ordering ambiguity of canonical variables
allows an additional term in the Laplacian, i.e. instead of $\nabla^{2}$
consider $\left(  \nabla^{2}-\xi R\right)  $ where $R$ is the curvature of the
metric in field space. In the following equations taken from \cite{barsJames}
the conformally exact choice $\xi=1/6$ was made, and then the equation was
simplified by rescaling the wavefunction $\Phi$ with a factor, $\Psi=\left(
\phi_{\gamma}^{2}-h_{\gamma}^{2}\right)  ^{1/2}\Phi$ to simplify it to the
form of Eq.(\ref{cosmoPsi}).}. In either the Minkowski-like $\left(
\phi_{\gamma},h_{\gamma}\right)  \leftrightarrow\left(  u,v\right)  $ or the
Rindler-like $\left(  \sigma,z\right)  $ coordinate systems, the WdWe was
constructed and analyzed in \cite{barsJames}, where the physical meaning of an
antigravity region behind cosmological singularities, as interpreted by
observers in the gravity regions, and the related issues of unitarity (no
problem), were discussed. Explicitly, the WdWe written in both coordinate
systems is given by
\begin{equation}%
\begin{array}
[c]{l}%
\left(
\begin{array}
[c]{c}%
\frac{1}{2}\left(  \partial_{\phi_{\gamma}}^{2}-\partial_{h_{\gamma}}%
^{2}\right)  -\frac{1}{2\left(  \phi_{\gamma}^{2}-h_{\gamma}^{2}\right)
}\left(  \partial_{\alpha_{1}}^{2}+\partial_{\alpha_{2}}^{2}\right)  +\rho
_{r}\\
+\frac{1}{2}\phi_{\gamma}^{4}f\left(  \frac{h_{\gamma}}{\phi_{\gamma}}\right)
-\frac{1}{2}\left(  \phi_{\gamma}^{2}-h_{\gamma}^{2}\right)  v\left(
\alpha_{1},\alpha_{2}\right)
\end{array}
\right)  \Psi\left(  \phi_{\gamma},h_{\gamma},\alpha_{1},\alpha_{2}\right)
=0,\\
\left(
\begin{array}
[c]{c}%
\partial_{z}^{2}+\frac{1}{4z^{2}}\left(  -\partial_{\alpha_{1}}^{2}%
-\partial_{\alpha_{2}}^{2}-\partial_{\sigma}^{2}+1\right) \\
+\frac{z}{2}v\left(  \sigma\right)  -\frac{1}{4}v\left(  \alpha_{1},\alpha
_{2}\right)  +\frac{\rho_{r}}{2z}%
\end{array}
\right)  \left(  z^{1/2}\Psi\left(  z,\sigma,\alpha_{1},\alpha_{2}\right)
\right)  =0.
\end{array}
\label{cosmoPsi}%
\end{equation}
Close to the singularity in the E-frame we have, $a_{E}^{2}\sim z\sim\left(
\phi_{\gamma}^{2}-h_{\gamma}^{2}\right)  \sim0,$ which is equivalent to being
close to the horizons in Fig.(10). In that neighborhood, assuming that the
terms ($\frac{z}{2}v\left(  \sigma\right)  -\frac{1}{4}v\left(  \alpha
_{1},\alpha_{2}\right)  $) can be neglected compared to the dominant and
subdominant $z^{-2},z^{-1}$ terms, the the wavefunction may be determined from
the approximate equation%
\begin{equation}%
\begin{array}
[c]{l}%
\left(  \partial_{z}^{2}+\frac{1}{4z^{2}}\left(  -\partial_{\alpha_{1}}%
^{2}-\partial_{\alpha_{2}}^{2}-\partial_{\sigma}^{2}+1\right)  +\frac{\rho
_{r}}{2z}\right)  \left(  z^{1/2}\Psi\left(  z,\sigma,\alpha_{1},\alpha
_{2}\right)  \right)  =0,\text{ or}\\
\left(  \partial_{z}^{2}+\frac{1}{4z^{2}}\left(  p_{1}^{2}+p_{2}^{2}+p_{3}%
^{2}+1\right)  +\frac{\rho_{r}}{2z}\right)  \left(  z^{1/2}\psi_{p}\left(
z\right)  \right)  =0,
\end{array}
\label{zE}%
\end{equation}
where the second equation applies to solutions of separable form, $\Psi\left(
z,\sigma,\alpha_{1},\alpha_{2}\right)  \sim e^{-i\left(  p_{1}\alpha_{1}%
+p_{2}\alpha_{2}+p_{3}\sigma\right)  }\psi_{p}\left(  z\right)  .$ The general
wavepacket has a form analogous to the $\varphi_{1,2,3,4}$ of Eq.(\ref{list})
in various regions I-IV, and continuity across horizons is required. For
example, for region I, the general solution is
\begin{equation}
\Psi_{1}\left(  z,\sigma,\alpha_{1},\alpha_{2}\right)  =\sum_{\pm}\int
d^{3}p\;\left[  a\left(  \vec{p}\right)  ~e^{-i\left(  p_{1}\alpha_{1}%
+p_{2}\alpha_{2}+p_{3}\sigma\right)  }\psi_{p}^{\pm}\left(  z\right)
+hc\right]  .
\end{equation}
where $\psi_{p}^{\pm}\left(  z\right)  $ are the two independent solutions of
the simplified equation in the single variable $z$. The exact solutions are
known in this case (see below), but it is useful to first intuitively
understand their physical behavior in the union of the four regions by
comparing the $\partial_{z}^{2}+\cdots$ equation to a non-relativistic
Schr\"{o}dinger equation, $\left(  -\partial_{z}^{2}+V\left(  z\right)
\right)  \psi_{0}\left(  z\right)  =0,$ with a potential energy, $V\left(
z\right)  =-\frac{1}{4z^{2}}\left(  p_{1}^{2}+p_{2}^{2}+p_{3}^{2}+1\right)
-\frac{\rho_{r}}{2z}$, and a wavefunction $\psi_{0}\left(  z\right)  \equiv
z^{1/2}\psi_{p}\left(  z\right)  $ for the 0 eigenvalue. The plot of the
potential $V\left(  z\right)  $ is given in Fig.(11)%
\begin{center}
\includegraphics[
height=1.4446in,
width=2.3191in
]%
{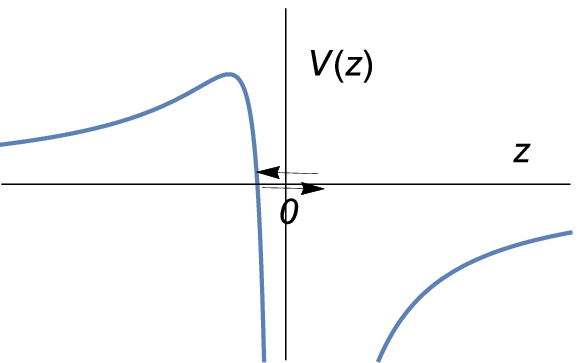}%
\\
Fig.(11) - Cosmological crunch and bang with antigravity in between.
\end{center}
The physical solution for $\psi_{0}\left(  z\right)  ,$ with correct boundary
conditions, can be described intuitively as a wavepacket approaching from the
region $z>0$ (a contracting universe in gravity region IV in Fig.10), passing
through $z=0$ (a cosmological crunch) and entering the antigravity region
where $z<0$, then necessarily reflecting from the barrier (that forms due to
radiation $\rho_{r}>0$) and unable to tunnel deep into negative values of $z$
(hence, spending little time in the antigravity region I or III in Fig. 10),
then passing through $z=0$ again (a cosmological big bang) and moving on to
the positive region $z>0$ (an expanding universe in gravity region II in
Fig.10). Thus, the exact wavefunction for the universe, which consists of
$\Psi_{1,2,3,4}\left(  z,\sigma,\alpha_{1},\alpha_{2}\right)  $ as described
above, should have appropriate boundary conditions that restrict the
coefficients $a_{1\pm}$ etc. to fit this physical behavior.

The exact analytic solution for the wavefunction $z^{1/2}\psi_{p}\left(
z\right)  $ confirms this expected behavior \cite{barsMiniSuper}. It should be
emphasized that this quantum behavior of a general wavepacket is in complete
agreement with the classical solution displayed in \cite{crunchBang} that
featured an attractor behavior for a cosmological bounce consisting of
Crunch-Bang transition with an antigravity region in between. As should be
expected, due to the fuzziness introduced by quantum mechanics, the passage
through the singularity in the E-frame at $a_{E}=0,$ is much softer in the
quantum version as compared to the classical version in \cite{crunchBang}.
This transition was managed in \cite{crunchBang} by using Weyl symmetry, while
in the quantum case here, it amounts to the continuity of the wavefunction at
the horizons just as discussed for the $\varphi_{1,2,3,4}$ in section
(\ref{fieldMR}).

Note that Fig.(11) is the same as Fig.(8) after replacing $z=-2y,$ and the
effective potential $V\left(  z\right)  $ is the same as $V_{eff}\left(
y\right)  $ in Eq.(\ref{VeffQ}) after renaming the parameters, $p_{1}%
^{2}+p_{2}^{2}+p_{3}^{2}=\omega^{2}$ and $\rho_{r}=\mu^{2}/2.$ Therefore, the
analytic solutions for the geodesically complete cosmological wavefunction
$z^{1/2}\psi_{p}\left(  z\right)  $ have exactly the same analyticity behavior
as the Rindler field $\varphi_{1,2,3,4}\left(  u,v\right)  $ given in
Eq.(\ref{list}). The physical boundary conditions (dying off wavefunction in
asymptotic antigravity regions I\&III) are reproduced by the horizon boundary
conditions (\ref{boundaA},\ref{F13asympt}) employed for the $\varphi
_{1,2,3,4}$ and can again be used here. We find that near $z=0,$ or
equivalently at the $u=0$ or $v=0$ horizons in Fig.(10), there are branch
points and associated branch cuts that lead to the same multiverse behavior
discussed in the main body of this paper.

What makes up a multiverse is the analytic properties of the wavefunction
that, via monodromy transformations, automatically contains different
coefficients on different levels of the multiverse resulting from the
canonical transformations like those in Eq.(\ref{can1},\ref{can2}). This
implies "discretized jumps" in probability for certain phenomena at different
levels of the multiverse. Further progress will be reported in
\cite{barsMiniSuper}.

In this way, we have demonstrated that there is the possibility of a new
cosmological multiverse in a geodesically complete cyclic-type cosmology. Now
there are interactions, so there remains to figure out if transitions between
the various levels of the cosmological multiverse can occur. In the context of
trying to determine the wavefunction for the universe, as in this appendix and
in \cite{barsMiniSuper}, the multiverse concept discussed in the main body of
the paper is more fitting and it is quite intriguing.

\end{document}